\documentclass[12pt,letterpaper]{article}

\usepackage{placeins}
\usepackage{subcaption}

\usepackage{setspace} 
\usepackage{footmisc}
\usepackage{amsfonts}
\usepackage{amsmath}
\usepackage{theorem}
\usepackage{float}

\usepackage{multirow}

\usepackage[top=2.5cm, bottom=2.5cm, left=2.5cm, right=2.5cm]{geometry}

\usepackage{amsmath}
\usepackage{xcolor}

\newcommand\bovermat[2]{%
	\makebox[0pt][l]{$\smash{\overbrace{\phantom{%
					\begin{matrix}#2\end{matrix}}}^{\text{#1}}}$}#2}

\usepackage{setspace}
\usepackage{footmisc}

\usepackage[natbibapa]{apacite}
\usepackage{amssymb}
\setcitestyle{aysep={}} 

\usepackage{mathtools}
\usepackage{amsmath}
\usepackage[latin2]{inputenc}
\usepackage{graphicx}
\usepackage{subcaption}
\usepackage{multirow}
\usepackage{esvect}
\usepackage{afterpage}

\usepackage{lipsum}
\usepackage{booktabs,array,dcolumn}

\let\ampersand\&
\renewcommand*\&{and}

\makeatletter
\def\@seccntformat#1{\@ifundefined{#1@cntformat}%
	{\csname the#1\endcsname\space}
	{\csname #1@cntformat\endcsname}}
\newcommand\section@cntformat{\thesection.\space}       
\newcommand\subsection@cntformat{\thesubsection.\space} 
\makeatother

\begin{document}
	
	
	
	\begin{titlepage}
		\date{}
		
		
		
		\title{\large 	{A new method for identifying what Cupid's invisible hand is doing. Is it spreading color blindness while turning  us more ``picky'' about spousal education?}}

		\singlespacing{\author{
				{Anna NASZODI*}  
				{Francisco MENDONCA}
		}	}

		\thanks{Corresponding author, email:  anna.naszodi@gmail.com\\
		This paper presents preliminary findings and is being distributed to economists and other interested readers solely to stimulate discussion and elicit comments. The views expressed in this paper are those of the authors and do not necessarily reflect the official views of the European Commission and the Central Bank of Hungary. Any errors or omissions are the responsibility of the authors.} 
		

		
		\maketitle
		\thispagestyle{empty}
		
		\noindent 
		\singlespacing{ 
		We develop a method suitable for detecting whether racial homophily is on the rise and also whether the economic divide (i.e., the gap between individuals with different education levels and thereby with different abilities to generate income) is growing in a society. We identify these changes with the changing aggregate marital preferences over the partners' race and education level through their effects on the share of inter-racial couples  and the share of educationally homogamous couples. These shares are shaped not only by preferences, but also by the distributions of marriageable men and women by traits. The method proposed is designed to control for changes in the trait distributions from one generation to another. By applying the method, we find the economic divide in the US to display a U-curve pattern between 1960 and 2010 followed by its slightly negative trend between 2010 and 2015. The identified trend of racial homophily suggests that the American society has become more and more permissive towards racial intermarriages since 1970. Finally, we refute the aggregate version of the status-cast exchange hypothesis based on the joint dynamics of the economic divide and the racial homophily.}


		\begin{flushleft}
		\end{flushleft}

	\end{titlepage}

	\newpage
	\setcounter{page}{1}
	\doublespacing
	\begin{center}
	\end{center}
	\begin{center}
		
		
		\LARGE{\textbf{A new method for identifying what Cupid's invisible hand is doing. Is it spreading color blindness while turning  us more ``picky'' about spousal education?}}

	\end{center}
	
	\newpage
	
	\begin{center}
		{\large Abstract
			
		}
	\end{center}
			We develop a method suitable for detecting if racial homophily is on the rise and also if the economic divide (i.e., the gap between individuals with different education levels and thereby with different abilities to generate income) is growing in a society. We identify these changes with the changing aggregate marital preferences over the partners' race and education level through their effects on the share of inter-racial couples  and the share of educationally homogamous couples. These shares are shaped not only by preferences, but also by the distributions of marriageable men and women by traits. The method proposed is designed to control for changes in the trait distributions from one generation to another. By applying the method, we find the economic divide in the US to display a U-curve pattern between 1960 and 2010 followed by its slightly negative trend between 2010 and 2015. The identified trend of racial homophily suggests that the American society has become more and more permissive towards racial intermarriages since 1970. Finally, we refute the aggregate version of the status-cast exchange hypothesis based on the joint dynamics of the economic divide and the racial homophily.

	\newpage
	
	\section{INTRODUCTION}
	
	There is a growing consensus in the literature over the historical trend in income and wealth inequality.
	   In particular, it is a widely held view that these dimensions of inequality exhibited a U-shaped pattern over the twentieth century in the US (see \citealp{PikettySaez2003},  \citealp{SaezZucman2016}).  
	   Even though there is still an ongoing debate about the exact shape of the trend, the stylized  U-curve pattern itself has not been challenged (see \citealp{Bricker2016}, \citealp{AutenSplinter2022},  \citealp{Geloso2022}). 
	   
	   The discourse in the educational assortative mating literature lags far behind the debate about wealth and income inequality. 
	   There is no consensus over the qualitative historical trends among those papers that identify the dynamics of inequality by analyzing changes in marital sorting along individual's income generating ability proxied by their final educational attainment (see \citealp{Rosenfeld2008}).  
	   
	   On the one hand, this is surprising, because unlike the studies on income and wealth inequality, the papers in the  assorative mating literature do not perform any perilous exercise of patching together data from different sources. Their main input,  the joint educational distribution of couples,  is provided by the statistical offices  ``packed up and parceled'' ready for analysis. 
Also, while under-reporting of income and wealth is a general concern of the researchers, under-reporting  of education level  is not.    	
On the other hand, the conflicting  findings in the assortative mating literature can be well explained by the  diversity of methods and indicators applied to identify what Cupid's invisible hand is doing. 

\textit{In this paper, we develop, apply and promote a new method that identifies changes in the degree of assortative mating along education and race jointly.}  
Our benchmark analysis is performed on US census data from IPUMS on marriages and cohabitations in 1980 and 1990.  
To complement the benchmark analysis, we also study more than five decades spanned by 1960 and 2015.  
We find the trend of the degree of sorting along the educational to be in accord with the U-shaped historical trend of income inequality irrespective of controlling for sorting along race, or not.  
Hopefully, our new method will contribute to the catching up of the assortative mating literature to the literature of income inequality.

	The conflicting nature of the empirical results in the educational assortative mating literature can easily be detected in studies analyzing the 1980s using US data. 
	For instance, \cite{Eika2019} find that   
	the degree of sorting has \textit{increased} gradually from 1940 to the 1980s, after which it has been \textit{stagnant}.
	\cite{Greenwood2014} find that it has continued to  \textit{increase} even after 1980.  

By contrast,  \cite{NaszodiMendonca2020} find the strength of aggregate preferences for educational homogamy  to have displayed a U-shape pattern over the second half of the twentieth century and the first decade of the twenty-first century:\footnote{We use the term 'aggregate preferences for homogamy'  interchangeably with the terms 'degree of sorting', 'degree of segmentation of the market', 'strength of social barriers to intermarriage', 'width of the social gap between different groups', because it is hardly possible to distinguish them empirically.} it has
	 \textit{decreased remarkably} over the 1980s,
	when the early boomers became less active on the marriage market and the late boomers entered this market.  
	They also document the \textit{substantial increase} in sorting in the 2000s, when the late generationX gradually replaced the early generationX on the market.	
	

The focus of the empirical analysis of this paper is on the most controversial decade, the 1980s. 
Our results support the findings of \cite{NaszodiMendonca2020}, although we apply a novel method relative to theirs. 
In particular, we control for sorting along race, while they do not.  
Due to this substantial difference, it is absolutely not straightforward that our empirical results should be even similar to theirs.

To gain insight on the importance of controlling for sorting along race, consider a hypothetical society with all men and all women marry someone from the opposite sex from their own generation when being young adults. 
	There are two generations in this society: the early generation and the late generation.     
	The education level of the late generation is not different from that of the early generation.  
	Moreover, people of a particular race  tend to have higher educational attainment than people of another race in both of the  generations.  
	
	In addition, we make the following assumptions: (i) those belonging to the late generation are exactly as much ``picky'' about spousal education as the members of the early generation were when being young adults;          
	and (ii) barriers to racial intermarriage have been reduced from one generation to the next, i.e.,  less members of the late generation than in the early generation have opposed to marry someone of a different race.       
	Under these assumptions one can mistakenly find the degree of educational sorting to be decreasing in this hypothetical society if not controlling for changes in sorting along race.\footnote{Conversely, 
		suppose that aggregate preferences for well-educated partners are stronger in the late generation than those were in the early generation. Further, let us assume that marital racial preferences are exactly the same in the two generations. 
	Then, one can mistakenly find that racial homophily is on the rise in this hypothetical society by studying inter-racial marriages without controlling for the increase in the degree of sorting along the educational.}
		



	Finally, an additional motive for studying sorting along race and education jointly is this. 
	The assumption in point (ii) seems realistic based on NORC's General Social Survey conducted in 2017.  
	The survey shows that those belonging to the generations born later, are less likely to oppose having a family member marry someone of a different race. 

In this paper, we find that the \textit{aggregate racial preferences revealed on the marriage market}  are similar to those stated in the General Social Survey: both rank the late boomers to be more permissive towards inter-racial marriages in comparison with the early boomers.  Also, we find that Cupid's invisible hand continued to spread color blindness even after 1990 when the members of the generationX gradually replaced the boomers on the marriage market; and well after 1990 when the early Millennials arrived to the market.  \cite{CaseDeaton2021} find one the three Fates, Decima to have acted similarly to Cupid: they identify the racial gap in life expectancy to have also narrowed between 1990 and 2018.

As to the economic divide, we find that the identified change in \textit{revealed preferences for spousal education} is not sensitive to controlling for changes in racial preferences: \textit{late boomers are found to be more permissive towards inter-educational marriages as well irrespective to the model specification.}   
We stress that it is far not a trivial finding in the light of the hypothetical society exemplifying the omitted variable bias with correlating racial and educational dimensions.


Assessing how the revealed marital preferences change over the spousal education is not straightforward for the following reason:  
we can rarely observe preferences directly.  
However, we can identify changes in preferences through their effects on the outcome of the matching process.   

	
	The equilibrium on 
	the 
	marriage market depends not only on the marital preferences, but also on the structural availability of potential 
	partners with various traits. 
	In reality, the race-specific and gender-specific educational distributions vary a lot across generations. 
Moreover, these changes can induce changes in preferences and vice versa.   
	So, to identify the revealed preferences for spousal education,  
	we have to control not only for racial preferences, 
	but also  for the traits' distributions, as well as for some interaction effects.\footnote{In addition, 
	identifying these preferences is even more challenging once the possibility of remaining \textit{single} is also taken into account.   
	About this point, see \cite{NaszodiMendonca2019}. They account for single people in a  sophisticated way by distinguishing  between ``singles by choice'' (who do not even look for a partner) and ``singles by chance'' (who are not successful at finding a partner acceptable by them).}   

	 
Some of the \textit{contributions} of this paper are methodological and conceptual. 
First, we identify the revealed preferences for spousal education by  controlling 
for all the confounding factors listed above.    
As we will see, it is a common but dubious practice in the literature to either fail to control for each of these factors or to control for them inadequately. 


Second, we do not apply the popular approach (followed by \citealp{BreenSalazar2011}, \citealp{Kremer1997}, \citealp{Eika2019} inter alia) of defining the \textit{marriage--inequality nexus} as the potential effect of changing assortativity on a monetary dimension of inequality.   
Rather, our paper joins the strand of the literature according to which changes in sorting itself is considered to reflect a change in a specific dimension of inequality.  
This specific dimension of inequality is seen by social stratification researchers as an indicator of the social gap between different education groups (see \citealp{Katrnak2012}).

A related point to the analyzed type of marriage--inequality nexus is this. 
Typically, the degree of marital sorting is characterized by a matrix-valued assortativity measure that is difficult to be used for inter-temporal and cross-country comparisons due to its multi-dimensional nature. 
We admit that it is useful to transform any matrix-valued assortativity measure, -- be it the matrix of odds-ratios forming the basis of the Althman's index  or any of its alternatives --,  
to a scalar-valued measure before any comparative analysis.  
However, the scalar-valued measure, that the matrix-valued assortativity measure is projected on, should not necessarily be a measure of income or wealth inequality.   

We find it much more appealing to project the sorting measure on the scalar-valued  share of homogamous couples.\footnote{This approach is also followed  by  \cite{NaszodiMendonca2020}.}    
This operation allows us to study a non-monetary dimension of inequality: the contribution of the changing marital preferences to the prevalence of homogamy.  
A positive (/negative) value of this indicator signals that the overall inequality (covering all dimensions relevant on the marriage market including income, wealth, health, etc) 
is growing (/diminishing) between the groups studied.    

Performing comparative analysis with this marital educational inequality indicator (henceforth, MEI-indicator) has many advantages.  
Most importantly, the marriage data are far more comparable both across countries and over time than the income data and the wealth data since the former are not subject to various measurement issues. 

The main methodological contribution of our paper is this. 
We develop a new method for quantifying changes in aggregate marital preferences over two traits, e.g.,  race and education level,  
	or any other pair of traits characterized by a dichotomous variable and an ordered categorical variable. 
The method we propose is a generalized version of the method developed by \cite{NaszodiMendonca2020}.  
The latter works under the assumption that people sort into marriages along a one-dimensional trait, e.g., education level.   
Henceforth, we refer to our generalized new method as the GNM-method, while  we refer to the original method as the NM-method.\footnote{See: \url{https://en.wikipedia.org/wiki/NM-method}.}

    

Generalizing the NM-method  is not as simple as adding a new explanatory variable to a (log-)linear regression model, because not even the NM-method is  (log-)linear. 
Also, its generalization is more complicated than applying the original NM for different racial groups separately. 
Such a simple approach can be limitedly suitable provided the marriage market is not segmented perfectly along race.

Our paper is not the first that aims at meeting the challenge of analyzing sorting along more than one dimension.  
	The papers by \cite{Chiappori2011}, \cite{GalichonSalanie2010},  \cite{NaszodiMendonca2019}  and \cite{Rosenfeld2008}  
	also belong to the \textit{multidimensional matching} strand of the assortative mating literature.

	\cite{Chiappori2011} provide a closed-form solution of a multidimensional matching model and then they test predictions of 
	how spouses trade off education and non-smoking.  
	In the model by \cite{GalichonSalanie2010}, the surplus from a marriage match depends on the partners' race,  education, 
	and some  other traits unobserved by the econometrician.       
In the full-fledged micro-founded model of  \cite{NaszodiMendonca2019},  matches are made with the Gale--Shapley (henceforth GS) algorithm and each individual is assumed to sort along two characteristics:  the marriageable person's educational attainment, and his or her reservation point (used as a proxy for the unobserved traits of the person and empirically identified by the search criteria of a group of dating site users). 
    Finally, \cite{Rosenfeld2008} examines sorting along three dimensions (race, education, and religion) in the US. 

	The  distinctive feature of this paper compared to most papers in the multidimensional marriage matching literature and also  
     	papers by \cite{Greenwood2014}  and \cite{Eika2019} in the single-dimensional matching literature  	is that our paper  builds on a \textit{different measure of aggregate marital preferences}.   
	 The measure we use was first proposed to be applied in the context of assortative mating  by \cite{LiuLu2006}. It is a slightly modified version  of the Coleman-index (see Eq.15 in \citealp{Coleman1958}).     
	 The Liu--Lu measure (henceforth LL-measure) forms the basis of the NM-method and the GNM-method as well. It is different from the \textit{conventional measures}, such as 
	the regression coefficient (applied by \citealp{Greenwood2014}),  
	the (generalized) marital sorting parameter (proposed and applied by \citealp{Eika2019}), 
	the marital surplus (developed by \citealp{ChooSiow2006}, while generalized and applied by \citealp{Chiappori2011}, \citealp{GalichonSalanie2010}), 
	and the odds-ratio (applied  by \citealp{Rosenfeld2008} inter alia).

What motivates us to use the LL-measure and not one of its alternatives?  
In part, the findings of our supplementary analysis (to be presented in Subsection \ref{sec:NM_Prop}) form the basis of our {measure-selection}. 
In the analysis, we impose the \textit{criterion} against each martial sorting measure to be  
 monotonously decreasing in intergenerational mobility. 
    The intuition behind our criterion is that a society, where the pauper's son has higher chance 
    to became the prince than in other societies,  cannot be less open to accept marriages between paupers and princesses in comparison with other societies.  
    As we will see, this \textit{monotonicity condition} is violated by many of the well-known measures of marital sorting, but the LL-measure.      



Our choice of characterizing aggregate marital preferences with the LL-measure  is  also \textit{motivated} by the  fact that this measure has already been shown to have \textit{attractive analytical and empirical properties} in the single-dimensional assorted trait framework.  
	In particular, \cite{LiuLu2006} claim that their measure can control for changes in the trait distribution, while other measures cannot.  
	
	
	In addition, \cite{NaszodiMendonca2020}  show that  among a comprehensive set of assortativity measures and models, including the odds-ratio and the \cite{ChooSiow2006} model,  only the LL-measure and the \cite{LiuLu2006} model characterize the revealed marital preferences consistently with a  survey  on individuals' declared marital preferences.   		
To wrap up this section,  we visit again the \textit{main empirical findings} together with the main  \textit{contributions} of our paper.      
(i) We find both the racial segmentation and the educational segmentation of the American marriage market to have been declined over the 1980s.       
(ii) Thereby, we confirm the finding of \cite{NaszodiMendonca2020}: 
young adults in  1990 (belonging to the generation of late boomers) are found to be  less ``picky'' with respect to their spouses' education level than young adults have been in 1980  (belonging to the early boomers) irrespective of controlling for sorting along race, or not.      
(iii) Consistently with the income inequality literature,  we find that marital educational inequality displayed a U-curve pattern between 1960 and 2010 in the US.
	 	
As to the contributions of this paper, 
(i) we propose and apply a new method  allowing us to identify the trend of sorting along spousal education by controlling for sorting along race (and also, the other way around).      
(ii) Our supplementary analysis presents some new considerations in favor of characterizing the degree of sorting  by the LL-measure.  
(iii) Finally, we interpret our empirical findings in a broad context. In the introduction, we already  shed  light on their relationship with the income inequality literature and the demography literature identifying social gaps by the differences in group-specific life expectancies. In Section \ref{sec:disc} we discuss their relationship with  the status-cast exchange hypothesis, some survey evidence, and a historian's narrative of the New Deal order and the neoliberal order.

	The rest of the paper is \textit{structured} as follows.   
	Section \ref{sec:eq} describes how we characterize the equilibrium in the marriage market.      
	Section \ref{sec:met}  presents the  method designed for identifying changes in marital preferences:  
	it introduces the original NM-method (applicable to study sorting along a single dimension), motivates its use,  
	 and develops the GNM-method 
	(suitable for studying sorting along multiple dimensions).\footnote{Readers already familiar with the NM may turn directly to Subsection \ref{sec:GNM}.} 
	Section \ref{sec:emp} applies the GNM-method using US census data.  
	In Section \ref{sec:disc}, we discuss the significance of the empirical findings.   
	Finally, Section \ref{sec:concl} concludes the paper.

	\section{CHARACTERIZING THE EQUILIBRIUM IN THE MARRIAGE MARKET}\label{sec:eq} 
	
	In this study, marriage is interpreted broadly: no distinction is made between married and unmarried couples.
	Accordingly, by  ``wives'' and ``husbands'' we also mean romantic heterosexual cohabiting partners. 

	In our analysis, the \textit{educational trait} variable is an ordered  categorical variable  that can take three possible values.    
	Its value $L$ stands for ``low level of education'' corresponding to not having completed the high school;  
	$M$ denotes ``medium level of education'' corresponding to having a high school degree, but 
	neither a college degree nor a university degree; 
	and $H$ stands for ``high level of education'' corresponding to holding at least a BA diploma.

	Our  \textit{race variable} can take two possible values.  
	In the benchmark specification,  race is either Black $(B)$, or White $(W)$.  The reason 
       for not considering more than two racial categories at a time is three-fold. 
       First,  we want to keep the number of model parameters reasonably low. Second, once a trait is allowed to be multinomial, the NM-method works only if this trait variable is ordered. It is definitely not the case with race. 
       Third, as it will be shown by the sensitivity analysis in Appendix B, our findings are robust to some alternative choices of the dichotomous race variable. 

	Accordingly, in the benchmark case, we  characterize the  matching outcome by Table \ref{tab:CT}.   
	We denote this \textit{contingency table}  by $K$.  
	Its element  $N_{h,w}$ is the number of  $h,w$-type marriages  with $h, w \in  \{WL, WM, WH, BL, BM, BH \}$, where $h$ denotes  the husbands' type and $w$ denotes  the wives' type. 
	One's type is given by one's race and education level.  
	
		\begin{center}
	\setlength{\tabcolsep}{2pt}
	\begin{table}[!htb]
		\caption{The $K$ contingency table}
		
						\begin{center}								
				\begin{tabular}{lllllllllll}	
					\hline	\hline							
					& &    & &   \multicolumn{6}{c}{Wife/female partner} &   \\ \cline{5-10}
					& &    & &   \multicolumn{3}{c}{Black} & \multicolumn{3}{c}{White} & \\ \cmidrule(lr){5-7} \cmidrule(lr){8-10} 
					& &    &Edu. & \multicolumn{1}{c}{$L$}   & \multicolumn{1}{c}{$M$} & \multicolumn{1}{c}{$H$} & \multicolumn{1}{c}{$L$}   & \multicolumn{1}{c}{$M$} & \multicolumn{1}{c}{$H$}&         $^{\text{{\normalsize{Total}}}}$ \\		\hline
					
					\multicolumn{1}{c}{\multirow{6}{*}{\rotatebox[origin=c]{90}{Husband/}}} &  \multicolumn{1}{c}{\multirow{6}{*}{\rotatebox[origin=c]{90}{\underline{male partner}}}}& \multicolumn{1}{c}{\multirow{3}{*}{\rotatebox[origin=c]{90}{\underline{Black}}}} & 
					$L$&	$N_{BL,BL}$	& 	$N_{BL,BM}$	& 	$N_{BL,BH}$	&	$N_{BL,WL}$	& 	$N_{BL,WM}$	& 	$N_{BL,WH}$& 	$N_{BL,\cdot}$	\\
					& &\multicolumn{1}{l}{}& 
					$M$&	$N_{BM,BL}$	& 	$N_{BM,BM}$	& 	$N_{BM,BH}$	&	$N_{BM,WL}$	& 	$N_{BM,WM}$	& 	$N_{BM,WH}$& 	$N_{BM,\cdot}$	\\
					& &\multicolumn{1}{l}{}&
					$H$&	$N_{BH,BL}$	& 	$N_{BH,BM}$	& 	$N_{BH,BH}$	&	$N_{BH,WL}$	& 	$N_{BH,WM}$	& 	$N_{BH,WH}$	& 	$N_{BH,\cdot}$	\\
					
					& & \multicolumn{1}{c}{\multirow{3}{*}{\rotatebox[origin=c]{90}{\underline{White}}}} & 
					$L$&	$N_{WL,BL}$	& 	$N_{WL,BM}$	& 	$N_{WL,BH}$&	$N_{WL,WL}$	& 	$N_{WL,WM}$	& 	$N_{WL,WH}$	& 	$N_{WL,\cdot}$	\\
					& &\multicolumn{1}{l}{}&
					$M$&	$N_{WM,BL}$	& 	$N_{WM,BM}$	& 	$N_{WM,BH}$	&	$N_{WM,WL}$	& 	$N_{WM,WM}$	& 	$N_{WM,WH}$& 	$N_{WM,\cdot}$	\\
					& &\multicolumn{1}{l}{}& 
					$H$&	$N_{WH,BL}$	& 	$N_{WH,BM}$	& 	$N_{WH,BH}$	&	$N_{WH,WL}$	& 	$N_{WH,WM}$	& 	$N_{WH,WH}$& 	$N_{WH,\cdot}$	\\
					& &\multicolumn{2}{c}{Total}&	$N_{\cdot,BL}$	& 	$N_{\cdot,BM}$	& 	$N_{\cdot,BH}$	&	$N_{\cdot,WL}$	& 	$N_{\cdot,WM}$	& 	$N_{\cdot,WH}$& $N_{\cdot,\cdot}$	\\
					\hline	\hline							
				\end{tabular}								
				
			
		\end{center}
	

		\label{tab:CT}
	\end{table} 
		\end{center}

	Knowing the contingency table $K$  (i.e.,  the joint distribution of  wives  and husbands by race and education level) 
	allows us to compute some \textit{descriptive statistics that characterize the equilibrium} on the marriage market.    
	In this paper, we use the \textit{share of educationally homogamous couples} that characterizes the equilibrium along the educational dimension. It is denoted by $\text{SEHC}(K)$ and calculated as   
	$(N_{BL,BL}+N_{BL,WL}+N_{WL,BL}+N_{WL,WL}+ 
	N_{BM,BM}+N_{BM,WM}+N_{WM,BM}+N_{WM,WM}+
	N_{BH,BH}+N_{BH,WH}+N_{WH,BH}+N_{WH,WH} ) /N_{\cdot,\cdot}$,   
	where $N_{\cdot,\cdot}$ denotes the total number of couples. 
	Similarly, we characterize the equilibrium along the racial dimension by  the \textit{share of inter-racial couples}: $\text{SIRC}(K)=1-(N_{B.,B.}+N_{W.,W.})/N_{\cdot,\cdot}$, where $N_{B.,B.}$ is the number of Black-Black couples, and 
	$N_{W.,W.}$ is the number of White-White couples.
	

	It is worth to note that $\text{SEHC}{(K)}$ and $\text{SIRC}(K)$  are indicators of the ``\textit{prevalence} of educational homogamy and racial exogamy''. 
	These indicators are not directly informative about the ``\textit{preferences} for educational homogamy and racial exogamy''. The next section describes how preferences can be identified.

	\section{IDENTIFYING AGGREGATE MARITAL PREFERENCES}\label{sec:met}
	
	Identifying changes in ``preference for educational homogamy'' and 
	``preference for racial exogamy'' is a challenging task. 
	The challenge stems mainly from the fact that preferences are not directly observable. 
	Therefore, we need to identify them through their effects on observed variables, such as 
	$\text{SEHC}{(K)}$ and $\text{SIRC}{(K)}$.\footnote{This approach is commonly applied to identify another directly unobservable phenomenon, discrimination (see \citealp{Oaxaca1973} and \citealp{Blinder1973}).} 
		
	Both $\text{SEHC}{(K)}$ and $\text{SIRC}{(K)}$ are driven by multiple factors. 
	In this paper, we control for the effects of the factors, other than preferences, with a \textit{decomposition}. 
	For the decomposition, we have to apply a {decomposition scheme} and construct {counterfactuals}. 
	
	As to the \textit{decomposition scheme}, we follow \cite{Biewen2014} and apply his additive decomposition formula with interaction effects. As to the \textit{counterfactual construction}, we follow \cite{NaszodiMendonca2020}. They propose a method in a simple set up.  Their method, the NM-method, is suitable for studying sorting along a \textit{one-dimensional categorical trait}.  Another precondition of the applicability of the NM-method is that the categorical assorted trait variable has to be either \textit{dichotomous} (e.g. Black/White, or Black/non-Black, or White/non-White),  
	or, if it can take more than two possible values, the trait has to be \textit{ordered}  
	(such as the level of educational attainment, or income quantile, or skier ability level).

	In Subsection \ref{sec:GNM}, we generalize the NM-method to \textit{two assorted traits} 
	after we introduce the \textit{decomposition scheme} in Subsection \ref{sec:decompscheme} and the NM-method in Subsection \ref{sec:NM}.

	\subsection{{Decomposition scheme}}\label{sec:decompscheme}

	For the empirical analysis, we use the additive decomposition scheme with interaction effects  promoted by  \cite{Biewen2014}. 
	In contrast to the popular \textit{sequential decomposition scheme},\footnote{The literature is not consistent with the terminology regarding the decomposition schemes. For example, the \textit{sequential decomposition} formula is referred to as a \textit{classical Oaxaca--Blinder decomposition} by \cite{Biewen2012}.} 		
		 this scheme  allows us to   
	identify the ceteris paribus effects of some factors net of their  interaction effects.\footnote{This feature of the Biewen decomposition scheme distinguishes itself from the decomposition scheme applied by \cite{Eika2019}.    
Their decomposition method abstracts away from the \textit{interaction-effects}, or as they refer to those, 
the ``general equilibrium conditions (e.g. simultaneous determination of education distributions and returns)''.}   



	For two factors ($A$ and $P$) and two time periods ($t=0$ and $t=1$), the decomposition scheme we apply is 
	\begin{multline}\label{eq:Bdecom2}
	f(A_1, P_1)-f(A_0, P_0)  = 
	\overbrace{[f(A_1, P_0)-f(A_0, P_0)]}^{\mbox{\textit{{\normalsize due to }}} \Delta \mbox{\textit{\normalsize A}}}+  
	\overbrace{[f(A_0, P_1)-f(A_0, P_0)]}^{\mbox{\textit{{\normalsize due to }}} \Delta \mbox{\textit{\normalsize  P}}}\\
	+\underbrace{[f(A_1, P_1)-f(A_1, P_0) - f(A_0, P_1)+ f(A_0, P_0)]}_{\mbox{\textit{\normalsize due to the joint effect of }} \Delta \mbox{\textit{\normalsize  A}} \mbox{\textit{\normalsize { and }}}  \Delta \mbox{\textit{\normalsize  P}}}  \;,
	\end{multline}
	where function $f(A_t, P_t)$ maps the space spanned by the two factors  into  $\mathbb{R}$.

	In one of our specific applications of this decomposition scheme,  function $f(A_t, P_t)$ is the $\text{SEHC}{(K_t)}$, i.e.,  the observed share of educationally homogamous couples  at time $t \in \{0 ,1\}$.   
	Under the assumption that the search and matching mechanism is frictionless, this share is the function of   
	(i) the observed availability $A_t$, i.e., the educational distributions of marriageable men and women at time $t$; 
	(ii)  the directly unobservable preferences over the partners' education level  $P_t$; and 
	(iii) the interaction of availability and preferences. 
		
	{If  frictions exist (e.g. in the form of school segregation limiting what potential partners can meet in school), those are natural to be thought of as the manifestations of certain social barriers reflecting social norms and preferences for segregation. We account  for them with our decomposition as part of the effects of the  directly unobservable preferences. It aggregates the effects of a bunch of empirically equivalent phenomena (e.g. changing social norms, changing social barriers, changing homophily, changing social gaps) that are all in the center of our interest.}

	In Eq. (\ref{eq:Bdecom2}), $f(A_1, P_0)$, and $f(A_0, P_1)$ represent the shares of educationally homogamous couples under the  \textit{counterfactuals} that the factors are measured at different points in time.  For instance, $f(A_1, P_0)$ is the share of educationally homogamous couples in an imaginary generation whose gender-specific educational distributions are identical to the gender-specific educational distributions of the generation most active on the marriage market at $t=1$ and whose marital  educational preferences are identical to the marital educational preferences of the generation most active on the market at $t=0$.


	Similarly to Eq. (\ref{eq:Bdecom2}), the decomposition scheme for three factors ($A$, $P\!R$ and $P\!E$)   is 
	\begin{multline}\label{eq1}
f(A_1, P\!R_1, P\!E_1)-f(A_0, P\!R_0, P\!E_0)  =  \overbrace{[f(A_1, P\!R_0, P\!E_0)-f(A_0, P\!R_0, P\!E_0)]}^{\text{due to } \Delta A \; (availability)}+\\
+\overbrace{[f(A_0, P\!R_1, P\!E_0)-f(A_0, P\!R_0, P\!E_0)]}^{\text{due to } \Delta P\!R \; \text{ (racial preferences)}} +\overbrace{[f(A_0, P\!R_0, P\!E_1)-f(A_0, P\!R_0, P\!E_0)]}^{\text{due to } \Delta P\!E \; \text{ (educational preferences)}}+\\ 
{+\footnotesize\underbrace{ f(A_1\!,\! P\!R_1\!,\! P\!E_0)\!-\!f(A_0\!,\! P\!R_0\!,\! P\!E_0)\! -\! f(A_1\!,\! P\!R_0\!,\! P\!E_0)\!+\!f(A_0, P\!R_0\!,\!P\!E_0)\! - \!f(A_0\!,\! P\!R_1\!,\! P\!E_0)\!+\!f(A_0\!,\! P\!R_0\!,\! P\!E_0)}_{\text{due to the joint effect of } \Delta P\!R \text{ and }  \Delta A}\!+}\\
{+\footnotesize\underbrace{  f(A_1\!,\! P\!R_0\!,\! P\!E_1)\!-\!f(A_0\!,\! P\!R_0\!,\! P\!E_0)\! -\! f(A_1\!,\! P\!R_0\!,\! P\!E_0)\!+\!f(A_0\!,\! P\!R_0\!,\! P\!E_0)\!    -\! f(A_0\!,\! P\!R_0\!,\! P\!E_1)\!+\!f(A_0\!,\! P\!R_0\!,\! P\!E_0)}_{\text{due to the joint effect of } \Delta P\!E \text{ and }  \Delta A}+}\\
{+\footnotesize\underbrace{  f(A_0\!,\! P\!R_1\!,\! P\!E_1)\!-\!f(A_0\!,\! P\!R_0\!,\! P\!E_0)\! -\! f(A_0\!,\! P\!R_1\!,\! P\!E_0)\!+\!f(A_0\!, \!P\!R_0\!,\! P\!E_0)\!    -\! f(A_0\!,\! P\!R_0\!,\! P\!E_1)\!+\!f(A_0\!,\! P\!R_0\!,\! P\!E_0)}_{\text{due to the joint effect of } \Delta P\!E \text{ and }  \Delta P\!R}+}\\
+\underbrace{\text{residuum}}_{\text{due to the joint effect of } \Delta P\!R, \; \Delta P\!E,  \text{ and }  \Delta A}\\
\end{multline}

	In the specific settings of our empirical analyses,  function $f(A_t, P\!R_t, P\!E_t)$ denotes either the observed share of educationally homogamous couples $\text{SEHC}{(K_t)}$, 
	or the observed share of inter-racial couples $\text{SIRC}{(K_t)}$ at time $t \in \{0 ,1\}$. 
	These shares are functions of the following three factors and their interactions:
	(i) the observed availability $A_t$, i.e., the educational and racial distributions of marriageable men and women at time $t$; 
	(ii) the directly unobservable preferences over the partners' race $P\!R_t$; 
	 (iii)  the directly unobservable preferences over the partners' education level  $P\!E_t$.   
	 The interaction effects of the factors represent joint, inseparable effects. 

	Finally,  
	$f(A_1, P\!R_1, P\!E_0)$, $f(A_1, P\!R_0, P\!E_1)$, $f(A_0, P\!R_1, P\!E_1)$, 
	$f(A_1, P\!R_0, P\!E_0)$, \\
	$f(A_0, P\!R_1, P\!E_0)$, $f(A_0, P\!R_0, P\!E_1)$ denote the shares (either $\text{SEHC}$, or $\text{SIRC}$) under different  counterfactuals,  
	where the three factors are not measured the same year.  For instance, $f(A_1, P\!R_1, P\!E_0)$ is the share of educationally homogamous couples (or the share of inter-racial couples) in an imaginary generation whose gender-specific educational and racial distributions are measured at $t=1$; whose marital racial  preferences  are measured at $t=1$; and whose marital educational preferences are measured at $t=0$.

	Apparently, the challenge of identifying the changes in the unobservable factors through their ceteris paribus effects on the observed shares 
	boils down to determining the counterfactual shares in the decomposition formula.  
	Calculating the values of these shares is a trivial task provided we know the corresponding counterfactual tables, i.e., 
	the joint distributions of husbands and wives under the counterfactuals.     

	Next, we will introduce how these counterfactual tables can be constructed with the NM-method in the two-factor ($A$, $P$) case and with the GNM-method in the three-factor ($A$, $P\!R$ and $P\!E$) case. Also, we motivate the choice of these methods.   


       \subsection{The original NM-method}\label{sec:NM}
	
	The NM-method  transforms a contingency table observed at time $t_p$ into another contingency table 
	representing the counterfactual equilibrium matching outcome.\footnote{
	The NM transformation method is implemented in Excel, Visual Basic, and R.  
	It can be downloaded  from \url{http://dx.doi.org/10.17632/x2ry7bcm95.2}} 
	 Under the counterfactual, the aggregate marital preferences are the same as at time $t_p$, while the structural availability  
	is measured at time $t_{a} (\neq t_p)$. 
	So, the counterfactual table is to be constructed from availability $A_{t_a}$ and preferences $P_{t_p}$.    

	We characterize  marital preferences at the aggregate level with the  LL-measure.  
	This  measure, as it was originally developed by \cite{LiuLu2006},  is a scalar-valued, ordinal measure that can be applied 
	if the \textit{assorted trait is a one-dimensional dichotomous variable} (e.g. taking the values $L$ or $H$). 
	The LL-measure was generalized by \cite{NaszodiMendonca2020} to characterize sorting along a one-dimensional  \textit{multinomial trait variable}. 
	The generalized LL-measure is matrix-valued. First, we define the original  LL-measure, before we introduce its generalized version. 
		
	The original LL-measure is identical to the value taken by a function ($f:\mathbb{N}^{2\times 2} \mapsto \mathbb{R}$) that assigns a scalar to a 2-by-2 contingency table, where the contingency table is of the form 
	\begin{equation}\label{Kmatrix2}
	Z^{2\text{-by-}2}= \begin{bmatrix}
	N_{L,L}    &  N_{L,H} \\
	N_{H,L}   &  N_{H,H}
	\end{bmatrix}   \;. 
	\end{equation}
	$N_{H,H}$ (/$N_{L,L}$) denotes the number of homogamous couples, where both spouses are $H$ (/$L$) type. 
	$N_{L,H}$ (/$N_{H,L}$) stands for the number of heterogamous couples, where the husbands (/wives) are $L$-type, while the wives (/husbands) are $H$-type. 
	
	Furthermore, we introduce the notations 
	$N_{H,\cdot}=N_{H,H}+N_{H,L}$,  $N_{\cdot,H}=N_{L,H}+N_{H,H}$,  $N_{\cdot,\cdot}=N_{\cdot,H}+N_{\cdot,L}$. 
	For a given triad of $\{N_{H,\cdot}, N_{\cdot,H}, N_{\cdot,\cdot}\}$,   $Q={N_{H,\cdot}N_{\cdot,H}}/ {N_{\cdot,\cdot}}$ denotes  the expected number of $H$,$H$-type couples under random matching.  We define $Q^-$ as the biggest integer 
	being smaller than, or equal to, $Q$.

	It is important to note that any actual realization of the joint distribution  $Z^{\text{act},{2\text{-by-}2}} \in \mathbb{N}^{2\times 2}$ with a given triad 
	can be represented by any of its cells. For instance, the actual value of the ${H,H}$ cell, i.e.,  $N^\text{act}_{H,H}$, can represent $Z^{\text{act},{2\text{-by-}2}}$, because all the other three cells' actual values are uniquely determined by the triad and $N^\text{act}_{H,H}$. 
	Therefore, there is a unique ranking of the joint distributions with the same triad. 
	This ranking is defined simply by the ranking of the ${H,H}$ cells: 
	that table ranks higher which has higher value in its ${H,H}$ cell.   
	  

	The original LL-measure defines a ranking among the joint distributions with the same, but also with different, triads by ranking their values at the ${H,H}$ cell relative to all possible values of $N_{H,H}$ conditional on the triad.  Under the assumption of non-negative sorting  (i.e., $Q^- \leq N^\text{act}_{H,H}$), the  original LL-measure is equivalent to the \textit{simplified LL-measure} defined as: 
	\begin{equation}\label{LiuLusimpl0}
	\text{LL}^{\text{sim}}(Z^{\text{act},2\text{-by-}2})= \frac{N^\text{act}_{H,H} -  \text{min}(N_{H,H}| N_{H,\cdot}, N_{\cdot,H}, N_{\cdot,\cdot} ) }{\text{max}(N_{H,H} |N_{H,\cdot}, N_{\cdot,H}, N_{\cdot,\cdot})-\text{min}(N_{H,H} | N_{H,\cdot}, N_{\cdot,H}, N_{\cdot,\cdot} ) }    \;. 
	\end{equation}
	The simplified LL-measure interprets as the ``actual minus minimum over maximum minus minimum''. 

	Under non-negative sorting,  $\text{min}(N_{H,H}| N_{H,\cdot}, N_{\cdot,H}, N_{\cdot,\cdot}  ) = Q^-$. And irrespective of the positive, negative, or random nature of sorting, $\text{max}(N_{H,H} | N_{H,\cdot}, N_{\cdot,H}, N_{\cdot,\cdot} )=\text{min}(N_{H,\cdot}, N_{\cdot,H} )$. By substituting these two equations to  Eq. (\ref{LiuLusimpl0}), we obtain  
	\begin{equation}\label{LiuLusimpl}
	\text{LL}^{\text{sim}}(Z^{\text{act},2\text{-by-}2})=  \frac{N^\text{act}_{H,H} -Q^- }{\text{min}(N_{H,\cdot}, N_{\cdot,H} )-Q^- } \;. 
	\end{equation} 
	Eq.(\ref{LiuLusimpl}) defines the original LL-measure under non-negative sorting, which is the empirically relevant type of sorting where the assorted trait is either the eduction level, or race. 


	 



Apparently, there are two extreme feasible matches that serve as benchmarks for the construction of the simplified LL-measure.   
One of the extreme matches is the \textit{perfectly positive match} generated with the following \textit{matching rule}: $H$-type individuals 
can marry $L$-type individuals only if no $H$-type individual from the opposite sex remains available.\footnote{Alternatively, if the assorted trait is race, then the corresponding matching rule is this: individuals with a given race  
can marry someone from a different race  only if nobody from the opposite sex and from their own race remains available.}  
Under this rule,   $\text{LL}^{\text{sim}}$  takes its maximum value, which is one. 
 
The other benchmark is obtained with the \textit{random matching rule}.    
Under random matching,  $N^\text{act}_{H,H}=\text{min}(N_{H,H}| N_{H,\cdot}, N_{\cdot,H}, N_{\cdot,\cdot})= Q^-$ and $\text{LL}^{\text{sim}}$ takes its minimum  value, which is zero. 
	


	Now, let us {relax the assumption that the assorted trait is dichotomous} by following \cite{NaszodiMendonca2020}.     	  
	In the \textit{multinomial case}, the one-dimensional assorted trait distribution can even be gender-specific. 
	For instance, it is possible that the market distinguishes between $n\geq2$ different education levels of men, and $m\geq2$ different education levels of women. 
	So, the aggregate market equilibrium at time $t$ is represented by the contingency table $Z_t$  of size  $n \times m$.

	If both the male-specific assorted trait variable and the female-specific assorted trait variable are {one-dimensional, ordered, categorical polytomous variables}, 
	then the aggregate marital preferences at time $t$ can be characterized  by the 
	\textit{matrix-valued  generalized Liu--Lu measure} (see \citealp{NaszodiMendonca2020}). 
	Its  $i,j$-th  element  is  
	\begin{equation}\label{LiuLugengen}
	\text{LL}^{\text{gen}}_{i,j} (Z_t)= 
	\text{LL}^{\text{sim}}( V_i  Z_t  W^T_j )    \;,
	\end{equation}
	where  $Z_t$ is the $n \times  m$ matrix representing the joint distribution;  
	   $V_i$ is the $2 \times  n$ matrix \vspace{6mm} \\ 
	   $V_i = \scriptsize{ \begin{bmatrix}
		\bovermat{i}{1    & \cdots &  1} & \bovermat{n-i}{ 0  & \cdots & 0}  \\
		0    & \cdots  & 0 & 1  & \cdots  & 1  	
		\end{bmatrix} }$   and  
	$W^T_j$ is the $m \times 2$ matrix given by the transpose of \vspace{6mm} \\
		$W_j = \scriptsize{ \begin{bmatrix}
		\bovermat{j}{1    & \cdots & 1} & \bovermat{m-j}{ 0  & \cdots  & 0}  \\
		0    & \cdots  & 0 & 1  & \cdots  & 1  	
		\end{bmatrix} }$ with   $ i \in \{1, \ldots, n-1 \} $, and  $j \in \{1, \ldots, m-1 \}$.  
	This is how the LL-measure is generalized for ordered, categorical, polytomous, one-dimensional assorted trait variables.

	Next, let us see how the (generalized) LL-measure is used by the \textit{NM-method for constructing counterfactual tables}. 
	We denote the NM-transformed contingency table by  
	$\text{NM}(Z_{t_p},Z_{t_a})=Z^*_{t_p, t_a}$,  
	where the preferences are measured at time $t_p$, while availability is measured at time $t_a$.   
	Unlike  $Z_{t_p}$ and $Z_{t_a}$,  $Z^*_{t_p, t_a}$  cannot be observed. 

	The counterfactual table $Z^*_{t_p, t_a}$ should meet the following conditions.    
	One condition is  $\text{LL}^{\text{gen}}(Z^*_{t_p, t_a})= \text{LL}^{\text{gen}}(Z_{t_p})$. 
	It makes the preferences the same under the counterfactual as at time $t_p$.   
	The other set of  conditions is on the availability which is given by a pair of restrictions of    
	$Z^*_{t_p, t_a}  e^T_{m}=  Z_{t_a}  e^T_{m}$, and  $ e_{n} Z^*_{t_p, t_a}= e_{n} Z_{t_a}$, where $e_{m}$ and  $e_{n}$ are all-ones row vectors of size $m$ and $n$, respectively.    
		
	First, we present the solution for $Z^*_{t_p, t_a}$  in the simplest case, where the 
	assorted trait variable  is dichotomous, 
	before we introduce the solution for the polytomous case. 

	In the \textit{dichotomous case}, the counterfactual table $Z^*_{t_p, t_a}$ to be determined is a 2-by-2 table, just like the observed tables  
	$Z_{t_p}=\begin{bmatrix}
	N^p_{L,L}    &  N^p_{L,H} \\
	N^p_{H,L}   &  N^p_{H,H}
	\end{bmatrix}$ and 
	$Z_{t_a}=\begin{bmatrix}
	N^a_{L,L}    &  N^a_{L,H} \\
	N^a_{H,L}   &  N^a_{H,H}
	\end{bmatrix} $.  The solution for its cell corresponding to the number of  ${H,H}$-type couples is:\footnote{For the derivation of Eq. (\ref{Solution}), see  \cite{NaszodiMendonca2020}.} 
	\small
	\begin{equation}\label{Solution}
	N^*_{H,H}   =
	\frac{\left[  N^p_{H,H} - \text{int}\left(\frac{N^p_{H,\cdot}N^p_{\cdot,H}} {N^p}\right)\right]  \left[{\text{min}\left(N^a_{H,\cdot}, N^a_{\cdot,H} \right)- \text{int}\left(\frac{N^a_{H,\cdot}N^a_{\cdot,H}} {N^a} \right) }\right]     }{\text{min}\left(N^p_{H,\cdot}, N^p_{\cdot,H} \right)- \text{int}\left(\frac{N^p_{H,\cdot}N^p_{\cdot,H}} {N^p} \right) }     
	+\text{int}\left(\frac{N^a_{H,\cdot}N^a_{\cdot,H}} {N^a}\right)   \;, 
	\end{equation}
	\normalsize 
	where $N^p_{H,H}$ is the number of ${H,H}$-type 
	couples observed at time $t_p$. Similarly,  
	$N^p_{H,\cdot}$ (the number of couples, where the husbands are $H$-type),   $N^p_{\cdot,H}$ (the number of couples, where the wives are $H$-type), and  ${N^p}$ (the total number of couples) are also observed at time $t_p$.  
	Whereas  $N^a_{H,\cdot}$, $N^a_{\cdot,H}$, and ${N^a}$ are observed at time  $t_a$. 
	So, Equation (\ref{Solution}) expresses $N^*_{H,H}$ as a function of variables with known values.  
	Regarding the values of all the other three cells of $Z^*_{t_p, t_a}$, those can be calculated from $N^*_{H,H}$  by using the condition on the row totals and column totals of $Z^*_{t_p, t_a}$.

	Next, let us see how the original NM-method works in the \textit{polytomous} case, where the counterfactual table  $\text{NM}(Z_{t_p},Z_{t_a})=Z^*_{t_p, t_a}$, as well as $Z_{t_p}$ and $Z_{t_a}$ are of size $n \times m$.   
	It is worth to note that $\text{NM}(Z_{t_p},Z_{t_a})$ depends on the row totals and column totals of $Z_{t_a}$, but not on  $Z_{t_a}$ itself.  
	So, instead of thinking of the  NM-method as a function mapping $\mathbb{N}^{n \times m} \times \mathbb{N}^{n \times m} \mapsto \mathbb{R}^{n \times m}$, we should rather think of it as a function mapping $\mathbb{N}^{n \times m} \times \mathbb{N}^{n} \times \mathbb{N}^{m} \mapsto \mathbb{R}^{n \times m}$. Accordingly, we will use the following alternative notation in the rest of this paper:    
	$\text{NM}(Z_{t_p}, Z_{t_a}  e^T_{m}, e_{n} Z_{t_a})$.  
	
	With this new notation,    
	the problem for the \textit{multinomial, one-dimensional assortative trait} can be formalized as follows.  
	Our goal is to determine the transformed contingency table $Z^*_{t_p, t_a}$ of size $n \times m$ under the restrictions 
	given by the target row totals and the target column totals observed at time ${t_a}$:  $R_{t_a}:= Z_{t_a} e^T_{m}= Z^*_{t_p, t_a} e^T_{m}$, and  $C_{t_a}:= e_{n} Z_{t_a}=e_{n} Z^*_{t_p, t_a}$.   
	The additional restriction is  $\text{LL}^{\text{gen}}(Z^*_{t_p, t_a})=\text{LL}^{\text{gen}}(Z_{t_p})$.

	By using Eq.(\ref{LiuLugengen}), we can rewrite the problem  as follows.   
	We look  for $Z^*_{t_p, t_a}$, where  
	$V_i  R_{t_a} = V_i  Z^*_{t_p, t_a} e^T_{m} $, and  $C_{t_a}  W^T_j = e_{n} Z^*_{t_p, t_a} W^T_j $;  and 
	$\text{LL}(V_i  Z_{t_p}  W^T_j)= \text{LL}(V_i  Z^*_{t_p, t_a}  W^T_j)$ 
	for all   $i \in \{1,..., n-1\} $ and $j \in \{1,..., m-1\} $. The matrices  
	$V_k$ and $W_p$ are defined the same as under Eq.(\ref{LiuLugengen}).   
	For each $i,j$-pairs, these equations define a problem of the 2-by-2 form.       
	Each problem  can be solved separately by applying Eq.(\ref{Solution}).      
	The solutions determine $(n-1) \times (m-1)$ entries of the $Z^*_{t_p, t_a}$ table. 
	The remaining $m+n-1$ elements of the $Z^*_{t_p, t_a}$ table can be determined with the help of the target row totals and target column totals.

  \subsubsection{A supplementary analysis}\label{sec:NM_Prop} 	

\cite{NaszodiMendonca2020} 
  visit many of the NM's attractive empirical properties and analytical properties, as well as the analytical properties of its transformed table. 
Here, we present a supplementary analysis with a toy model providing further support for applying the LL-measure and the NM. 
In our {toy model there is achieved status  and ascribed status}. 
The ascribed status represents  heterogeneity unobserved by the econometrician.        
In this model, the GS-algorithm matches men and women along the two statuses.\footnote{Matching with the GS-algorithm allows us to model interesting and complex interactions in a manageable way. Unlike in the standard multinomial logit model of discrete choice, the chances of a man of type x to be matched with a woman of type y rather than with z may dependent not only on the numbers of type x men and  type y women in the GS model, but also on the population size of the type z women inter alia.}  
	
	It is intuitive to impose a \textit{monotonicity condition}:  the degree of marital sorting along the achieved status cannot be low in those societies, where the intergenerational mobility  is low (i.e., the association between individuals' achieved status and ascribed status is strong) since   
	both high degree of marital sorting and low intergenerational mobility are the manifestations of the same phenomenon. This phenomenon is the general lack of ``openness'' of societies (see \citealp{Katrnak2012}). 
        As we will see, this {monotonicity condition} is violated if marital sorting is quantified by any of its well-known measures, but the LL-measure.


	In our toy model, young people finalize their education determining their \textit{achieved status}.  
	Then,  their \textit{ascribed status} is determined when they inherit wealth from their parents as adults.
	Finally,  they get married.  They sort into couples along the pair of status traits.  
	Although both of the assorted traits can be observed by the potential partners of each individual, it is only the achieved status that is observed by the econometrician.     
	
	For the sake of simplicity, we make the following assumptions. 
	Each of the trait variables is either low ($L$), or high ($H$).  
	The two assorted traits are perfect substitutes in this model: somebody with high ascribed status and low achieved status is just as attractive as somebody with low ascribed status and high achieved status. This assumption governs how men with various traits are ranked by women and also how women with various traits are ranked by men. 
	
	Now, let us assume that people match using the GS-algorithm. 
	We apply the aggregate version of this algorithm that was proposed by \cite{Hsieh2012}.    
	The aggregate GS-algorithm characterizes the matching outcome at an aggregate level by determining the joint distribution of couples directly without determining who will be matched with whom.        
	
	The aggregate GS-algorithm works as follows in our model.  
	First, it matches the $H,H$-type men with the $H,H$-type women. 
	If these groups are not equally large, e.g., the number of $H,H$-type men exceeds the number of $H,H$-type women, then 
	some $H,H$-type men will be matched in the next step of the algorithm with some women who are of the second most attractive types as being either $L,H$-type, or $H,L$-type.    
	If there are too few $L,H$-type and $H,L$-type women relative to the 
	$H,H$-type men unmatched in the first step, then those $H,H$-type men who remained single after the first two steps of the algorithm will be 
	matched either with $L,L$-type women, or remain single depending on the relative size of these two groups. 	
	
	Men who are of the second most attractive type, i.e., being  $H,L$-type or  $L,H$-type, are matched similarly to the $H,H$-type men. However, they can enter the marriage market once all the $H,H$-type men have left the market with their spouses. So, the $H,L$-type and the $L,H$-type men select their spouses from the group of those women, who remained unmatched after all the $H,H$-type men have already engaged with their partners.   
	Finally, the  $L,L$-type men enter the market once the $H,H$-type, $H,L$-type and $L,H$-type  men and their partners have already left the market. 
	So, this is how the aggregate GS-algorithm determines the joint distribution of couples in our toy model.       
	
	Next, let us take a numerical example with two \textit{hypothetical societies}. 
	In the {first society}, the 
	distributions of the marriageable men and women are given by the last column and the last row of the $K^{GS}_1$ table below.\\ 
\resizebox{\textwidth}{!}{\begin{tabular}{ cc }   
	The $K^{GS}_1$ contingency table & The $K^{GS}_2$ contingency table \\  
	\setlength{\tabcolsep}{3pt} 
	\renewcommand{\arraystretch}{1} 
	\begin{tabular}{llll ccccc}								
		\hline	\hline							
		& &    & &   \multicolumn{4}{c}{Wife} &   \\ \cline{5-8}
		& &    & &   \multicolumn{2}{c}{Low inheritance} & \multicolumn{2}{c}{High inheritance} & \\ \cmidrule(lr){5-6} \cmidrule(lr){7-8} 
		& &    &Education & \multicolumn{1}{c}{$L$}    & \multicolumn{1}{c}{$H$} & \multicolumn{1}{c}{$L$}   & \multicolumn{1}{c}{$H$}&         $^{\text{{\normalsize{Total}}}}$ \\		\hline
		
		\multicolumn{1}{c}{\multirow{4}{*}{\rotatebox[origin=c]{90}{\underline{$\;\;\;\;\;$Husband$\;\;\;\;\;$}}}} &     \multicolumn{1}{c}{\multirow{2}{*}{\rotatebox[origin=c]{90}{\underline{Low i.}}}} &    &
		$L$&	$\;\;\;400$	& 	$\;\;\;\;\;\;\;0$	&	$\;\;\;\;\;\;\;0$	& 	$\;\;\;\;\;\;\;0$ & 	$\;\;\;400$	\\
		& &\multicolumn{1}{l}{}&
		$H$&	$\;\;\;\;\;\;\;0$	& 	$\;\;\;\;\;\;\;0$	&	$\;\;\;\;\;\;\;0$	& 	$\;\;\;\;\;\;\;0$	& 	$\;\;\;\;\;\;\;0$	\\
		
		&  \multicolumn{1}{c}{\multirow{2}{*}{\rotatebox[origin=c]{90}{\underline{High i.}}}} &   &
		$L$&	$\;\;\;\;\;\;\;0$	& 	$\;\;\;\;\;\;\;0$ &	$\;\;\;\;\;\;\;0$	& 	$\;\;\;\;\;\;\;0$	& 	$\;\;\;\;\;\;\;0$	\\
		& &\multicolumn{1}{l}{}& 
		$H$&	$\;\;\;100$	& 	$\;\;\;\;\;\;\;0$	&	$\;\;\;\;\;\;\;0$	& 	$\;\;\;500$ & 	$\;\;\;600$	\\
		& &\multicolumn{2}{c}{Total}&	$\;\;\;500$	& 	$\;\;\;\;\;\;\;0$	&	$\;\;\;\;\;\;\;0$	& 	$\;\;\;500$ & 1,000	\\
		\hline	\hline							
	\end{tabular} &  
		\setlength{\tabcolsep}{3pt} 
	\renewcommand{\arraystretch}{1} 
	
	\begin{tabular}{llllccccc}								
		\hline	\hline							
		& &    & &   \multicolumn{4}{c}{Wife} &   \\ \cline{5-8}
		& &    & &   \multicolumn{2}{c}{Low inheritance} & \multicolumn{2}{c}{High inheritance} & \\ \cmidrule(lr){5-6} \cmidrule(lr){7-8} 
		& &    &Education & \multicolumn{1}{c}{$L$}    & \multicolumn{1}{c}{$H$} & \multicolumn{1}{c}{$L$}   & \multicolumn{1}{c}{$H$}&         $^{\text{{\normalsize{Total}}}}$ \\		\hline
			
	\multicolumn{1}{c}{\multirow{4}{*}{\rotatebox[origin=c]{90}{\underline{$\;\;\;\;\;$Husband$\;\;\;\;\;$}}}} &     \multicolumn{1}{c}{\multirow{2}{*}{\rotatebox[origin=c]{90}{\underline{Low i.}}}} &    &
	$L$&	$\;\;\;396$	&   $\;\;\;\;\;\;\;0$		&	$\;\;\;\;\;\;\;0$	& 	$\;\;\;\;\;\;\;0$ & 	$\;\;\;396$	\\
		& &\multicolumn{1}{l}{}&
		$H$&	$\;\;\;\;\;\;\;4$	& 	$\;\;\;\;\;\;\;0$	&	$\;\;\;\;\;\;\;0$	& 	$\;\;\;\;\;\;\;0$	& 	$\;\;\;\;\;\;\;4$	\\
		
	&  \multicolumn{1}{c}{\multirow{2}{*}{\rotatebox[origin=c]{90}{\underline{High i.}}}} &   &
		$L$&	$\;\;\;\;\;\;\;6$	& 	$\;\;\;\;\;\;\;0$ &	$\;\;\;\;\;\;\;0$	& 	$\;\;\;\;\;\;\;0$	& 	$\;\;\;\;\;\;\;6$	\\
		& &\multicolumn{1}{l}{}& 
		$H$&	$\;\;\;\;89$	& 	$\;\;\;\;\;\;\;5$	&	$\;\;\;\;\;\;\;5$	& 	$\;\;\;495$ & 	$\;\;\;594$	\\
		& &\multicolumn{2}{c}{Total}&	$\;\;\;495$	& 	$\;\;\;\;\;\;\;5$	&	$\;\;\;\;\;\;\;5$	& 	$\;\;\;495$ & 1,000	\\
		\hline	\hline							
	\end{tabular} \\
\end{tabular}}
\vspace{2mm}

	In this society there is absolutely no intergenerational mobility no matter how we quantify this phenomenon:  
	those who have low inheritance have low education level, while those who have high inheritance have high education level.     
	Once individuals are matched with the aggregate GS-algorithm, 
	the joint distribution of husbands and wives is given by the inner part of table $K^{GS}_1$.

	In the {second hypothetical society} of $K^{GS}_2$, there is intergenerational mobility: 
	1\% of those who have low inheritance are highly educated, and 1\% of those who have high inheritance are low educated both among men and women. 
	So, the intergenerational mobility makes the distributions of the marriageable men and women in this society different from those in the first example.  
	Accordingly, $K^{GS}_2$ has row totals and column totals that are different from those of the $K^{GS}_1$ table. 
	We assume that men and women in this second society are also matched with the GS-algorithm. The aggregate outcome of the matching is given by the inner part of table $K^{GS}_2$.

	The econometrician cannot observe the inheritance, only the education levels.    
	Accordingly, the contingency table observed in the two societies are\\ 
	$K^{GS, \; obs}_1= \begin{bmatrix}
	400    &  $\;\;\;\;0$ \\
	100   &  500
	\end{bmatrix}   \;, $  and  
	$K^{GS, \; obs}_2= \begin{bmatrix}
	402    &  $\;\;\;\;0$ \\
	$\;\;98$   &  500
	\end{bmatrix}   \;. $ 
	
	It is insightful to calculate some \textit{conventional measures of marital sorting} both for  $K^{GS, \; obs}_1$ and $K^{GS, \;  obs}_2$. 
	These are reported by Table \ref{tab:MS}. 
	Apparently, most of the results in this table are counter-intuitive:  
	almost all the indicators that supposed to quantify the degree of marital sorting are either lower in the first society, where there is no intergenerational mobility, or, the indicator is not defined.\footnote{\cite{Mood2010} raises a related issue with the odds-ratio based  logistic regression analysis: it cannot adequately control for unobserved heterogeneity.}    

	\setlength{\tabcolsep}{2pt}
	\begin{table}[!htb]
		\caption{Some conventional measures of marital sorting and the LL-measure}
			\begin{flushleft}		
				\begin{tabular}{lcccccc}								
					\hline	\hline
					&Regression coef.       &Regression coef.     &Correlation coef. &Aggregate&Odds-ratio &Liu--Lu \\
					&(husbands'             &(wives'              &(between          &marital  &(cross-    &value \\
 					&education is           &education is         &wives' and        &sorting  &product-   &    \\
					&explained by           &explained by         &husbands'         &parameter&ratio)     &          \\
					&wives' edu.)           &husbands' edu.)      &education)	     &         &           &         \\ \cmidrule(lr){2-7}
					$K^{GS, \; obs}_1$           &  0.800  		& 0.833			& 0.816          &  1.815 	 &  not 	& 1  \\
					$K^{GS, \; obs}_2$           &  0.804		& 0.836			& 0.820		 &  1.818	 &  defined	& 1  \\
					
					\hline	\hline	
				
												\end{tabular}								
		
	\end{flushleft}
	
				\textit{Notes}: The regression coefficient was applied by \cite{Greenwood2014}, the aggregate 
				 marital sorting parameter was proposed  by \cite{Eika2019},  	while the odds-ratio was  
				 applied  by \cite{Rosenfeld2008} inter alia. The marital surplus indicator developed by \cite{ChooSiow2006} cannot be calculated without data on single individuals.


		\label{tab:MS}
	\end{table}

The only exception is the LL-measure: it quantifies marital sorting as being the same in the two societies.  
So, unlike the conventional measures, the LL-measure does not violate the monotonicity condition.  
This finding also supports the application of the LL-measure as an aggregate measure of marital sorting.   
Thereby it is also in favor of the NM. 

\subsection{The GNM-method}\label{sec:GNM}
	
	In this section, we generalize the NM-method for the case where 
	individuals sort along two traits (that are assumed to be neither perfect complementers, nor perfect substitutes).    
	The generalized NM-method, the GNM-method,\footnote{The GNM method is implemented in Stata, Matlab and R. It can be downloaded  from [link to Mendeley] \url{}.}  works under the assumption that sorting along the two dimensions is sequential: 
	each individual is assumed to sort along the same dimension first and then along the other dimension. In other words, individuals are assumed to have lexicographic preferences over the assorted traits.   
	
	To be consistent with the empirical part of this paper, let us 
	call  one of the traits  race (taking the dichotomous nominal values $W$, or  $B$), while we refer to the other assorted trait  as education (taking the ordered values $L$, $M$, or $H$). 
			
	Let us denote by $t_r$ the year at which marital preferences over partners' \textit{race} is measured.     
	Similarly, we denote by $t_e$ the year when  preferences over partners' \textit{education} level  is measured, and 
	we denote by $t_a$  the year when \textit{availability} is measured.  
	We observe the matching outcomes at $t_r$,  $t_e$, and $t_a$  by observing  the 2n-by-2m tables $K_r$,  $K_e$, and $K_a$, respectively. 
	These tables are of the form of Table \ref{tab:CT} and represent the joint traits distributions of husbands and wives.

	The $\text{GNM}(K_r, K_e, K_a)$  would map $\mathbb{N}^{2n \times 2m} \times \mathbb{N}^{2n \times 2m} \times \mathbb{N}^{2n\times 2m}  \mapsto \mathbb{R}^{2n \times 2m}$ provided the GNM-method can be defined perfectly analogously to the  $\text{NM}(Z_{t_p},Z_{t_a})$ function. However, as we will see, 
	the GNM does not assign a unique table to each element in its domain. 
	In fact, $\text{GNM}(K_r, K_e, K_a)$ assigns a finite set of 2n-by-2m tables to each triad of $\{K_r, K_e, K_a\}$. 
		
	Next, we present how the $\text{GNM}(K_r, K_e, K_a)$ works under the assumption that individuals sort along the dichotomous race variable first.  
	As the first step of the GNM-method, the NM-method is applied in order to obtain the \textit{racial distribution} of couples under the counterfactual. In particular,  we calculate   
	$\text{NM}(X K_r Y^T, X K_a Y^T)$ and denote it by $Z^*_{t_r, t_a}$, where  \vspace{3mm}	\\  
	\vspace{3mm}
		matrix  $X = \scriptsize{ \begin{bmatrix}
		\bovermat{n}{1    & \cdots &  1} & \bovermat{n}{ 0  & \cdots & 0}  \\
		0    & \cdots  & 0 & 1  & \cdots  & 1  	
		\end{bmatrix} }$   and  
	$Y^T$ is the $2m \times 2$ matrix given by the transpose of \\
	
	$Y = \scriptsize{ \begin{bmatrix}
		\bovermat{m}{1    & \cdots & 1} & \bovermat{m}{ 0  & \cdots  & 0}  \\
		0    & \cdots  & 0 & 1  & \cdots  & 1  	
		\end{bmatrix} }$.  
	The 2-by-2 table $Z^*_{t_r, t_a}$ represents the racial distribution of couples under the counterfactual.

	As a second step of the GNM-method,  the \textit{educational distribution} of couples is calculated under the counterfactual set of conditions that includes the racial distribution 
	$Z^*_{t_r, t_a}$. 
	The challenge here is that 
	we know only the gender-specific educational target distributions of the population. 
	However, we do not know the educational target distributions of husbands and wives for each of the four marriage types $BB, BW, WB, WW$  defined by race.  
	By using a new notation, what we do not know is  $v^{*,\text{race}\; i,j}_{\text{male}}$, a column vector of size $n$, which is the educational target distribution of husbands of race $i$ married to women of race $j$  ($i,j \in \{W,B\}$). Also, we do not know $v^{*,\text{race} \; j,i}_{\text{female}}$, a row vector of size $m$, which is the educational target distribution of wives of race $i$ married to men of race $j$  ($i,j \in \{W,B\}$).

	
	If we knew the target distributions $v^{*,\text{race}\; i,j}_{\text{male}}$ and $v^{*,\text{race} \; j,i}_{\text{female}}$ 
	then  
	applying the NM-method for the four racial types of marriages separately, 
	could provide us a unique counterfactual joint distribution of wives and husbands along both race and education. 
	
	The corresponding four NM transformations are defined as follows.  
	For couples, where both the wives and the husbands are of race $B$, it is 
	\begin{equation}\label{eq_BB}
	\text{NM}(K_{e,1..n,1..m},  v^\text{*,race  B,B}_{\text{male}}, v^\text{*,race B,B}_{\text{female}})=Z^*_{\text{race  B,B}} \;,
	\end{equation}
	with  the constraints defined by the target marginals:  
	$Z^*_{\text{race  B,B}}  e^T_{m}=  v^\text{*,race  B,B}_{\text{male}}$,   $ e_{n} Z^*_{\text{race  B,B}}= v^\text{*,race  B,B}_{\text{female}}$, and the constraint defined by the counterfactual racial distribution:    
	$ e_n  v^\text{*,race  B,B}_{\text{male}}= v^\text{*,race  B,B}_{\text{female}}  e^T_m = Z^*_{t_r, t_a, 1, 1}$.

	For couples, where both the wives and the husbands are of race $W$, it is 
	\begin{equation}\label{eq_WW}
	\text{NM}(K_{e,n+1..2n,m+1..2m},  v^\text{*,race  W,W}_{\text{male}}, v^\text{*,race   W,W}_{\text{female}})=Z^*_{ \text{race  W,W}}\;,
	\end{equation}
	with  the constraints    
	$Z^*_{\text{race  W,W}}  e^T_{m}=  v^\text{*,race   W,W}_{\text{male}}$,   
	$ e_{n} Z^*_{\text{race  W,W}}= v^\text{*,race  W,W}_{\text{female}}$, and 
	$ e_n  v^\text{*,race  W,W}_{\text{male}}= v^\text{*,race  W,W}_{\text{female}}  e^T_m = Z^*_{t_r, t_a, 2, 2}$.

	For inter-racial marriages, where the husbands are of race $W$  and the wives are of race $B$, it is 
	\begin{equation}\label{eq_WB}
	\text{NM}(K_{e,n+1..2n,m+1..2m},  v^\text{*,race  W,B}_{\text{male}}, v^\text{*,race  W,B}_{\text{female}})=Z^*_{\text{race  W,B}}\;,
	\end{equation}
	with  the constraints  
	$Z^*_{\text{race W,B}}  e^T_{m}=  v^\text{*,race  W,B}_{\text{male}}$,   
	$ e_{n} Z^*_{\text{race  W,B}}= v^\text{*,race  W,B}_{\text{female}}$, and 
	$ e_n  v^\text{*,race  W,B}_{\text{male}}= v^\text{*,race  W,B}_{\text{female}}  e^T_m = Z^*_{t_r, t_a, 2, 1}$.    
	
	Finally, for inter-racial marriages,  where the husbands are of race $B$  and the wives are of race $W$, it is
	\begin{equation}\label{eq_BW}
	\text{NM}(K_{e,n+1..2n,m+1..2m},  v^\text{*,race  B,W}_{\text{male}}, v^\text{*,race  B,W}_{\text{female}})=Z^*_{ \text{race  B,W}}\;,
	\end{equation}
	with  the constraints   
	$Z^*_{\text{race B,W}}  e^T_{m}=  v^\text{*,race  B,W}_{\text{male}}$,   
	$ e_{n} Z^*_{\text{race  B,W}}= v^\text{*,race  B,W}_{\text{female}}$ , and 
	$ e_n  v^\text{*,race  B,W}_{\text{male}}= v^\text{*,race  B,W}_{\text{female}}  e^T_m = Z^*_{t_r, t_a, 1, 2}$.

	These four transformation problems are under-determined: 
	in general,  there is a set of $v^{*,\text{race}\; i,j}_{\text{male}}$,  $v^{*,\text{race}\; i,j}_{\text{female}}$ pairs that fulfill all the constraints, while each pair define a unique set of 	
	counterfactual tables of 
	${Z^*_{\text{race B,B}}, Z^*_{\text{race  B,W}}, Z^*_{\text{race  W,B}}, Z^*_{\text{race  W,W}}}$.  So, in general, there are multiple solutions for the counterfactual table. 

	Fortunately, lack of uniqueness of $\text{GNM}(K_r, K_e, K_a)$ does not prevent us to perform  counterfactual decompositions.  
	For the decompositions, we do not need to know what would be the joint distribution of traits under the counterfactual.   
	It is sufficient to know a particular moment of it:       
	this moment is either the share of educationally homogamous couples $\text{SEHC}(\text{GNM}(K_r, K_e, K_a))$, or  the share of inter-racial marriages $\text{SIRC}(\text{GNM}(K_r, K_e, K_a))$.

	The set of counterfactual constraints determine a finite number of possible counterfactual tables.   
	So, these constraints determine an interval for any scalar-valued moment of the counterfactual joint distribution. This interval contains all possible values of the given moment.      
	Each interval of the outcome variable SEHC, or SIRC define the intervals of the components of our interest, 
	i.e., the contributions of changing preferences over the partners' race and education 
	to the share of educationally homogamous couples, or the share of inter-racial marriages. 
	


	
	
	We obtain the component-intervals by solving a maximization problem and a minimization problem with 
	a scalar-valued objective function of either $\text{SEHC}(\text{GNM}(K_r, K_e, K_a))$, or $\text{SIRC}(\text{GNM}(K_r, K_e, K_a))$. If SEHC is to be maximized under the counterfactual,  the optimization problem is:\footnote{In the empirical application of the GNM in this paper, $n=m=3$ making the parameter space 8 dimensional. Any extension of the model that 
		increases the number of racial categories, or the number of traits considered increases 
		the number of parameters to be estimated and  makes  
		the optimization problems challenging to solve.}
	\begin{equation}\label{eq_maxSEHC}
	\text{max}_{ v^{*,\text{race}\; i,j}_{\text{male}},  v^{*,\text{race}\; i,j}_{\text{female}}} \text{SEHC}(\text{GNM}(K_r, K_e, K_a)) \;,
	\end{equation}
    with $i,j \in \{W,B\}$, subject to the constraints on  the independent elements of the vectors     
	$v^{\text{*,race}\; i,j}_{\text{male}}$ and $v^{\text{*,race} \; j,i}_{\text{female}}$ in the $\mathbb{N}^{2(n-1+m-1)}$ space. 	These constraints are defined below Equations (\ref{eq_BB}), (\ref{eq_WW}), (\ref{eq_WB}), and (\ref{eq_BW}).
	The other three optimization problems can be formalized similarly to Equation (\ref{eq_maxSEHC}).


	In principle, the GNM-method can be applied not only when sorting along race precedes sorting along education, 
	but also when sorting along these two traits is in the reverse sequence.   
	The choice of sequence depends on which of the traits is considered to be more important in the population studied. 
	Our benchmark empirical analysis in this paper relies on the assumption that Americans' primary trait of sorting was race when being young adults. 
	However, we perform our analysis with the alternative assumption as well (see Appendix C).

	\subsubsection{A numerical example with the GNM-method}\label{sec:NEWmethods_num}

	We illustrate the application of the GNM-method with a simple numerical example. To ease the exposition, 
we consider two dichotomous traits. Individuals can either be Black ($B$) or White ($W$) and have attained a low ($L$) or high ($H$) education level. 
Suppose that the table to be transformed, i.e, the seed table, is given by\\ $K_e= K_r = \begin{bmatrix}
	$46$ & $\;\;\;\;142$ & $\;\;\;\;\;\;\;\;7$ & $\;\;\;\;\;\;29$ \\
	$98$ & 2,453 & $\;\;\;\;198$ & $\;\;\;\;442$ \\
	$\;\;5$ & $\;\;\;\;\;\;\;8$ & $\;1,532$ & $\;\;1,667$ \\
	$\;\;3$ & $\;\;\;\;181$ & $\;\;\;\;949$ & $30,735$
	\end{bmatrix}$ in the form of Table \ref{tab:CT}, where $\text{SEHC}(K_e)=92\%$ and $\text{SIRC}(K_e)=2.3\%$. 	
	The target column totals and row totals are given by  $C = \begin{bmatrix}
	127 & 2,665 & 1,987 & 32,661
	\end{bmatrix}$ and the transpose of $R^T = \begin{bmatrix}
	170 &  3,008 & 	2,627 &	31,635	\end{bmatrix}$.

	As the first step of the GNM, we aggregate the table $K_r$, the row totals $R$, and the column totals $C$ along the educational dimension.   
	It results in the following seed table $K^{Race}$, target row total vector $R^{Race}$, and column total vector $C^{Race}$:
	        $K^{Race}= X K_r Y^T = \begin{bmatrix}
	$2,733$ & $\;\;\;\;\;676$ \\
	$\;\;\;197$ & $34,883$
	\end{bmatrix}$, $R^{Race} = X R= \begin{bmatrix}
	$\;\;3,178$ \\
	$34,262$
	\end{bmatrix}$ and $C^{Race} = C Y^T = \begin{bmatrix}
	2,792 & 34,648 
	\end{bmatrix}$.


Next, we apply the NM-method using formula (\ref{Solution}). It yields    
$\text{NM}(K^{Race}, R^{Race}, C^{Race}) = \begin{bmatrix}
	$2,605$ & $\;\;\;\;\;573$ \\
	$\;\;\;187$ & $34,075$
	\end{bmatrix}$ that we denote by $Z^*_{t_r, t_a}$. 
	This matrix represents the joint racial distribution of husbands and wives under the counterfactual.

As a second step, we take the following four  2-by-2 sub-matrices of $K_e$:  $K^{BB}_e = \begin{bmatrix}
	$46$ & $\;\;\;142$ \\
	$98$ & $2,453$
	\end{bmatrix}$, $K^{BW}_e = \begin{bmatrix}
	$\;\;\;\;\;7$ & $\;\;29$ \\
	$198$ & $442$
	\end{bmatrix}$, $K^{WB}_e = \begin{bmatrix}
	$5$ & $\;\;\;8$ \\
	$3$ & $181$
	\end{bmatrix}$, 
	$K^{WW}_e = \begin{bmatrix}
	$1,532$ & $\;\;1,667$ \\
	$\;\;\;949$ & $30,735$
	\end{bmatrix}$  and solve for each separate problem defined by Equations (\ref{eq_BB}), (\ref{eq_WW}), (\ref{eq_WB}),  (\ref{eq_BW}) and the constraints below them.

We set the target marginals to make the educational distribution over race consistent with the counterfactual racial distribution $Z^*_{t_r, t_a}$ calculated in the first step: 
$R^{BB} = \begin{bmatrix}
	\beta_{1} \\
	K^{Race}_{1,1} - \beta_{1}
	\end{bmatrix}$, $C^{BB} = \begin{bmatrix}
	\beta_{2} & K^{Race}_{1,1} - \beta_{2}
	\end{bmatrix}$, $R^{BW} = \begin{bmatrix}
	R_{1} - \beta_{1} \\
	R_{2} - R_{1} + \beta_{1}
	\end{bmatrix}$, $C^{BW} = \begin{bmatrix}
	C_{1} - \beta_{2} & C_{2} - C_{1} + \beta_{2}
	\end{bmatrix}$, $R^{WB} = \begin{bmatrix}
	R_{3} - \beta_{3} \\
	R_{4} - K^{Race}_{2,2} + \beta_{3}
	\end{bmatrix}$, $C^{WB} = \begin{bmatrix}
	C_{3} - \beta_{4} & C_{4} - K^{Race}_{2,2} + \beta_{4}
	\end{bmatrix}$,	$R^{WW} = \begin{bmatrix}
	\beta_{4} \\
	K^{Race}_{2,2} - \beta_{3}
	\end{bmatrix}$, $C^{WW} = \begin{bmatrix}
	\beta_{4} & K^{Race}_{2,2} - \beta_{4}
	\end{bmatrix}$.

Obviously, the counterfactual table to be obtained with the GNM depends on the parameter vector
 $\beta=[\beta_1, \beta_2, \beta_3, \beta_4 ]$.     
In this numerical example presented with an illustrative purpose, we choose $\beta$ to maximize SEHC. 
The resulting counterfactual is given by $K^*_{\text{max, SEHC} } = \begin{bmatrix}
	$\;\;33$ & $\;\;\;122$      & $\;\;12$    & $\;\;\;\;\;\;2$ \\
	$\;\;73$ & $2,377$            & $\;\;\;8$ & $\;\;\;\;165$ \\
	$185$      & $\;\;\;\;\;\;2$  & $862$       & $\;1,578$ \\
	$386$      & $\;\;\;\;\;\;0$  & $554$       & $31,081$
	\end{bmatrix}$, where  $\beta_{\text{max, SEHC} }=[155 \; 106 \; 1,416 \; 2,440]$.

We can calculate two descriptive statistics of $K^*_{\text{max, SEHC} }$:  $\text{SEHC}(K^*_{\text{max, SEHC}})=92.7\%$ and $\text{SIRC}(K^*_{\text{max, SEHC}})=2\%$. 
To recall, the same statistics of the table $K_e$ were  $\text{SEHC}(K_e)=92\%$ and $\text{SIRC}(K_e)=2.3\%$. 
So, 0.7\% increase in SEHC and 0.3\% decease in SIRC can be attributed to the change in the distributions of males and females along race and education provided SEHC is maximized under the counterfactual of no change in preferences. 

As opposed to this numerical example,  in the empirical part of the paper we construct the counterfactual and perform the decomposition not only by maximizing SEHC with the choice of $\beta$, but also by minimizing SEHC, maximizing SIRC, and minimizing SIRC. These four alternative specifications together represent model uncertainty. By performing the decompositions with each of the four alternative model specifications, we obtain  interval estimates for the contributions of the factors. 


\section{EMPIRICAL ANALYSIS}\label{sec:emp}
	
In the empirical analysis, we use decennial census data of the United States from IPUMS.   
Our data covers heterosexual young couples with  male partners aged 30 to 34 years. 
The variable on the highest educational attainment can take three values: 
``less than high school",  ``high school completed", and ``tertiary education completed".\footnote{So, we work with the same three educational categories as \cite{ChooSiow2006}, \cite{NaszodiMendonca2020}.} The variable on race is dichotomous, taking the values of ``Black" or ``White". 

Our benchmark analysis focusing on the 1980s is presented in Subsection	\ref{sec:dec}.   
In Subsection \ref{sec:dec19602015}, we study a much longer period spanned by 1960 and 2015.   


	Our motivation for using the census data only from 1980 and 1990 in our benchmark analysis is twofold.
	First, as already mentioned in the introduction, there is a disagreement  about the trend of economic inequality in the 1980s among the papers identifying the trend from the joint educational distributions of couples.   
	Second, the share of individuals who identify with more than one group in response to the race question in the census has grown substantially since 2000, 
	when the multi-race option was offered to the respondents the first time. 
	The modification of racial categories in 2000 does not affect the comparison of the data from 1980 and 1990. 

\subsection{{Benchmark analysis of the 1980s}}\label{sec:dec}

This section presents the main results of decomposing changes in 
the share of educational homogamous couples (SEHC) and the share of inter-racial couples (SIRC) in the US between 1980 and 1990. 
The census wave-specific contingency tables are presented in Tables \ref{tab:CT1980} and \ref{tab:CT1990}.

\setlength{\tabcolsep}{2pt}																										
\begin{table}[!htb]
	                \caption{Joint educational and racial distribution of young American  couples in \textbf{1980}}        
																											
	
					\begin{center}																					
			\begin{tabular}{llllrrrrrrr}																							
				\hline	\hline																					
				& &    & &   \multicolumn{6}{c}{Wife/female partner} &   \\ \cline{5-10}																						
				& &    & &   \multicolumn{3}{c}{Black} & \multicolumn{3}{c}{White} & \\ \cmidrule(lr){5-7} \cmidrule(lr){8-10} 																						
				& &    &Edu. & \multicolumn{1}{c}{$L$}   & \multicolumn{1}{c}{$M$} & \multicolumn{1}{c}{$H$} & \multicolumn{1}{c}{$L$}   & \multicolumn{1}{c}{$M$} & \multicolumn{1}{c}{$H$}&         $^{\text{{\normalsize{Total}}}}$ \\		\hline																				
				
				\multicolumn{1}{c}{\multirow{6}{*}{\rotatebox[origin=c]{90}{Husband/}}} &  \multicolumn{1}{c}{\multirow{6}{*}{\rotatebox[origin=c]{90}{\underline{male partner}}}}& \multicolumn{1}{c}{\multirow{3}{*}{\rotatebox[origin=c]{90}{\underline{Black}}}} & 																						
				$L$&	52,624	& 	57,135	& 	2,882	&	1,261	& 	1,862	& 	161	&	115,925	\\								
				& &\multicolumn{1}{l}{}& 																						
				$M$&	40,183	& 	215,109	& 	26,318	&	1,920	& 	8,428	& 	1,341	&	293,299	\\								
				& &\multicolumn{1}{l}{}&																						
				$H$&	2,101	& 	29,703	& 	30,252	&	220	& 	2,364	& 	2,080	& 	66,720	\\								
				
				& & \multicolumn{1}{c}{\multirow{3}{*}{\rotatebox[origin=c]{90}{\underline{White}}}} & 																						
				$L$&	420	& 	380	& 	20	&	391,812	&	355,816	& 	11,933	& 	760,381	\\								
				& &\multicolumn{1}{l}{}&																						
				$M$&	400	& 	2,082	& 	241	&	366,104	& 	2,473,432	& 	253,857	&	3,096,116	\\								
				& &\multicolumn{1}{l}{}& 																						
				$H$&	120	& 	881	& 	722	&	26,231	& 	831,013	& 	834,386	&	1,693,353	\\								
				& &\multicolumn{2}{c}{Total}&	95,848	& 	305,290	& 	60,435	&	787,548	& 	3,672,915	& 	1,103,758	&	6,025,794	\\								
				\hline	\hline
			\end{tabular}
	\end{center}																				
		\textit{Source}: IPUMS, US census 1980.\\ 		
		\textit{Note}: 	The educational categories are  $L$ corresponding to not having completed the high school; $M$ corresponding to having a high school degree, but no tertiary level degree;  																	
		 and  $H$ corresponding to holding a tertiary education diploma. 
		 Age of husbands/male partners is between 30 and 34. 	
	 																									
	\label{tab:CT1980}																									
\end{table}

\begin{table}[!htb]
	                \caption{Joint educational and racial distribution of young American  couples in \textbf{1990}}        
																											
	
					\begin{center}																					
			\begin{tabular}{llllrrrrrrr}																							
				\hline	\hline																					
				& &    & &   \multicolumn{6}{c}{Wife/female partner} &   \\ \cline{5-10}																						
				& &    & &   \multicolumn{3}{c}{Black} & \multicolumn{3}{c}{White} & \\ \cmidrule(lr){5-7} \cmidrule(lr){8-10} 																						
				& &    &Edu. & \multicolumn{1}{c}{$L$}   & \multicolumn{1}{c}{$M$} & \multicolumn{1}{c}{$H$} & \multicolumn{1}{c}{$L$}   & \multicolumn{1}{c}{$M$} & \multicolumn{1}{c}{$H$}&         $^{\text{{\normalsize{Total}}}}$ \\		\hline																				
				
				\multicolumn{1}{c}{\multirow{6}{*}{\rotatebox[origin=c]{90}{Husband/}}} &  \multicolumn{1}{c}{\multirow{6}{*}{\rotatebox[origin=c]{90}{\underline{male partner}}}}& \multicolumn{1}{c}{\multirow{3}{*}{\rotatebox[origin=c]{90}{\underline{Black}}}} & 																						
				$L$&	16,979	& 	31,727	& 	1,831	&	1,020	& 	2,084	& 	24	&	53,665	\\								
				& &\multicolumn{1}{l}{}& 																						
				$M$&	23,065	& 	266,494	& 	37,943	&	2,427	& 	19,553	& 	3,225	&	352,707	\\								
				& &\multicolumn{1}{l}{}&																						
				$H$&	747	& 	32,024	& 	37,569	&	172	& 	3,458	& 	3,843	& 	77,813	\\

				& & \multicolumn{1}{c}{\multirow{3}{*}{\rotatebox[origin=c]{90}{\underline{White}}}} & 																						
				$L$&	522	& 	733	& 	51	&	288,490	&	326,968	& 	13,244	& 	630,008	\\								
				& &\multicolumn{1}{l}{}&																						
				$M$&	221	& 	6,594	& 	1,375	&	273,737	& 	3,012,395	& 	425,235	&	3,719,557	\\								
				& &\multicolumn{1}{l}{}& 																						
				$H$&	197	& 	1,349	& 	2,163	&	12,978	& 	628,489	& 	877,171	&	1,522,347	\\								
				& &\multicolumn{2}{c}{Total}&	41,731	& 	338,921	& 	80,932	&	578,824	& 	3,992,947	& 	1,322,742	&	6,356,097	\\								
				\hline	\hline
			\end{tabular}																							
					\end{center}
			\textit{Source}: IPUMS, US census 1990.\\
			\textit{Note}: same as below Table \ref{tab:CT1980}.																		


	\label{tab:CT1990}																									
\end{table}

\begin{figure}[H]
			\caption{Decomposition of changing prevalence of \textit{inter-racial} couples in the US between 1980 and 1990.} 
			
		\begin{subfigure}{0.49\textwidth}
			\includegraphics[width=\linewidth]{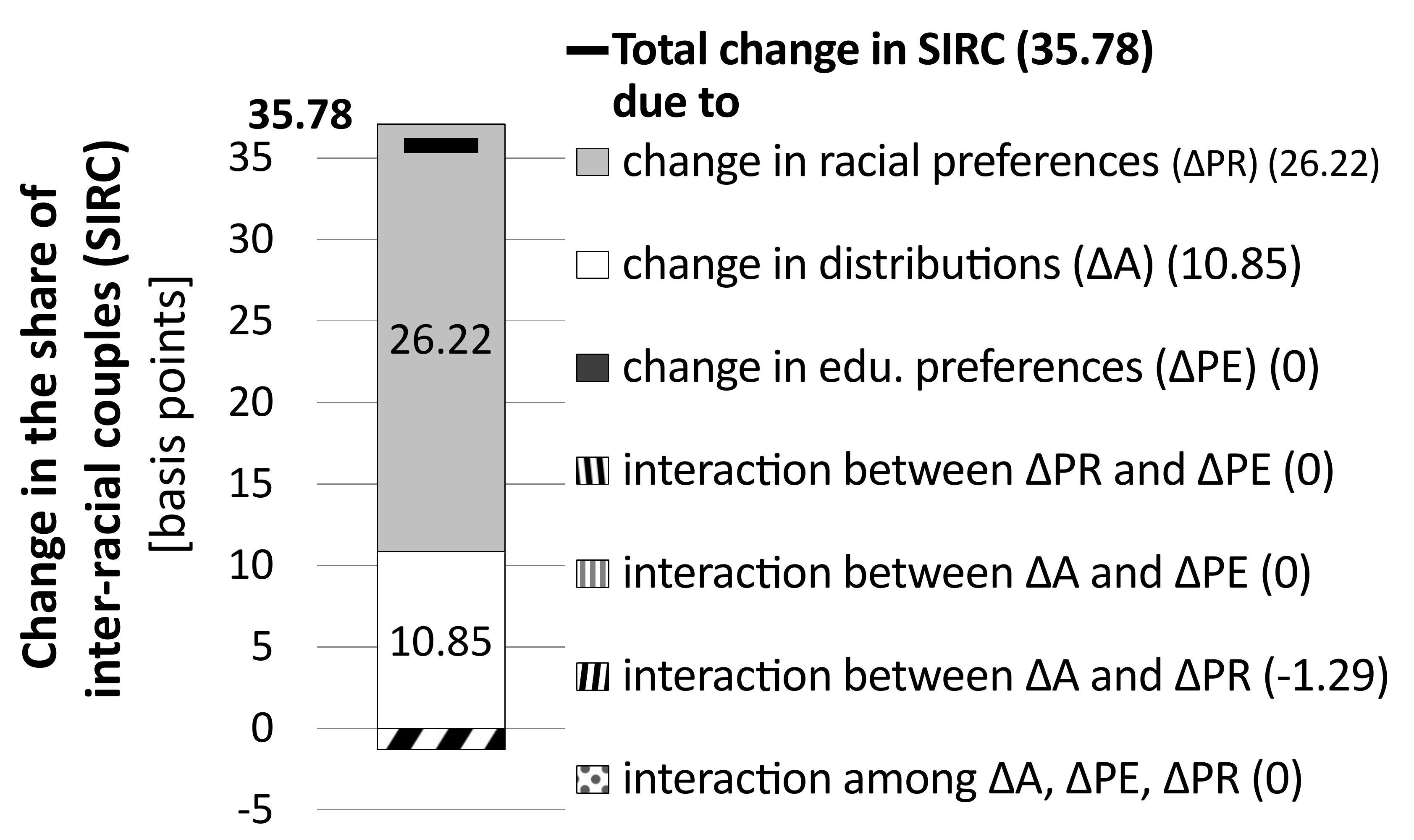}
			\caption{Decomposition by maximizing SIRC}
			\label{fig:aSIRC}
		\end{subfigure}%
		\hspace*{\fill}   
		\begin{subfigure}{0.49\textwidth}
			\includegraphics[width=\linewidth]{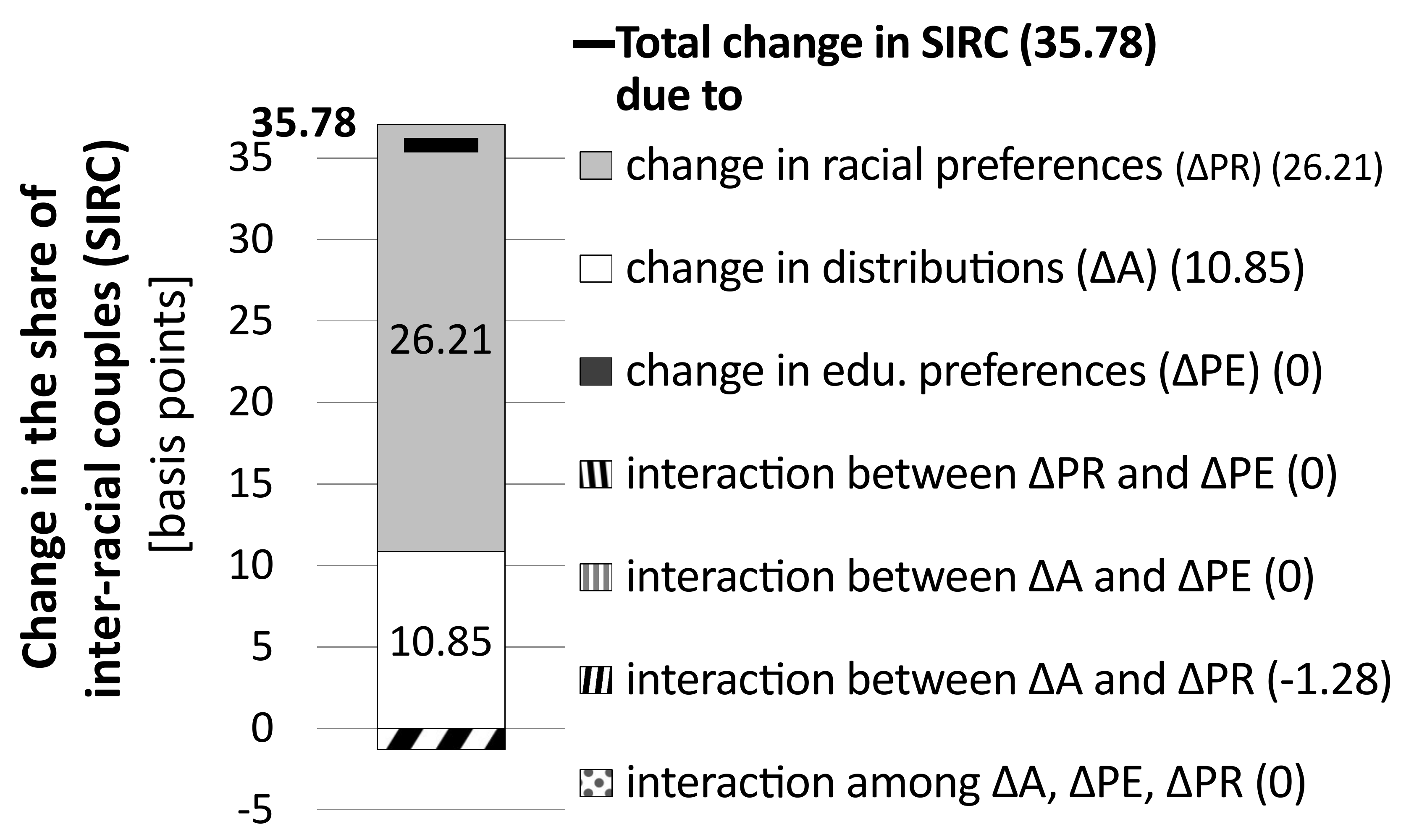}
			\caption{Decomposition by minimizing SIRC}
			\label{fig:bSIRC}
		\end{subfigure}%

		\textit{Notes}: The decomposition is conducted by using  the 
		additive decomposition scheme with interaction effect (see Eq.\ref{eq1}), while the counterfactual contingency tables are constructed with the GNM-method (see Subsection \ref{sec:GNM}) using data in Tables \ref{tab:CT1980} and \ref{tab:CT1990}.  Individuals are assumed to sort along the racial dimension first. 
		
		\label{fig:Decomp_GNM_SIRC_19801990}
		
	\end{figure}
\begin{figure}[H]
			\caption{Decomposition of changing prevalence of \textit{educational} marital homogamy in the US between 1980 and 1990.} 
				\begin{subfigure}{0.49\textwidth}
			\includegraphics[width=\linewidth]{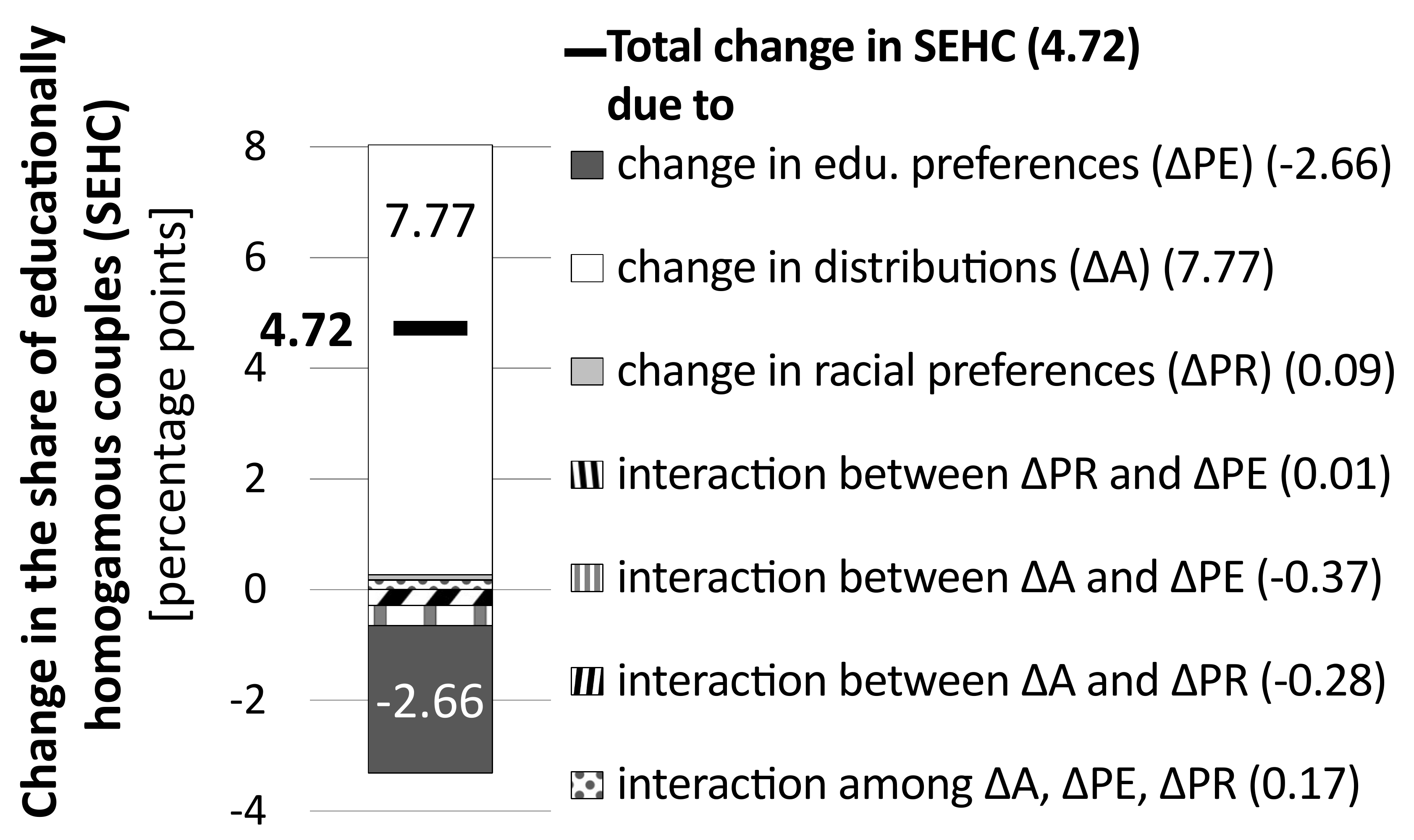}
			\caption{Decomposition by maximizing SEHC}
			\label{fig:aSEHC}
		\end{subfigure}%
		\hspace*{\fill}   
		\begin{subfigure}{0.49\textwidth}
			\includegraphics[width=\linewidth]{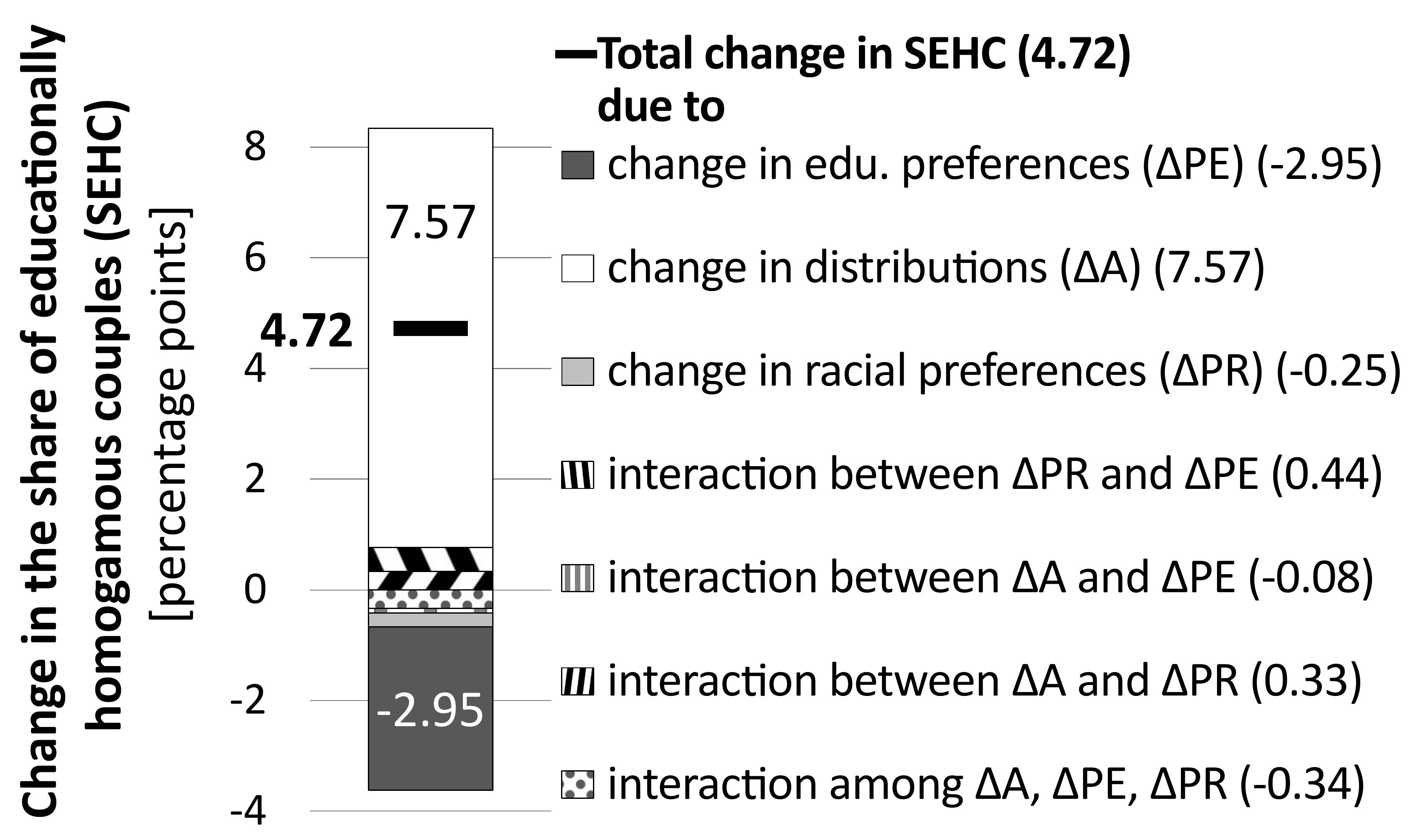}
			\caption{Decomposition by minimizing SEHC}
			\label{fig:bSEHC}
		\end{subfigure}%

		\textit{Notes}: same as under Figure \ref{fig:Decomp_GNM_SIRC_19801990}.   
		
		\label{fig:Decomp_GNM_SEHC_19801990}
		
	\end{figure}

As it is reported by Figures \ref{fig:Decomp_GNM_SIRC_19801990} and \ref{fig:Decomp_GNM_SEHC_19801990},  the SIRC increased by 35.78 basis points, while the SEHC increased by more than 4 percentage points (see the bold black markers in Figures \ref{fig:Decomp_GNM_SIRC_19801990} and \ref{fig:Decomp_GNM_SEHC_19801990}) over the 1980s. So, both the prevalence of educational homogamy and the prevalence of inter-racial marriages increased over the analyzed decade. These  facts themselves are hardly indicative about the changes in the factors of our interest, i.e., the preferences for racial exogamy and educational homogamy.   

By decomposing the 35.78 bps change in SIRC, we find the racial preferences to be the most important driver of inter-racial marriages (see the light gray bars on Figures  \ref{fig:Decomp_GNM_SIRC_19801990} a, or b in the positive range). 
In other words, the increasing prevalence of inter-racial marriages among the young American adults in the 1980s can be attributed primarily to the fact that one generation (the early boomers) with given racial marital preferences were replaced on the marriage market by another generation  (the late boomers) with more permissive racial preferences towards exogamy.   
This finding is robust to estimating the eight-dimensional $\beta$ parameter vector by minimizing the SIRC under the counterfactual, or maximizing it. 

In addition, our finding  is consistent with the results of the NORC's General Social Survey  conducted in 2017.\footnote{See: 
	\scriptsize{\url{https://www.pewresearch.org/social-trends/2017/05/18/2-public-views-on-intermarriage/}}.}  According to the survey, the share of those who 
would oppose an intermarriage in their family, is substantially lower among the late boomers than among the early boomers.\footnote{Although 
	we interpret the difference between the responses of various cohorts interviewed in the same year as evidence for the difference between the  
	attitudes of their \textit{generations}, it has an alternative interpretation as well. 
	According to the alternative view, as someone gets \textit{older}, the person is more likely 
	to oppose inter-racial marriages irrespective of which generation he or she belongs to.	 
	The choice between the 'generation-effect interpretation' and the 'age-effect interpretation' can be 
	facilitated with data from the same survey to be repeated in the coming years.}

Let us turn to the decomposition of $\Delta$SEHC.  	
We find that in the 1980s SEHC's  main driver was the changing structural availability of men and women. 
If no other factors had changed, then the SEHC would have increased by more than 7 percentage points (see the white bars of Figures  \ref{fig:Decomp_GNM_SEHC_19801990} a, or b).   
By contrast, \textit{the  preferences for spousal education are calculated to have played a negative effect on the SEHC}: if only these preferences had changed over the 1980s, then the SEHC would have decreased by almost 3 percentage points. 
This finding is robust to estimating the eight-dimensional $\beta$ parameter vector by minimizing, or maximizing the SEHC 
  (see the dark bars of Figures \ref{fig:Decomp_GNM_SEHC_19801990} a and  b). 

How can we interpret these components?   
First, \textit{the generation of the late boomers} (whose matches were observed in 1990 when being young adults) \textit{was more permissive towards inter-educational marriages than the generation of the early boomers} (whose matches were observed in 1980 when being young adults). This result is consistent with the finding of \cite{NaszodiMendonca2020}. However, unlike them, in this paper we control for the racial dimension (i.e., the changing racial composition, racial preferences and the interaction of these two factors with any other factors). Therefore, our finding is free of the criticism that the identified factor-contributions are potentially biased by omitting the racial factors.    

Second, the decompositions of $\Delta$SEHC and $\Delta$SIRC over the 1980s suggest that \textit{controlling for the racial dimension} when identifying changes in preferences for spousal education \textit{can be relevant in principle}. This is because preferences for educational homogamy and preferences for inter-racial marriages can change simultaneously. 
This is not only a theoretical possibility occurring in a hypothetical society, but an empirical fact illustrated by the American society in the 1980s.      

However,  \textit{controlling for the racial dimension does not seem to be essential in the context of the specific empirical problem studied in this paper}.  
The identified effect of changing preferences for spousal education in the 1980s is not particularly sensitive to the model specification. \cite{NaszodiMendonca2020} identify this effect on the SEHC to be  -3 percentage points without controlling for the racial dimension.\footnote{See: {Figures\_Naszodi\_Mendonca2021\_Decomposition\_US\_1980\_2010\_age3034.xlsm}  available  from \url{http://dx.doi.org/10.17632/x2ry7bcm95.2}.} In this paper, we identify the effect  to be of the same magnitude as being between -2.95 percentage points (see the dark bar of Figure \ref{fig:Decomp_GNM_SEHC_19801990} b) and -2.66 percentage points (see the dark bar of Figure \ref{fig:Decomp_GNM_SEHC_19801990}  a) in our benchmark analysis.  
Obviously, the outcome of this specific sensitivity analysis does not imply that taking into account sorting along two dimensions would be unnecessary in any other empirical application. 

\subsection{{Analyzing multiple decades}}\label{sec:dec19602015}
	
In this subsection, we study the  
long-term trend in preferences for educational homogamy, as well as the long-term trend in preferences for inter-racial marriages, between 1960 and 2015. 
We replicate the  analysis in Subsection \ref{sec:dec} for six periods (1960s, 1970s, 1980s, 1990s, 2000s and the five-years period of 2010--2015) to quantify  the period-specific contributions of the changing educational preferences to the changes in the SEHC and the 
	 period-specific contributions of the changing racial  preferences to the changes in the SIRC.\footnote{Appendix A presents the contingency tables for the census years 1960, 1970, 2000, 2010 and 2015 together with the detailed results of the decompositions.}  
	 The former defines the time series of the MEI-indicator, while the latter defines the time series of the marital racial inclusiveness-indicator (henceforth MRI-indicator).\footnote{The time series of the MRI-indicator and that of the MEI-indicator are reported by Tables \ref{tab:aggr_SIRC_BW}  and \ref{tab:aggr_SEHC_BW} in Appendix A.}



\begin{figure}[H]
		
		\caption{Historical trends of the economic divide and the racial inclusiveness based on the MEI-indicator and the MRI-indicator, respectively - sorting along education is assumed to follow sorting along race}										
\begin{center}
			
		\includegraphics[width=\linewidth]{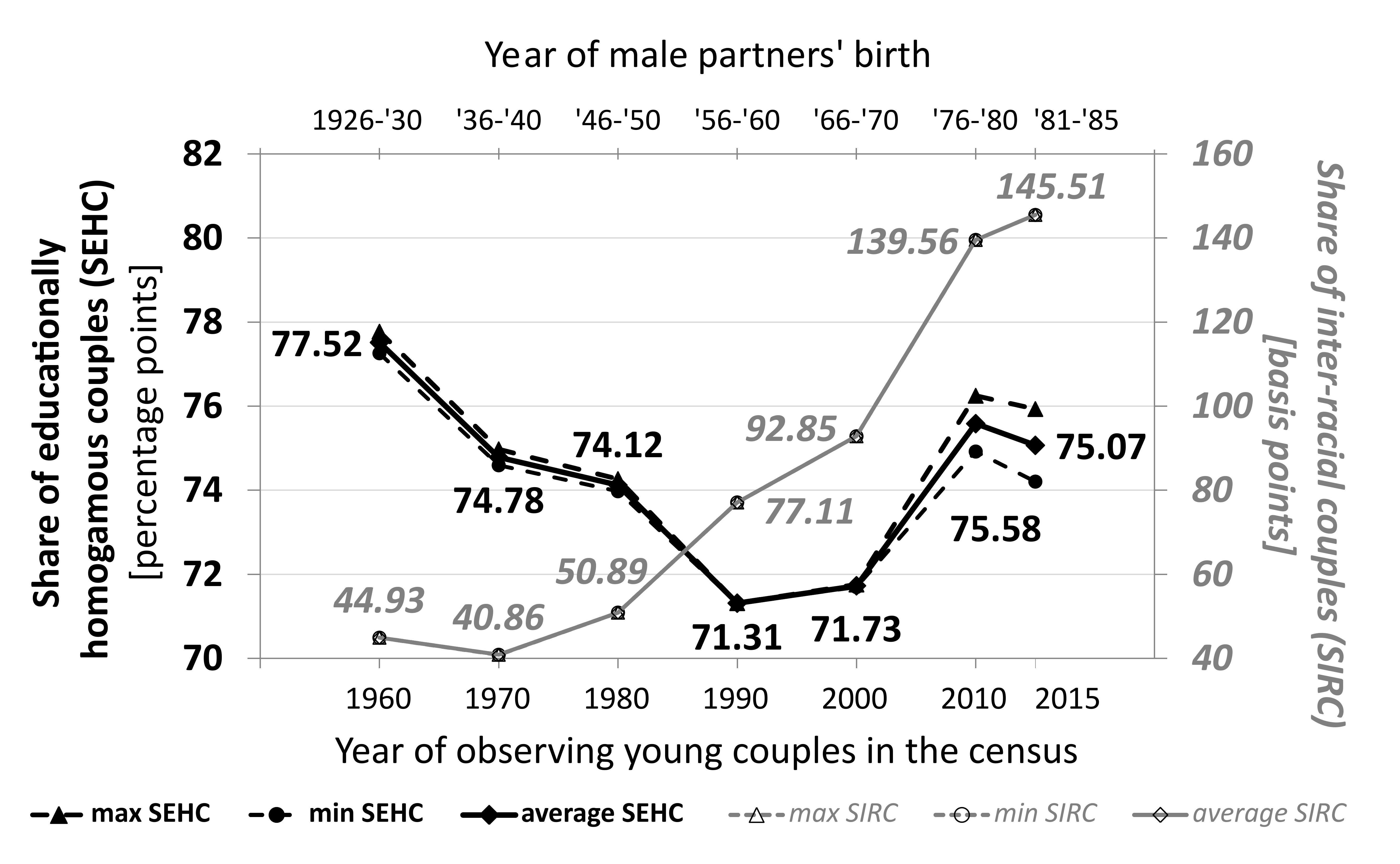}

\end{center}		
		
		\label{fig:aggr}
	\textit{Notes}: the benchmark year is 1990. The maximum and the minimum change in SIRC and SEHC attributed to changing racial or educational preferences over each decade (or 5 years in case of the period 2010--2015) is obtained by performing the decompositions with maximizing and  minimizing SEHC and SIRC under the counterfactuals. The average SEHC and average SIRC are given by the  averages  of the corresponding  maximum and minimum values. The three gray lines are coinciding, because the minimum and maximum counterfactual series  of SIRC, as well as their average series,  are hardly different from each.  
	\end{figure}

We use the MEI-indicator to study what would have been the share of educationally homogamous couples (SEHC) between 1960 and 2015 if only the educational preferences had changed across the consecutive generations relative to its value observed in the benchmark year of 1990. 
Similarly, we use the MRI-indicator to study what would have been the share of inter-racial couples (SIRC) between 1960 and 2015 if only the racial preferences had changed  across the consecutive generations  relative to the same  benchmark year. These counterfactual time series are presented by Figure \ref{fig:aggr}.

The trend of the economic divide together with the trend of the racial divide trace some social and economic changes in the US. 
Over the 1960s, Cupid's invisible hand used a double edged sword:     
while one edge could substantially decrease the economic segmentation of the market,   
the other edge shaping racial preferences has increased  the market segmentation along the racial dimension. 

In the 1990s, as well as in the first decade following the turn of the Millennium, Cupid's invisible hand used again a double edged sword.          
While it substantially decreased the racial segmentation of the market,    
it increased the market segmentation along the economic dimension. 
This finding is perfectly in line  with the main result of \cite{CaseDeaton2021}, who identify the educational gap in life expectancy to have widened while racial gap has narrowed after 1990 in the US.  

Among the analyzed periods, the 1970s and the five-years period between 2010 and 2015 are the most similar to the 1980s.  
Over these three periods, the marriage market has become less segmented both along the economic dimension and the racial dimension according to our point estimates. However, these changes are less pronounced relative to the changes in the 1980s (see the relative steepness of the gray line segments in the corresponding periods and the relative slope of the black continuous line segments in the corresponding periods in Figure \ref{fig:aggr}). In addition, the model uncertainty reflected by our interval estimates suggests that the trend of the economic divide is ambiguous after 2010.

\section{{Discussion of the empirical results}}\label{sec:disc}

In this section, we interpret our empirical findings in a broad context. In particular, we shed light on their relationship with the status-cast exchange hypothesis,  
some survey evidence, and a historian's narrative of the New Deal order and the neoliberal order.
	
\subsection{{Discussion of the results and the status-cast exchange hypothesis}}\label{sec:disc_status-cast}

As a by-product of our analysis, we can reject the dynamic, aggregate version of the \textit{status-cast exchange hypothesis}.   
The static and individual level version of this hypothesis was originally postulated by classic sociologists such as   
\cite{Davis1941} and \cite{merton1941intermarriage}. 
According to their hypothesis,  
individuals from a given race with lower social status who racially intermarry typically ``exchange'' their higher achieved socioeconomic positions for their partners' higher ascribed social status.

Similar to \cite{Rosenfeld2005}, we find even the implicit  assumption of this hypothesis highly problematic,  because race is not hierarchical. 
Accordingly,  we do not treat it as an ordered variable in our analysis. 
However,  we cannot avoid imposing the assumption that race is hierarchical, otherwise, we can neither formalize, nor test, nor reject the status-cast exchange hypothesis.    

The dynamic and aggregate version of the hypothesis is the following. When aggregate preferences for educational homogamy become weaker (i.e., the social gap between different educational groups closes, or, in other words, the high educational trait  depreciates on the marriage market), the aggregate preferences for racial endogamy should become stronger (i.e., the social gap should increase between different racial  groups) 
provided the ``terms of trade'' between the achieved status and the ascribed status is constant (as it is in our toy model in Subsection \ref{sec:NM_Prop}). 

Actually, we see just the opposite in our data: over the 1980s both types of aggregate preferences became weaker (i.e., both the social gap between different racial groups and the social gap between different educational groups were about to close. Or, in other words, the segmentation of the marriage market has decreased both along the racial dimension and the economic dimension).

The status-cast exchange hypothesis has never been popular.  
There have been many theoretical critiques of the hypothesis.  
Also, several attempts have been made to empirically refute it with mixed success (see \citealp{Rosenfeld2005} and \citealp{Kalmijn2010}). 
This is not surprising in light of the following fact. By measuring marital sorting with any of the commonly applied indicators used in  Subsection \ref{sec:NM_Prop}, one cannot reject the dynamic, aggregate version of the status-cast exchange hypothesis. 
This is because contrary to our result, the late boomers are found to be less permissive towards inter-educational marriages than the early boomers if these generations are compared on the basis of the odds-ratio, or any other measure from the same family of indicators  (see: \citealp{NaszodiMendonca2020}).

\subsection{{Discussion of the results and some survey evidence}}\label{sec:disc_survey}

It is shown in the literature that the U-shaped pattern of educational homophily, also found in this paper, is robust to being identified from certain survey data or  the census data ``interrogated'' by the NM. 
In particular, by using a survey from the Pew Research Center, \cite{NaszodiMendonca2020}  find that the strength of aggregate preferences also displays a U-shaped pattern if being identified from   individuals' self-reported  preferences on spousal education. 

\cite{Naszodi2024} and \cite{Naszodi2023} confirm this  finding. In particular, \cite{Naszodi2024} shows with a pseudo panel analysis of survey data from two waves   that the differences between the  responses of the generations studied by \cite{NaszodiMendonca2020} remain significant after controlling for potential changes in marital preferences over the course of individuals' lives.  \cite{Naszodi2023} also presents survey-based supporting evidence for the U-shaped pattern of educational homophily:  
she finds the share of the most permissive survey respondents for educational intermarriage to have displayed a hump-shaped pattern. 

As to the racial preferences,  we could use  the NORC's General Social Survey in this paper to validate our finding on the decreasing racial endogamy after 1980.   
  

Related to this approach, a question arises naturally.  
If surveys are able to track reliably the trends in the center of our interest then why do we need any model or method for the same purpose? 

We have three answers. First, the US is unique in this respect: for most of the other countries in the world,  no similar surveys are available. 
So, researchers with an international focus have to rely on population data of  couples and a method fit for the purpose.  

Second, even if there are surveys available for studying changes in preferences related to spousal education and barriers to inter-racial marriages, it is difficult to construct survey questions suitable for studying either of these two  phenomenon in itself.             

Third, small sample surveys 
are typically not suitable for studying little but important changes in the population. An example for such a phenomenon is the change in the share of inter-racial marriages (and its driving factors) of the magnitude of being  expressed in basis points. {Another example  is the moderately changing trend of the economic gap in the US between 2010 and 2015 (see Fig. \ref{fig:aggr}). 
	We could hardly identify these small changes from any survey with the usual sample size.}  

\subsection{{Discussion of the results and a historian's narrative}}\label{sec:disc_hist}

The historical trends identified with the MEI-indicator and the MRI-indicator  
are  perfectly  in line with the narrative of \cite{Gerstle2022}. 
He also finds racism and economic inequality to have had opposite  trends  before the late 1960s. 
He explains it as follows. The New Deal, while successfully addressing economic inequality, failed to address racial equality. Otherwise President Roosevelt could not have get his programme through the Congress, where the congressmen from the white South had said ``We'll let your progressive economic policy pass. As long as you don't interfere with the racial hierarchies of Southern life''.

Some decades later, President Johnson committed the Democratic party to civil rights and racial equality by passing the Civil Rights Act of 1964, the Voting Rights Act of 1965, and by deciding by the US Supreme Court in 1967 that ``anti-miscegenation'' laws (forbidding people of different races to marry) are unconstitutional. While all these three measures are important milestones for racial equality, these have contributed to gradually losing support of the Democrats by the white South. The oil price shocks in the 1970s and the war in Vietnam are named by \cite{Gerstle2022} as two additional blows on top of the racial divide that explain why the election in 1980 was lost by President Carter, and why President Reagan could gain popularity with his economic programme based on the neoliberal  paradigm. 

While the first turning point of the dark line in Figure \ref{fig:aggr} reflects the neoliberal turn, the most recent turning point, -- provided there was one around 2010 --, is probably the sign of a rising new economic paradigm. 
Future research using new data will tell whether the year 2010 can mark the beginning of a new era.  
	\section{CONCLUSION}\label{sec:concl}

	In this paper, 
	we generalized the NM-method developed by \cite{NaszodiMendonca2020}.  
	While the original NM-method is suitable for studying marital sorting along one single trait (e.g. education level, or race), 
	our generalized method, the GNM-method can be used to study sorting along two traits jointly. 
	The GNM-method can be applied, for instance,  to quantify changes in sorting along the educational trait, while controlling for sorting along the racial trait, or the other way around. 

	In the empirical part of this paper, we studied the  intergenerational changes  in Americans' marital \textit{racial} preferences between 1960 and 2015. 
	We quantified the changes in the directly unobservable preferences  through  their  effects on the share of inter-racial couples by applying a series of decompositions.   
	The effects we identified with the GNM are net of other simultaneously emerging effects, such as the effects of changes in the educational marital preferences, 
	and  the changes in availability of potential partners with various traits, 
	 as well as the joint effects of changing preferences and availability.  

	In addition, we applied the GNM-method to study the intergenerational changes in  Americans' marital preferences over spousal \textit{education}.  
	Similarly to the changes in racial preferences, we identified the changes in  preferences for spousal education with a series of decompositions by quantifying   
	their ceteris paribus effects on the share of educationally homogamous couples. 

	Our results obtained with the GNM are threefold. 
	First, by decomposing the change in the share of inter-racial couples over the 1980s, we found that 
	American late boomers were typically more permissive towards inter-racial marriages in 1990 than the early boomers were in 1980.      
	In other words, our decomposition supports the view that \textit{racial segmentation of the marriage market has moderated during the 1980s}.  
	
	Second, and most importantly, we found that after controlling for the difference in the aggregate racial preferences between the early boomers and the late boomers, as well as for the difference between their structural availability,    
	changing preferences for partners' education  exerted a downward effect on the share of educationally homogamous couples over the 1980s.    
	This finding suggests  that the \textit{social divide along the economic dimension (proxied by the education level) was less pronounced among the late boomers in 1990 than it was among the early boomers in 1980}. To sum, Cupid's invisible hand made the marriage market less segmented both along the economic dimension and the racial dimension over the 1980s.

	Third, by analyzing the marriage market between 1960 and 2015, we found that \textit{the 1980s was a unique decade}.      
	In particular, it was very different from the 1960s, 1990s, and 2000s, while it was somewhat different from the 1970s and the 2010--2015 period.  In the 1960s, 1990s,  as well as in the first decade following the turn of the Millennium,  Cupid's invisible hand used a double edged sword: when one of its edges decreased the economic segmentation of the market,   
	its other edge shaping racial preferences increased  the market segmentation along the racial dimension, or the other way around. 
	In the 1970s and  between 2010 and 2015, although the marriage market became less segmented both along the economic dimension and the racial dimension, the changes were less pronounced relative to that in the 1980s.
	Still,  we can refute the dynamic, aggregate version of the status-cast exchange hypothesis with data covering any of these three periods. 

	Our empirical findings have the following messages for policy-making.  
	Reinventing those policies that made the late boomers different from the early boomers has the potential to reverse the trend of growing economic inequality without deepening the racial divide in the US society. 
	The steadily enhancing inclusiveness along the racial dimension we identified makes it unlikely that the twenty-first century's US voters would buy a mix of a New Deal-like programme and racial hierarchies despite the fact that such a mix has been successfully sold to the twentieth century's voters by President Roosevelt.  

		\newpage
	
	\bibliography{Bib_Race}
	 	
	\newpage
	
	\setcounter{page}{1}

	\begin{center}
		
	\textbf{\LARGE{Online appendix}}\\
	   \large{of the paper entitled}\\
	 \vspace{5mm}
				\textbf{{\LARGE{A new method for identifying what Cupid's invisible hand is doing. Is it spreading color blindness while turning  us more ``picky'' about spousal education?}}}\\
				 \vspace{5mm}
	           \large{by}\\
	           \vspace{5mm}
	           \LARGE {A. Naszodi and F. Mendonca}
	
	\end{center}

 \newpage

\section*{{Appendix A: contingency tables from 1960--2015 and the detailed results of the decompositions}}\label{sec:longPer}


In this appendix, we present the joint educational and racial distributions of young American couples for 1960 (see Table \ref{tab:CT1960}), 1970 (see Table \ref{tab:CT1970}), 1980 (see Table \ref{Atab:CT1980}), 1990 (see Table \ref{Atab:CT1990}), 2000 (see Table \ref{tab:CT2000}), 2010 (see Table \ref{tab:CT2010}) and 2015 (see Table \ref{tab:CT2015}).  
In addition, we present the detailed results of the decompositions (see Tables \ref{tab:aggr_SIRC_BW} and  \ref{tab:aggr_SEHC_BW}).

%
%

\FloatBarrier
\setlength{\tabcolsep}{2pt}																										
\begin{table}[!htb]
	\caption{Joint educational and racial distribution of young American  couples in \textbf{1960}}        
	
	
					\begin{center}																					
			\begin{tabular}{llllrrrrrrr}																							
				\hline  \hline                                                                                                                                                                  
				& &    & &   \multicolumn{6}{c}{Wife/female partner} &   \\ \cline{5-10}                                                                                                                                                                                
				& &    & &   \multicolumn{3}{c}{Black} & \multicolumn{3}{c}{White} & \\ \cmidrule(lr){5-7} \cmidrule(lr){8-10}                                                                                                                                                                          
				& &    &Edu. & \multicolumn{1}{c}{$L$}   & \multicolumn{1}{c}{$M$} & \multicolumn{1}{c}{$H$} & \multicolumn{1}{c}{$L$}   											& \multicolumn{1}{c}{$M$} & \multicolumn{1}{c}{$H$}&         $^{\text{{\normalsize{Total}}}}$ \\               \hline                                                                                                                                                          
				
				\multicolumn{1}{c}{\multirow{6}{*}{\rotatebox[origin=c]{90}{Husband/}}} &  \multicolumn{1}{c}{\multirow{6}{*}										{\rotatebox[origin=c]{90}{\underline{male partner}}}}& \multicolumn{1}{c}{\multirow{3}{*}{\rotatebox[origin=c]{90}{\underline{Black}}}} &                                                                                                                                                                          
				$L$&   204,208   &   61,758   &   3,990   &   1,394   &   896   &   0   &   272,246   \\   
				& &\multicolumn{1}{l}{}&                                                                                                                                                                                
				$M$&   30,279   &   51,913   &   5,782   &   498   &   797   &   199   &   89,468   \\   
				& &\multicolumn{1}{l}{}&                                                                                                                                                                                
				$H$&   1,792   &   6,674   &   8,475   &   0   &   100   &   0   &   17,041   \\   
				
				& & \multicolumn{1}{c}{\multirow{3}{*}{\rotatebox[origin=c]{90}{\underline{White}}}} &                                                                                                                                                                          
				$L$&   1,296   &   699   &   0   &   1,133,043   &   707,071   &   13,251   &   1,855,360   \\   
				& &\multicolumn{1}{l}{}&                                                                                                                                                                                
				$M$&   100   &   399   &   0   &   420,587   &   1,277,885   &   66,433   &   1,765,404   \\   
				& &\multicolumn{1}{l}{}&                                                                                                                                                                                
				$H$&   0   &   298   &   200   &   32,461   &   427,101   &   223,858   &   683,918   \\   
				& &\multicolumn{2}{c}{Total}&   237,675   &   121,741   &   18,447   &   1,587,983   &   2,413,850   &   303,741   &   4,683,437   \\
				\hline  \hline
			\end{tabular}																					
			\end{center}
	\textit{Source}: IPUMS, US census.\\ 		
	\textit{Note}: 	The educational categories are  $L$ corresponding to not having completed the high school; $M$ corresponding to having a high school degree, but no tertiary level degree;  																	
	and  $H$ corresponding to holding a tertiary education diploma. 
	Age of husbands/male partners is between 30 and 34. 	
	
	\label{tab:CT1960}																									
\end{table}

\setlength{\tabcolsep}{2pt}																										
\begin{table}[!htb]
	\caption{Joint educational and racial distribution of young American  couples in \textbf{1970}}        
	
	
					\begin{center}																					
			\begin{tabular}{llllrrrrrrr}																							
		\hline  \hline                                                                                                                                                                  
		& &    & &   \multicolumn{6}{c}{Wife/female partner} &   \\ \cline{5-10}                                                                                                                                                                                
		& &    & &   \multicolumn{3}{c}{Black} & \multicolumn{3}{c}{White} & \\ \cmidrule(lr){5-7} \cmidrule(lr){8-10}                                                                                                                                                                          
		& &    &Edu. & \multicolumn{1}{c}{$L$}   & \multicolumn{1}{c}{$M$} & \multicolumn{1}{c}{$H$} & \multicolumn{1}{c}{$L$}   & \multicolumn{1}{c}{$M$} & \multicolumn{1}{c}{$H$}&         $^{\text{{\normalsize{Total}}}}$ \\               \hline                                                                                                                                                          
		
		\multicolumn{1}{c}{\multirow{6}{*}{\rotatebox[origin=c]{90}{Husband/}}} &  \multicolumn{1}{c}{\multirow{6}{*}{\rotatebox[origin=c]{90}{\underline{male partner}}}}& \multicolumn{1}{c}{\multirow{3}{*}{\rotatebox[origin=c]{90}{\underline{Black}}}} &                                                                                                                                                                          
		$L$&   122,395   &   67,428   &   3,305   &   905   &   901   &   0   &   194,934   \\   
		& &\multicolumn{1}{l}{}&                                                                                                                                                                                
		$M$&   43,778   &   105,066   &   7,516   &   600   &   2,308   &   500   &   159,768   \\   
		& &\multicolumn{1}{l}{}&                                                                                                                                                                                
		$H$&   1,106   &   9,610   &   10,516   &   300   &   300   &   300   &   22,132   \\   
		
		& & \multicolumn{1}{c}{\multirow{3}{*}{\rotatebox[origin=c]{90}{\underline{White}}}} &                                                                                                                                                                          
		$L$&   100   &   100   &   0   &   623,133   &   472,580   &   12,817   &   1,108,730   \\   
		& &\multicolumn{1}{l}{}&                                                                                                                                                                                
		$M$&   200   &   802   &   100   &   409,333   &   1,641,091   &   96,840   &   2,148,366   \\   
		& &\multicolumn{1}{l}{}&                                                                                                                                                                                
		$H$&   100   &   502   &   100   &   22,745   &   451,532   &   318,353   &   793,332   \\   
		& &\multicolumn{2}{c}{Total}&   167,679   &   183,508   &   21,537   &   1,057,016   &   2,568,712   &   428,810   &   4,427,262   \\
		\hline  \hline
			\end{tabular}																					
			\end{center}

	\label{tab:CT1970}																									
\end{table}

\setlength{\tabcolsep}{2pt}																										
\begin{table}[!htb]
	\caption{Joint educational and racial distribution of young American  couples in \textbf{1980}}        
	
	
				\begin{center}																					
			\begin{tabular}{llllrrrrrrr}																							
				\hline	\hline																					
				& &    & &   \multicolumn{6}{c}{Wife/female partner} &   \\ \cline{5-10}																						
				& &    & &   \multicolumn{3}{c}{Black} & \multicolumn{3}{c}{White} & \\ \cmidrule(lr){5-7} \cmidrule(lr){8-10} 																						
				& &    &Edu. & \multicolumn{1}{c}{$L$}   & \multicolumn{1}{c}{$M$} & \multicolumn{1}{c}{$H$} & \multicolumn{1}{c}{$L$}   & \multicolumn{1}{c}{$M$} & \multicolumn{1}{c}{$H$}&         $^{\text{{\normalsize{Total}}}}$ \\		\hline																				
				
				\multicolumn{1}{c}{\multirow{6}{*}{\rotatebox[origin=c]{90}{Husband/}}} &  \multicolumn{1}{c}{\multirow{6}{*}{\rotatebox[origin=c]{90}{\underline{male partner}}}}& \multicolumn{1}{c}{\multirow{3}{*}{\rotatebox[origin=c]{90}{\underline{Black}}}} & 																						
				$L$&	52,624	& 	57,135	& 	2,882	&	1,261	& 	1,862	& 	161	&	115,925	\\								
				& &\multicolumn{1}{l}{}& 																						
				$M$&	40,183	& 	215,109	& 	26,318	&	1,920	& 	8,428	& 	1,341	&	293,299	\\								
				& &\multicolumn{1}{l}{}&																						
				$H$&	2,101	& 	29,703	& 	30,252	&	220	& 	2,364	& 	2,080	& 	66,720	\\								
				
				& & \multicolumn{1}{c}{\multirow{3}{*}{\rotatebox[origin=c]{90}{\underline{White}}}} & 																						
				$L$&	420	& 	380	& 	20	&	391,812	&	355,816	& 	11,933	& 	760,381	\\								
				& &\multicolumn{1}{l}{}&																						
				$M$&	400	& 	2,082	& 	241	&	366,104	& 	2,473,432	& 	253,857	&	3,096,116	\\								
				& &\multicolumn{1}{l}{}& 																						
				$H$&	120	& 	881	& 	722	&	26,231	& 	831,013	& 	834,386	&	1,693,353	\\								
				& &\multicolumn{2}{c}{Total}&	95,848	& 	305,290	& 	60,435	&	787,548	& 	3,672,915	& 	1,103,758	&	6,025,794	\\								
				\hline	\hline
			\end{tabular}																					
			\end{center}
	
	\label{Atab:CT1980}																									
\end{table}

\begin{table}[!htb]
	\caption{Joint educational and racial distribution of young American  couples in \textbf{1990}}        
	
	
					\begin{center}																					
			\begin{tabular}{llllrrrrrrr}																							
				\hline	\hline																					
				& &    & &   \multicolumn{6}{c}{Wife/female partner} &   \\ \cline{5-10}																						
				& &    & &   \multicolumn{3}{c}{Black} & \multicolumn{3}{c}{White} & \\ \cmidrule(lr){5-7} \cmidrule(lr){8-10} 																						
				& &    &Edu. & \multicolumn{1}{c}{$L$}   & \multicolumn{1}{c}{$M$} & \multicolumn{1}{c}{$H$} & \multicolumn{1}{c}{$L$}   & \multicolumn{1}{c}{$M$} & \multicolumn{1}{c}{$H$}&         $^{\text{{\normalsize{Total}}}}$ \\		\hline																				
				
				\multicolumn{1}{c}{\multirow{6}{*}{\rotatebox[origin=c]{90}{Husband/}}} &  \multicolumn{1}{c}{\multirow{6}{*}{\rotatebox[origin=c]{90}{\underline{male partner}}}}& \multicolumn{1}{c}{\multirow{3}{*}{\rotatebox[origin=c]{90}{\underline{Black}}}} & 																						
				$L$&	16,979	& 	31,727	& 	1,831	&	1,020	& 	2,084	& 	24	&	53,665	\\								
				& &\multicolumn{1}{l}{}& 																						
				$M$&	23,065	& 	266,494	& 	37,943	&	2,427	& 	19,553	& 	3,225	&	352,707	\\								
				& &\multicolumn{1}{l}{}&																						
				$H$&	747	& 	32,024	& 	37,569	&	172	& 	3,458	& 	3,843	& 	77,813	\\

				& & \multicolumn{1}{c}{\multirow{3}{*}{\rotatebox[origin=c]{90}{\underline{White}}}} & 																						
				$L$&	522	& 	733	& 	51	&	288,490	&	326,968	& 	13,244	& 	630,008	\\								
				& &\multicolumn{1}{l}{}&																						
				$M$&	221	& 	6,594	& 	1,375	&	273,737	& 	3,012,395	& 	425,235	&	3,719,557	\\								
				& &\multicolumn{1}{l}{}& 																						
				$H$&	197	& 	1,349	& 	2,163	&	12,978	& 	628,489	& 	877,171	&	1,522,347	\\								
				& &\multicolumn{2}{c}{Total}&	41,731	& 	338,921	& 	80,932	&	578,824	& 	3,992,947	& 	1,322,742	&	6,356,097	\\								
				\hline	\hline
			\end{tabular}																							
					\end{center}
		

	\label{Atab:CT1990}																									
\end{table}

\setlength{\tabcolsep}{2pt}                                                                                                                                                                                                             
        \begin{table}[!htb]                                                                                                                                                                                     \caption{Joint educational and racial distribution of young American  couples in \textbf{2000}}        
        	                                        
                
                                               \begin{center}                                                                                                                                                                  
                                \begin{tabular}{llllrrrrrrr}                                                                                                                                                                                    
                                        \hline  \hline                                                                                                                                                                  
                                        & &    & &   \multicolumn{6}{c}{Wife/female partner} &   \\ \cline{5-10}                                                                                                                                                                                
                                        & &    & &   \multicolumn{3}{c}{Black} & \multicolumn{3}{c}{White} & \\ \cmidrule(lr){5-7} \cmidrule(lr){8-10}                                                                                                                                                                          
                                        & &    &Edu. & \multicolumn{1}{c}{$L$}   & \multicolumn{1}{c}{$M$} & \multicolumn{1}{c}{$H$} & \multicolumn{1}{c}{$L$}   & \multicolumn{1}{c}{$M$} & \multicolumn{1}{c}{$H$}&         $^{\text{{\normalsize{Total}}}}$ \\               \hline                                                                                                                                                          
                                        
                                        \multicolumn{1}{c}{\multirow{6}{*}{\rotatebox[origin=c]{90}{Husband/}}} &  \multicolumn{1}{c}{\multirow{6}{*}{\rotatebox[origin=c]{90}{\underline{male partner}}}}& \multicolumn{1}{c}{\multirow{3}{*}{\rotatebox[origin=c]{90}{\underline{Black}}}} &                                                                                                                                                                          
                                        $L$&   6,904   &   17,186   &   1,105   &   913   &   2,126   &   315   &   28,549   \\   
                                        & &\multicolumn{1}{l}{}&                                                                                                                                                                                
                                        $M$&   15,440   &   225,551   &   40,729   &   2,390   &   25,304   &   4,577   &   313,991   \\   
                                        & &\multicolumn{1}{l}{}&                                                                                                                                                                                
                                        $H$&   839   &   24,122   &   39,414   &   98   &   3,897   &   4,868   &   73,238   \\   
                                        
                                        & & \multicolumn{1}{c}{\multirow{3}{*}{\rotatebox[origin=c]{90}{\underline{White}}}} &                                                                                                                                                                          
                                        $L$&   195   &   463   &   45   &   157,436   &   187,298   &   12,903   &   358,340   \\   
                                        & &\multicolumn{1}{l}{}&                                                                                                                                                                                
                                        $M$&   469   &   8,023   &   1,367   &   140,842   &   1,902,425   &   439,786   &   2,492,912   \\   
                                        & &\multicolumn{1}{l}{}&                                                                                                                                                                                
                                        $H$&   61   &   1,576   &   2,801   &   6,929   &   383,492   &   929,669   &   1,324,528   \\   
                                        & &\multicolumn{2}{c}{Total}&   23,908   &   276,921   &   85,461   &   308,608   &   2,504,542   &   1,392,118   &   4,591,558   \\
                                        \hline  \hline
                                \end{tabular}           
                                               \end{center}

                \label{tab:CT2000}
        \end{table}

\setlength{\tabcolsep}{2pt}                                                                                                                                                                                                             
        \begin{table}[!htb]                                                                                                                                                                                     \caption{Joint educational and racial distribution of young American  couples in \textbf{2010}}        
        	                                        
                
                                               \begin{center}                                                                                                                                                                  
                                \begin{tabular}{llllrrrrrrr}                                                                                                                                                                                    
                                        \hline  \hline                                                                                                                                                                  
                                        & &    & &   \multicolumn{6}{c}{Wife/female partner} &   \\ \cline{5-10}                                                                                                                                                                                
                                        & &    & &   \multicolumn{3}{c}{Black} & \multicolumn{3}{c}{White} & \\ \cmidrule(lr){5-7} \cmidrule(lr){8-10}                                                                                                                                                                          
                                        & &    &Edu. & \multicolumn{1}{c}{$L$}   & \multicolumn{1}{c}{$M$} & \multicolumn{1}{c}{$H$} & \multicolumn{1}{c}{$L$}   & \multicolumn{1}{c}{$M$} & \multicolumn{1}{c}{$H$}&         $^{\text{{\normalsize{Total}}}}$ \\               \hline                                                                                                                                                          
                                        
                                        \multicolumn{1}{c}{\multirow{6}{*}{\rotatebox[origin=c]{90}{Husband/}}} &  \multicolumn{1}{c}{\multirow{6}{*}{\rotatebox[origin=c]{90}{\underline{male partner}}}}& \multicolumn{1}{c}{\multirow{3}{*}{\rotatebox[origin=c]{90}{\underline{Black}}}} &                                                                                                                                                                          
                                        $L$&   4,558   &   12,631   &   1,528   &   654   &   2,187   &   687   &   22,245   \\   
                                        & &\multicolumn{1}{l}{}&                                                                                                                                                                                
                                        $M$&   9,542   &   140,427   &   38,904   &   1,642   &   25,624   &   6,777   &   222,916   \\   
                                        & &\multicolumn{1}{l}{}&                                                                                                                                                                                
                                        $H$&   232   &   18,790   &   47,176   &   239   &   5,269   &   6,538   &   78,244   \\   
                                        
                                        & & \multicolumn{1}{c}{\multirow{3}{*}{\rotatebox[origin=c]{90}{\underline{White}}}} &                                                                                                                                                                          
                                        $L$&   482   &   608   &   237   &   153,199   &   152,920   &   13,759   &   321,205   \\   
                                        & &\multicolumn{1}{l}{}&                                                                                                                                                                                
                                        $M$&   275   &   8,858   &   3,281   &   90,567   &   1,376,757   &   482,266   &   1,962,004   \\   
                                        & &\multicolumn{1}{l}{}&                                                                                                                                                                                
                                        $H$&   0   &   1,367   &   4,567   &   4,327   &   265,891   &   948,595   &   1,224,747   \\   
                                        & &\multicolumn{2}{c}{Total}&   15,089   &   182,681   &   95,693   &   250,628   &   1,828,648   &   1,458,622   &   3,831,361   \\
                                        \hline  \hline
                                \end{tabular}           
                                               \end{center}

                \label{tab:CT2010}
        \end{table}

\setlength{\tabcolsep}{2pt}                                                                                                                                                                                                             
        \begin{table}[!htb]                                                                                                                                                                                     \caption{Joint educational and racial distribution of young American  couples in \textbf{2015}}        
        	                                        
                
                                               \begin{center}                                                                                                                                                                  
                                \begin{tabular}{llllrrrrrrr}                                                                                                                                                                                    
                                        \hline  \hline                                                                                                                                                                  
                                        & &    & &   \multicolumn{6}{c}{Wife/female partner} &   \\ \cline{5-10}                                                                                                                                                                                
                                        & &    & &   \multicolumn{3}{c}{Black} & \multicolumn{3}{c}{White} & \\ \cmidrule(lr){5-7} \cmidrule(lr){8-10}                                                                                                                                                                          
                                        & &    &Edu. & \multicolumn{1}{c}{$L$}   & \multicolumn{1}{c}{$M$} & \multicolumn{1}{c}{$H$} & \multicolumn{1}{c}{$L$}   & \multicolumn{1}{c}{$M$} & \multicolumn{1}{c}{$H$}&         $^{\text{{\normalsize{Total}}}}$ \\               \hline                                                                                                                                                          
                                        
                                        \multicolumn{1}{c}{\multirow{6}{*}{\rotatebox[origin=c]{90}{Husband/}}} &  \multicolumn{1}{c}{\multirow{6}{*}{\rotatebox[origin=c]{90}{\underline{male partner}}}}& \multicolumn{1}{c}{\multirow{3}{*}{\rotatebox[origin=c]{90}{\underline{Black}}}} &                                                                                                                                                                          
                                        $L$&   4,234   &   9,589   &   709   &   219   &   1,360   &   855   &   16,966   \\   
                                        & &\multicolumn{1}{l}{}&                                                                                                                                                                                
                                        $M$&   6,742   &   121,614   &   46,744   &   1,186   &   25,536   &   8,370   &   210,192   \\   
                                        & &\multicolumn{1}{l}{}&                                                                                                                                                                                
                                        $H$&   877   &   21,103   &   47,801   &   305   &   7,192   &   13,375   &   90,653   \\   
                                        
                                        & & \multicolumn{1}{c}{\multirow{3}{*}{\rotatebox[origin=c]{90}{\underline{White}}}} &                                                                                                                                                                          
                                        $L$&   214   &   229   &   0   &   115,950   &   131,152   &   15,198   &   262,743   \\   
                                        & &\multicolumn{1}{l}{}&                                                                                                                                                                                
                                        $M$&   586   &   9,363   &   3,317   &   71,361   &   1,226,193   &   508,341   &   1,819,161   \\   
                                        & &\multicolumn{1}{l}{}&                                                                                                                                                                                
                                        $H$&   50   &   1,799   &   4,236   &   9,708   &   270,053   &   1,058,477   &   1,344,323   \\   
                                        & &\multicolumn{2}{c}{Total}&   12,703   &   163,697   &   102,807   &   198,729   &   1,661,486   &   1,604,616   &   3,744,038   \\
                                        \hline  \hline
                                \end{tabular}  
                                                                                                                                                                                 
                                               \end{center}

                \label{tab:CT2015}
        \end{table}




		\setlength{\tabcolsep}{2pt}                                                                                                                                                                                                             																				
\begin{table}[!htb]                                                                                                                                                                                                             																				
	\caption{Results of the decompositions - outcome variable: \textit{share of inter-racial couples} (SIRC), period: 1960--2015, racial categories: Black and White, sorting is along the racial dimension first}        																				
	
	\begin{center}                                                                                                                                                                                               																				
		\begin{tabular}{lllrrrrrr}                                                                                                                                                                                    																				
			\hline  \hline                                                                                                                                                                  																				
			&	&	&	  $\;\;$ 1960-$\;$	&	 $\;\;$1970-$\;$	&	 $\;\;$1980-$\;$	&	 $\;\;$1990-$\;$	&	 $\;\;$2000-$\;$	&	 $\;\;$2010-$\;$  \\                                                                                                                                                                             				
			&	&	&	$\;\;$'70$\;\;$	&	$\;\;$'80$\;\;$	&	$\;\;$'90$\;\;$	&	$\;\;$'00$\;\;$	&	$\;\;$'10$\;\;$	&	$\;\;$'15$\;\;$   \\ \cline{4-9}                                                                                                                                                                                				
			& & \multicolumn{1}{l}{Total change in SIRC}  							&	3.65	&	22.99	&	35.78	&	52.45	&	51.30	&	27.99	\\			
			& &  {$\;\;$due to}							&		&		&		&		&		&		\\			
			\multicolumn{1}{c}{\multirow{14}{*}{\rotatebox[origin=c]{90}{Identification:}}} &   \multicolumn{1}{c}{\multirow{7}{*}{\rotatebox[origin=c]{90}{\underline{$\;\;\;$maximization$\;\;\;$}}}} &	$\;$ $\Delta$  distributions ($\Delta$A)  						&	7.87	&	13.80	&	10.85	&	34.50	&	8.47	&	22.19	\\			
			& &	$\;$ $\Delta$  racial preferences ($\Delta$PR) = MRI						&	-4.06	&	10.03	&	26.22	&	15.74	&	46.72	&	5.95	\\			
			& &	$\;$ $\Delta$  educational preferences ($\Delta$PE)						&	0.00	&	0.00	&	0.00	&	0.00	&	0.00	&	0.00	\\			
			& &	$\;$ interaction between $\Delta$A and $\Delta$PR						&	-0.16	&	-0.84	&	-1.29	&	2.21	&	-3.89	&	-0.16	\\			
			& &	$\;$ interaction between $\Delta$A and $\Delta$PE						&	0.00	&	0.00	&	0.00	&	-0.01	&	0.00	&	0.01	\\			
			& &	$\;$ interaction between $\Delta$PR and $\Delta$PE						&	0.00	&	0.00	&	0.00	&	0.00	&	0.00	&	0.00	\\			
			& &	$\;$ interaction among $\Delta$A, $\Delta$PE, $\Delta$PR						&	0.00	&	0.00	&	0.00	&	0.01	&	-0.01	&	-0.01	\\ \cline{3-9}   			
			
			&\multicolumn{1}{c}{\multirow{7}{*}{\rotatebox[origin=c]{90}{\underline{$\;\;\;$minimization$\;\;\;$}}}} &	$\;$ $\Delta$  distributions ($\Delta$A)  						&	7.87	&	13.80	&	10.85	&	34.50	&	8.47	&	22.19	\\			
			& &	$\;$ $\Delta$  racial preferences ($\Delta$PR) = MRI						&	-4.07	&	10.03	&	26.21	&	15.74	&	46.72	&	5.95	\\			
			& &	$\;$ $\Delta$  educational preferences ($\Delta$PE)						&	0.00	&	0.00	&	0.00	&	0.00	&	0.00	&	0.00	\\			
			& &	$\;$ interaction between $\Delta$A and $\Delta$PR						&	-0.15	&	-0.84	&	-1.28	&	2.22	&	-3.89	&	-0.16	\\			
			& &	$\;$ interaction between $\Delta$A and $\Delta$PE						&	0.00	&	0.01	&	0.00	&	0.00	&	0.00	&	0.00	\\			
			& &	$\;$ interaction between $\Delta$PR and $\Delta$PE						&	0.00	&	0.00	&	0.00	&	0.00	&	-0.01	&	0.00	\\			
			& &	$\;$ interaction among $\Delta$A, $\Delta$PE, $\Delta$PR						&	0.00	&	0.00	&	0.00	&	0.00	&	0.01	&	0.00	\\	\hline	\hline	
		\end{tabular}                                                                                                                                                                                   																				
	\end{center}																					
	
	\textit{Source}: Authors' calculation using US census data from IPUMS.\\																	
	\textit{Notes}: The decompositions are conducted by using the 																			
	additive decomposition scheme with interaction effects (see Eq.\ref{eq1}), while the counterfactual contingency tables are constructed with the GNM-method (see Subsection \ref{sec:GNM}). 																	
	The educational categories are  ``low level of education'' corresponding to not having completed the high school;  																	
	``medium level of education'' corresponding to having a high school degree;  																	
	and  ``high level of education'' corresponding to holding a tertiary education diploma. 																	
	Age of husbands/male partners is between 30 and 34.

	\label{tab:aggr_SIRC_BW}																		
\end{table}

	
		\setlength{\tabcolsep}{2pt}                                                                                                                                                                                                             																			
\begin{table}[!htb]                                                                                                                                                                                                             																			
	\caption{Results of the decompositions - outcome variable: \textit{share of educationally homogamous couples} (SEHC), period: 1960--2015, racial categories: Black and White, sorting is along the racial dimension first}        																	
	
	\begin{center}                                                                                                                                                                  																			
		\begin{tabular}{lllrrrrrr}                                                                                                                                                                                    																			
			\hline  \hline                                                                                                                                                                  																			
			&	&	&	  $\;\;$ 1960-$\;$	&	 $\;\;$1970-$\;$	&	 $\;\;$1980-$\;$	&	 $\;\;$1990-$\;$	&	 $\;\;$2000-$\;$	&	 $\;\;$2010-$\;$  \\                                                                                                                                                                             			
			&	&	&	$\;\;$'70$\;\;$	&	$\;\;$'80$\;\;$	&	$\;\;$'90$\;\;$	&	$\;\;$'00$\;\;$	&	$\;\;$'10$\;\;$	&	$\;\;$'15$\;\;$   \\ \cline{4-9}                                                                                                                                                                                			
			& & \multicolumn{1}{l}{Total change in SEHC}  							&	1.82	&	2.78	&	4.72	&	0.63	&	-1.02	&	-0.76	\\		
			& &  {$\;\;$due to}							&		&		&		&		&		&		\\		
			\multicolumn{1}{c}{\multirow{14}{*}{\rotatebox[origin=c]{90}{Identification:}}} &   \multicolumn{1}{c}{\multirow{7}{*}{\rotatebox[origin=c]{90}{\underline{$\;\;\;$maximization$\;\;\;$}}}} &	$\;$ $\Delta$  distributions ($\Delta$A)  						&	4.28	&	3.74	&	7.77	&	-0.35	&	-4.93	&	0.25	\\		
			& &	$\;$ $\Delta$  racial preferences ($\Delta$PR) 						&	0.06	&	0.14	&	0.09	&	-0.26	&	0.08	&	0.35	\\		
			& &	$\;$ $\Delta$  edu. preferences ($\Delta$PE) = MEI						&	-2.67	&	-0.62	&	-2.66	&	0.44	&	4.49	&	-0.32	\\		
			& &	$\;$ interaction between $\Delta$A and $\Delta$PR						&	0.14	&	-0.21	&	-0.28	&	0.50	&	0.02	&	0.02	\\		
			& &	$\;$ interaction between $\Delta$A and $\Delta$PE						&	0.30	&	-0.15	&	-0.37	&	0.97	&	-0.80	&	-0.67	\\		
			& &	$\;$ interaction between $\Delta$PR and $\Delta$PE						&	-0.10	&	-0.24	&	0.01	&	0.43	&	-1.21	&	-0.18	\\		
			& &	$\;$ interaction among $\Delta$A, $\Delta$PE, $\Delta$PR						&	-0.20	&	0.12	&	0.17	&	-1.11	&	1.31	&	-0.20	\\ \cline{3-9}   		
			
			&\multicolumn{1}{c}{\multirow{7}{*}{\rotatebox[origin=c]{90}{\underline{$\;\;\;$minimization$\;\;\;$}}}} &	$\;$ $\Delta$  distributions ($\Delta$A)  						&	4.41	&	3.68	&	7.57	&	-0.75	&	-5.43	&	-0.42	\\		
			& &	$\;$ $\Delta$  racial preferences ($\Delta$PR) 						&	-0.04	&	-0.11	&	-0.25	&	-0.45	&	-0.51	&	-0.44	\\		
			& &	$\;$ $\Delta$  edu. preferences ($\Delta$PE) = MEI						&	-2.80	&	-0.71	&	-2.95	&	0.38	&	3.22	&	-0.72	\\		
			& &	$\;$ interaction between $\Delta$A and $\Delta$PR						&	-0.05	&	0.11	&	0.33	&	0.54	&	0.65	&	0.46	\\		
			& &	$\;$ interaction between $\Delta$A and $\Delta$PE						&	0.16	&	-0.26	&	-0.08	&	0.06	&	0.58	&	0.06	\\		
			& &	$\;$ interaction between $\Delta$PR and $\Delta$PE						&	0.02	&	-0.03	&	0.44	&	0.50	&	0.54	&	0.13	\\		
			& &	$\;$ interaction among $\Delta$A, $\Delta$PE, $\Delta$PR						&	0.12	&	0.10	&	-0.34	&	0.35	&	-0.08	&	0.18	\\	\hline	\hline
		\end{tabular}                                                                                                                                                                                   																			
	\end{center}																				
	
	\textit{Source}: Authors' calculation using US census data from IPUMS.\\																
	\textit{Notes}: Same as under Table \ref{tab:aggr_SIRC_BW}.

	\label{tab:aggr_SEHC_BW}																	
\end{table}                                     																		
	
	\FloatBarrier

	\section*{{Appendix B: Sensitivity analysis with respect to the racial categories}}\label{sec:rcheck}
	

	In this appendix, we check the sensitivity of the decomposition results to our choice of the racial categories in the benchmark analysis. 
	Here, we use the racial categories White/non-White and Black/non-Black (rather than Black/White). 
	
\begin{figure}[H]
		\caption{Decomposition of changing prevalence \textit{inter-racial} couples in the US between 1980 and 1990  (Racial categories used:  \textbf{White and non-White})} 
		
		\begin{subfigure}{0.49\textwidth}
			\includegraphics[width=\linewidth]{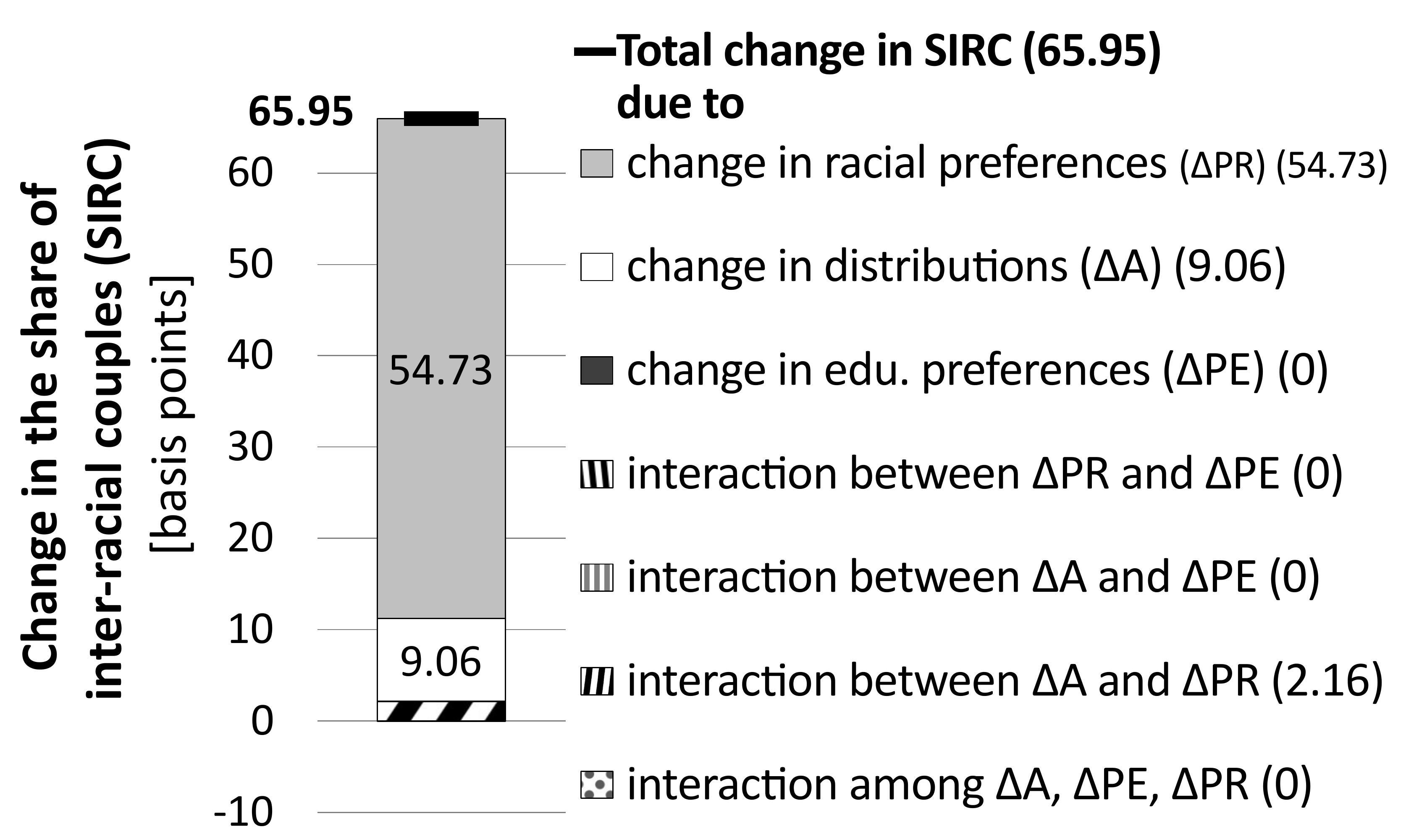}
			\caption{Decomposition by maximizing SIRC}
			\label{figApp:aSIRC}
		\end{subfigure}%
		\hspace*{\fill}   
		\begin{subfigure}{0.49\textwidth}
			\includegraphics[width=\linewidth]{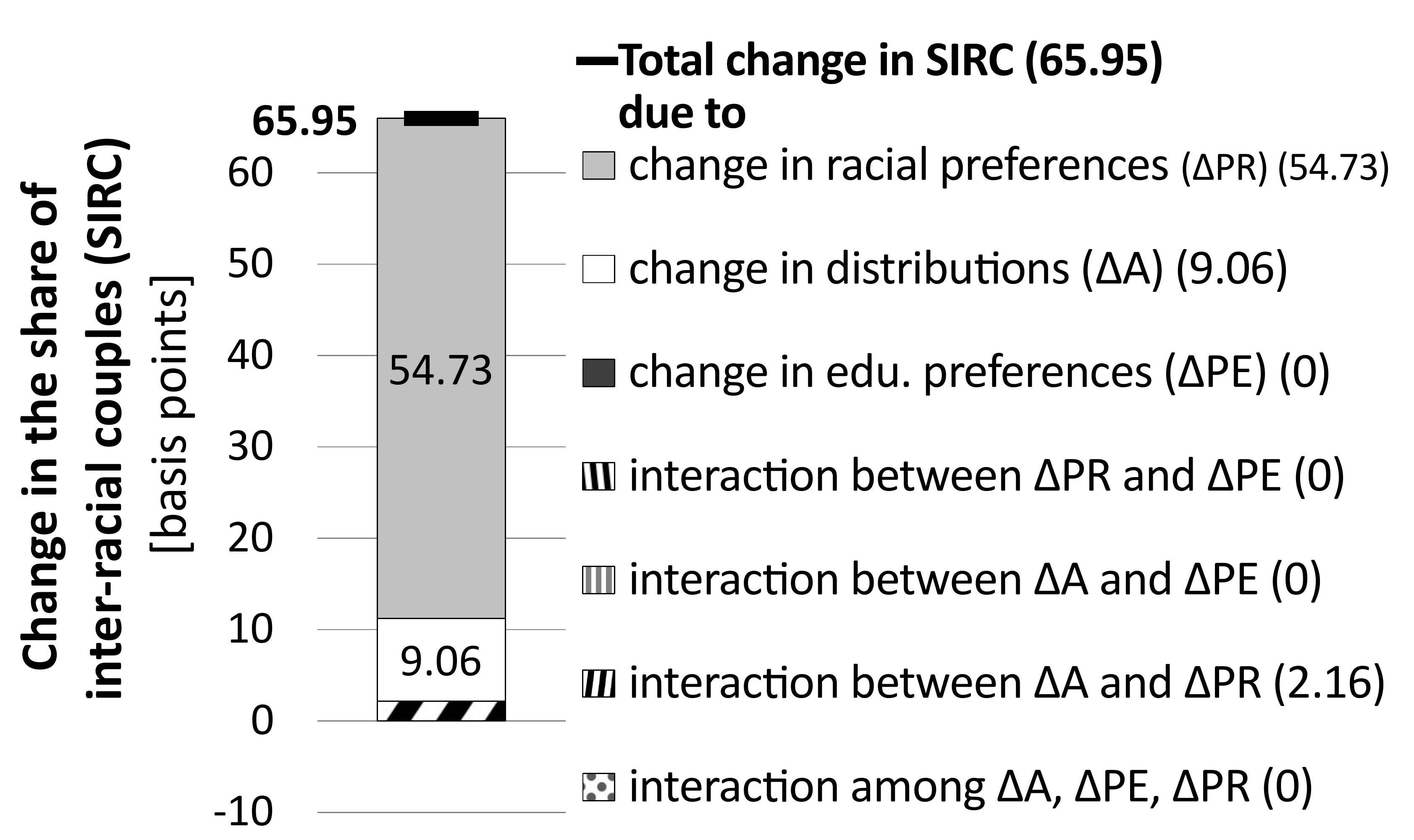}
			\caption{Decomposition by minimizing SIRC}
			\label{figApp:bSIRC}
		\end{subfigure}%
			
					\textit{Notes}: The decompositions are conducted by using  the 
			additive decomposition scheme with interaction effects (see Eq.\ref{eq1}), while the counterfactual contingency tables are constructed with the GNM-method (see Subsection \ref{sec:GNM}) using data in Tables \ref{tab:CT1980_WNW} and \ref{tab:CT1990_WNW}.  Individuals are assumed to sort along the racial dimension first.

		\label{fig:Decomp_GNM_SIRC_WNW_19801990}
		
	\end{figure}

\begin{figure}[H]
	\caption{Decomposition of changing prevalence \textit{inter-racial} couples in the US between 1980 and 1990 
		(Racial categories used:  \textbf{Black and non-Black})} 
	
		\begin{subfigure}{0.49\textwidth}
			\includegraphics[width=\linewidth]{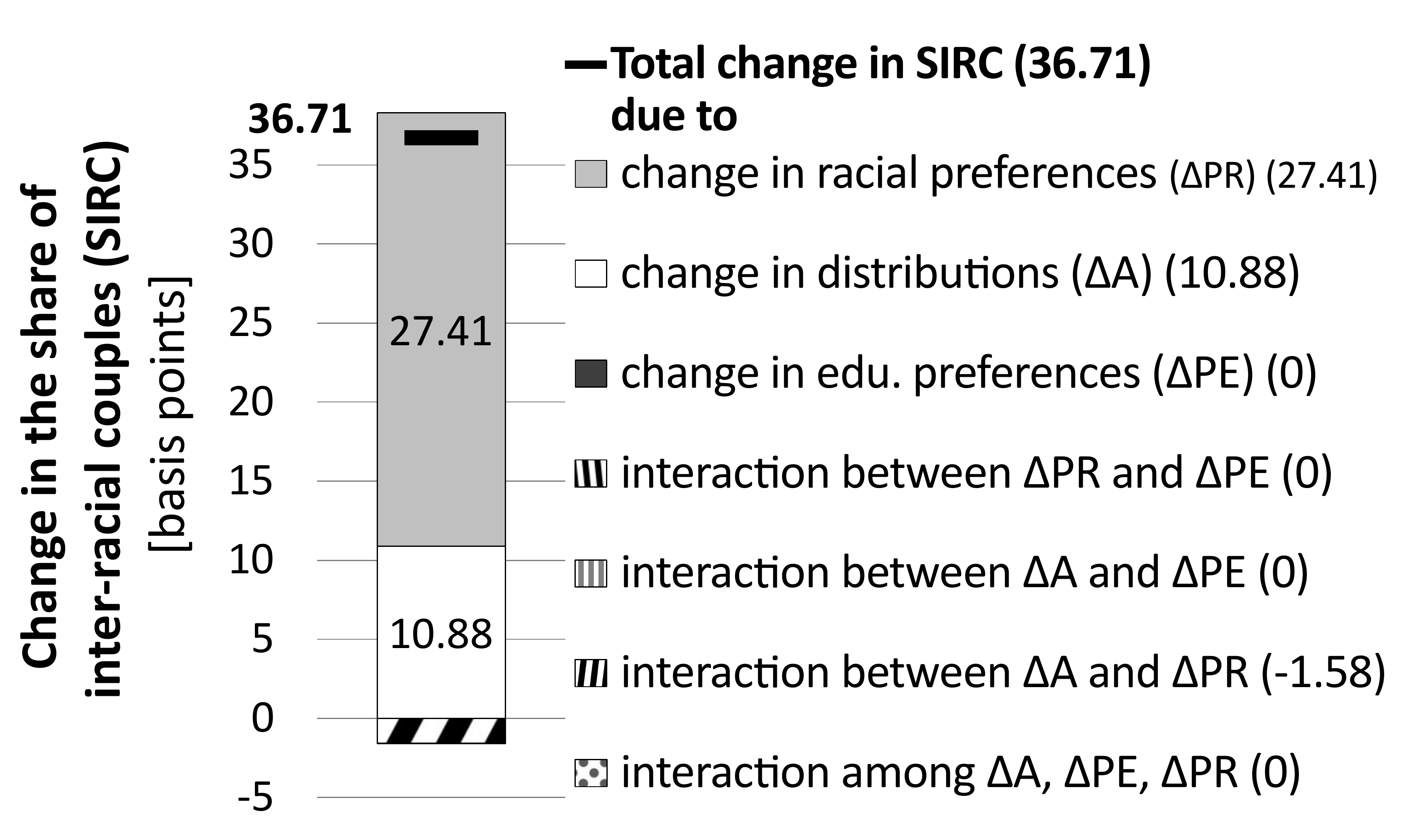}
			\caption{Decomposition by maximizing SIRC}
			\label{figApp:aSIRC}
		\end{subfigure}%
		\hspace*{\fill}   
		\begin{subfigure}{0.49\textwidth}
			\includegraphics[width=\linewidth]{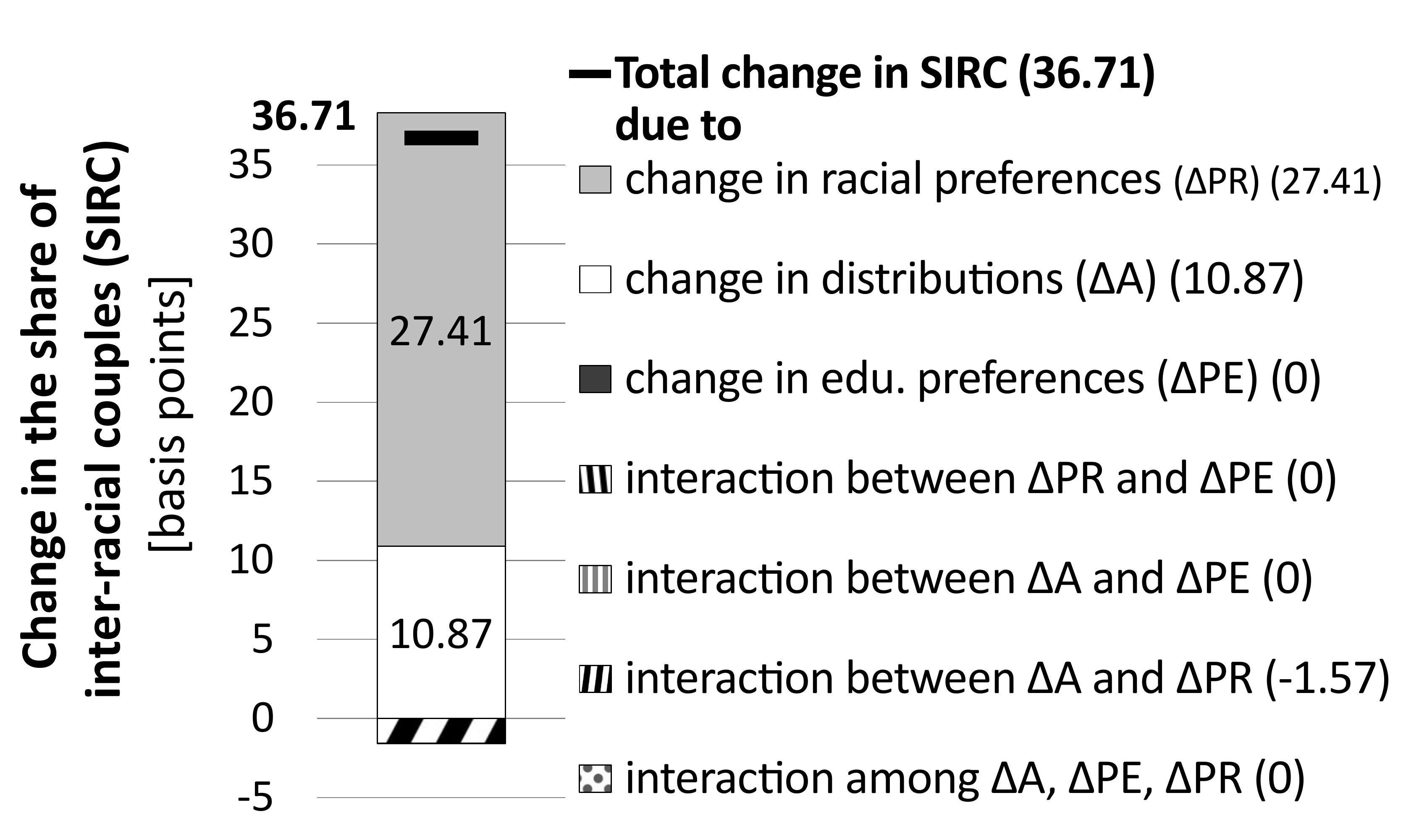}
			\caption{Decomposition by minimizing SIRC}
			\label{figApp:bSIRC}
		\end{subfigure}%

				\textit{Notes}: The decompositions are conducted by using  the 
		additive decomposition scheme with interaction effects (see Eq.\ref{eq1}), while the counterfactual contingency tables are constructed with the GNM-method (see Subsection \ref{sec:GNM}) using data in Tables \ref{tab:CT1980_BNB} and \ref{tab:CT1990_BNB}.  Individuals are assumed to sort along the racial dimension first.

		\label{fig:Decomp_GNM_SIRC_BNB_19801990}
		
	\end{figure}

	Figures \ref{fig:Decomp_GNM_SIRC_WNW_19801990} and \ref{fig:Decomp_GNM_SIRC_BNB_19801990}  present the results of our robustness checks for the SIRC with the  White/non-White categories, and with the Black/non-Black categories, respectively. In both cases, the results are qualitatively the same as in our benchmark analysis:
	the MRI-indicator has increased over the 1980s, irrespective of the racial categories used. 

\begin{figure}[H]
				\caption{Decomposition of changing prevalence of \textit{educational} marital homogamy in the US between 1980 and 1990 (Racial categories used:  \textbf{White and non-White})} 
		
		\begin{subfigure}{0.49\textwidth}
			\includegraphics[width=\linewidth]{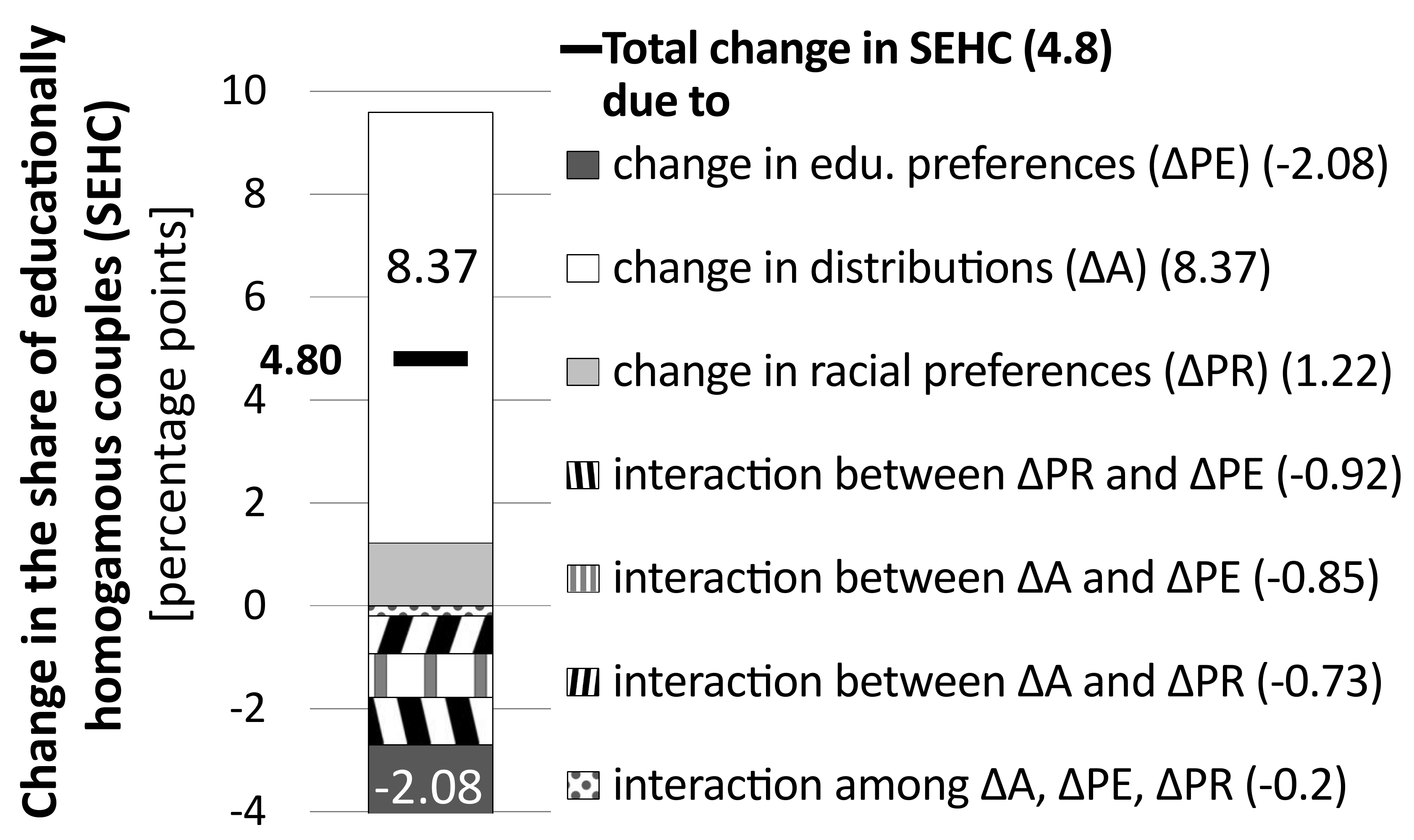}
			\caption{Decomposition by maximizing SEHC}
			\label{fig:Decomp_GNM_SEHC_WNW_19801990_a}
		\end{subfigure}%
		\hspace*{\fill}   
		\begin{subfigure}{0.49\textwidth}
			\includegraphics[width=\linewidth]{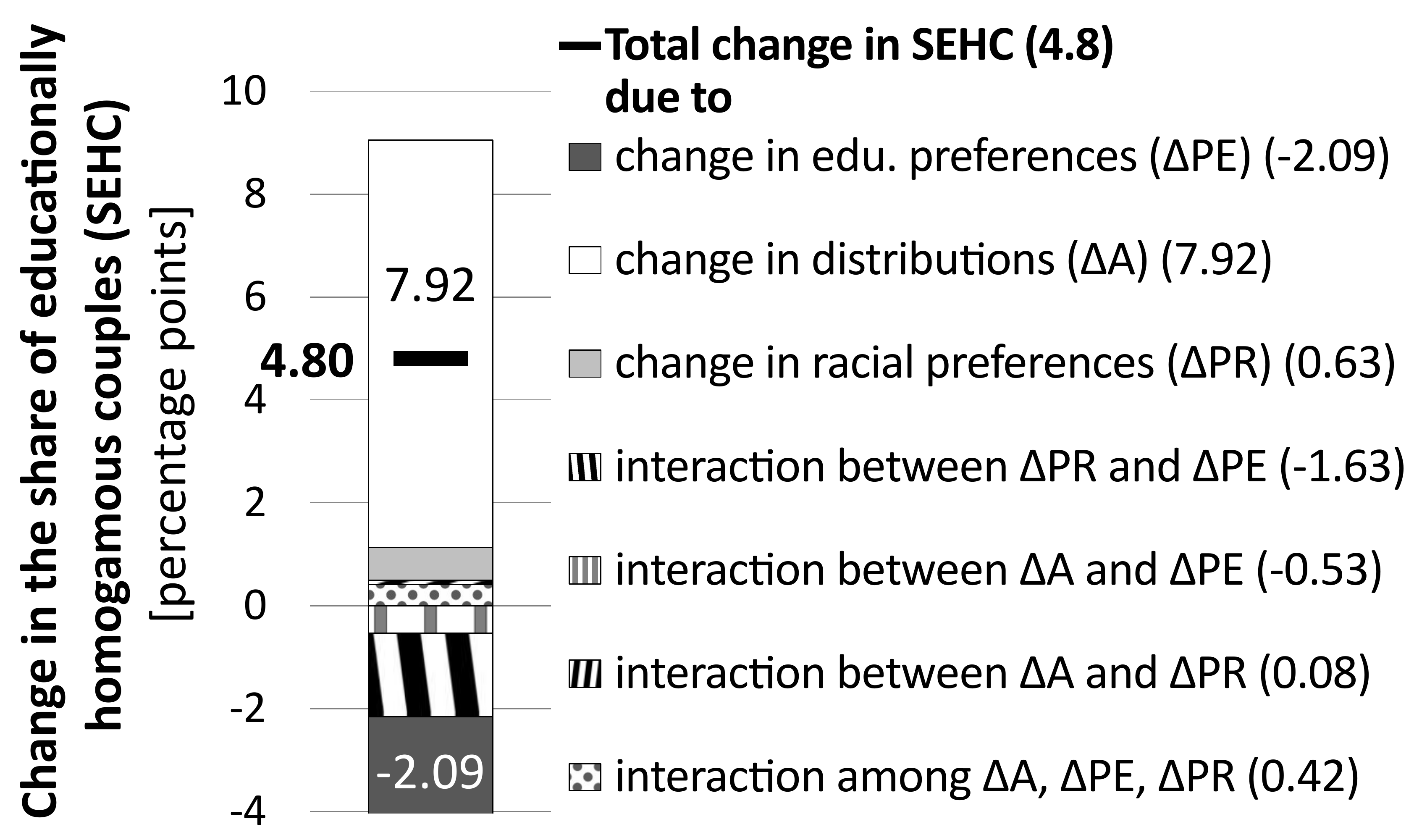}
			\caption{Decomposition by minimizing SEHC}
			\label{fig:Decomp_GNM_SEHC_WNW_19801990_b}
		\end{subfigure}%

		\textit{Notes}: The decompositions are conducted by using  the 
		additive decomposition scheme with interaction effects (see Eq.\ref{eq1}), while the counterfactual contingency tables are constructed with the GNM-method (see Subsection \ref{sec:GNM}) using data in Tables \ref{tab:CT1980_WNW} and \ref{tab:CT1990_WNW}.  Individuals are assumed to sort along the racial dimension first. 
				
		\label{fig:Decomp_GNM_SEHC_WNW_19801990}
		
	\end{figure}

\begin{figure}[H]
			\caption{Decomposition of changing prevalence of \textit{educational} marital homogamy in the US between 1980 and 1990 
			(Racial categories used:  \textbf{Black and non-Black})} 
		
		\begin{subfigure}{0.49\textwidth}
			\includegraphics[width=\linewidth]{Decomp_GNM_19801990_BNB_SEHC_Max.pdf}
			\caption{Decomposition by maximizing SEHC}
			\label{figApp:aSEHCBNB}
		\end{subfigure}%
		\hspace*{\fill}   
		\begin{subfigure}{0.49\textwidth}
			\includegraphics[width=\linewidth]{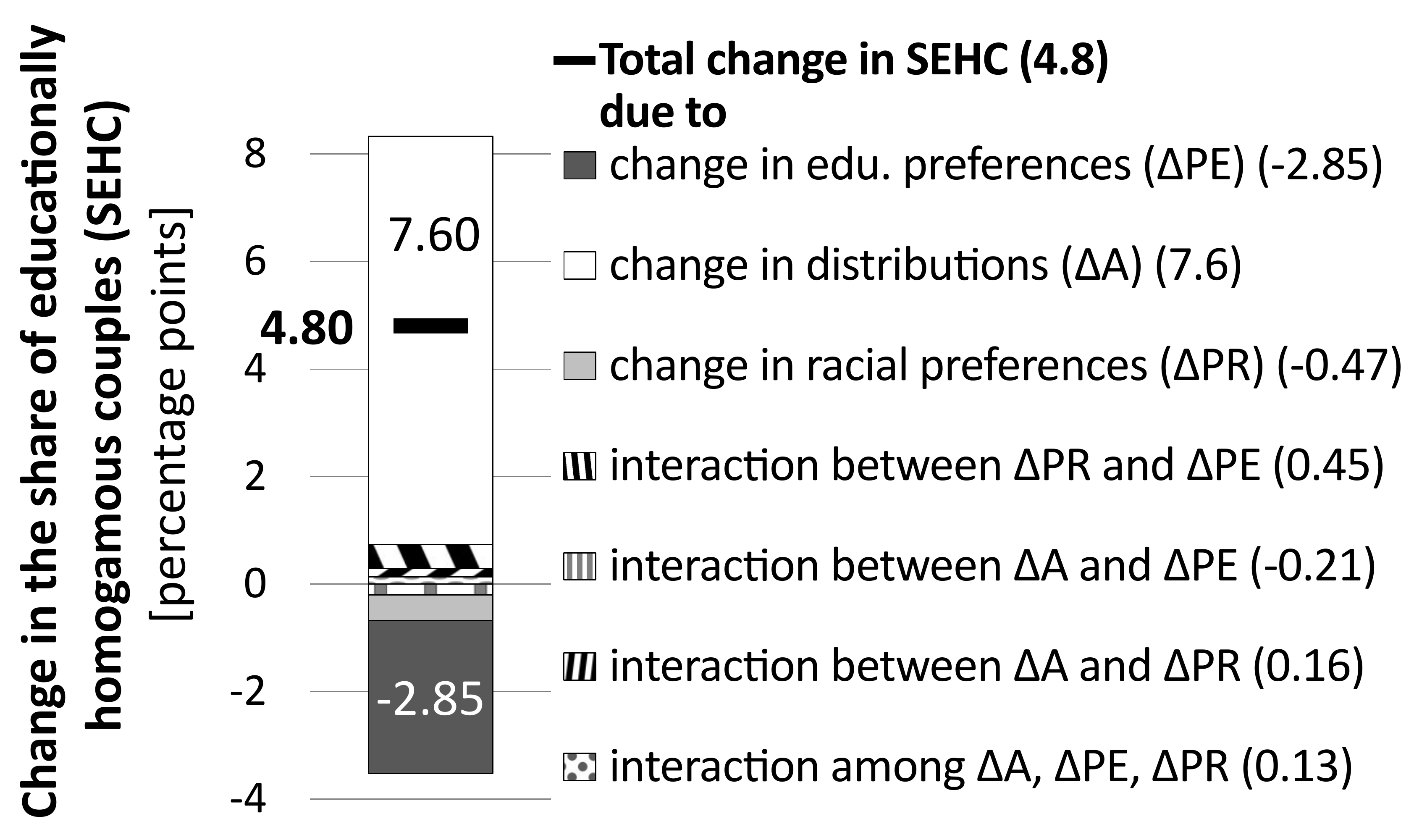}
			\caption{Decomposition by minimizing SEHC}
			\label{figApp:bSEHCBNB}
		\end{subfigure}%

				\textit{Notes}: The decompositions are conducted by using  the 
additive decomposition scheme with interaction effects (see Eq.\ref{eq1}), while the counterfactual contingency tables are constructed with the GNM-method (see Subsection \ref{sec:GNM}) using data in Tables \ref{tab:CT1980_BNB} and \ref{tab:CT1990_BNB}.  Individuals are assumed to sort along the racial dimension first.

		\label{fig:Decomp_GNM_SEHC_BNB_19801990}
		
	\end{figure}

	
	The robustness check for the SEHC is plotted in Figures \ref{fig:Decomp_GNM_SEHC_WNW_19801990} and \ref{fig:Decomp_GNM_SEHC_BNB_19801990}. The estimated contribution of changing educational preferences to the prevalence of educational homogamy is of the same magnitude and has the same sign as in the benchmark analysis. It is in the [-2.09; -2.08] percentage points range if the White/non-White categories are used, while it is calculated to be -2.7  percentage points if the Black/non-Black categories are used. To recall, the same effect was calculated to be in the [-2.66; -2.95] percentage points range in the benchmark analysis.  All in all, the MEI-indicator has decreased over the 1980s, irrespective of the racial categories used.  
	
		\FloatBarrier

\setlength{\tabcolsep}{2pt}																		
\begin{table}[!htb]																		
	\caption{Joint educational (L,M,H) and racial \textbf{(White, non-White)} distribution  of young American  couples in \textbf{1980}}        																	
	
	
	\begin{center}														
		\begin{tabular}{llllrrrrrrr}															
			\hline	\hline													
			& &    & &   \multicolumn{6}{c}{Wife/female partner} &   \\ \cline{5-10}														
			& &    & &   \multicolumn{3}{c}{non-White} & \multicolumn{3}{c}{White} & \\ \cmidrule(lr){5-7} \cmidrule(lr){8-10} 														
			& &    &Edu. & \multicolumn{1}{c}{$L$}   & \multicolumn{1}{c}{$M$} & \multicolumn{1}{c}{$H$} & \multicolumn{1}{c}{$L$}   & \multicolumn{1}{c}{$M$} & \multicolumn{1}{c}{$H$}&         $^{\text{{\normalsize{Total}}}}$ \\		\hline												
			
			\multicolumn{1}{c}{\multirow{6}{*}{\rotatebox[origin=c]{90}{Husband/}}} &  \multicolumn{1}{c}{\multirow{6}{*}{\rotatebox[origin=c]{90}{\underline{male partner}}}}& \multicolumn{1}{c}{\multirow{3}{*}{\rotatebox[origin=c]{90}{\underline{non-W}}}} & 														
			$L$&	67,771	& 	63,599	& 	3,922	&	4,463	& 	5,726	& 	341	&	145,822	\\
			& &\multicolumn{1}{l}{}& 														
			$M$&	51,108	& 	255,632	& 	36,127	&	5,983	& 	27,546	& 	3,162	&	379,558	\\
			& &\multicolumn{1}{l}{}&														
			$H$&	4,463	& 	52,378	& 	70,538	&	620	& 	8,774	& 	7,603	&	144,376	\\
			
			& & \multicolumn{1}{c}{\multirow{3}{*}{\rotatebox[origin=c]{90}{\underline{White}}}} & 														
			$L$&	4,105	& 	3,984	& 	221	&	391,812	& 	355,816	& 	11,933	&	767,871	\\
			& &\multicolumn{1}{l}{}&														
			$M$&	13,264	& 	29,803	& 	3,304	&	366,104	& 	2,473,432	& 	253,857	&	3,139,764	\\
			& &\multicolumn{1}{l}{}& 														
			$H$&	1,921	& 	9,667	& 	9,129	&	26,231	& 	831,013	& 	834,386	&	1,712,347	\\
			& &\multicolumn{2}{c}{Total}&	142,632	& 	415,063	& 	123,241	&	795,213	& 	3,702,307	& 	1,111,282	&	6,289,738	\\
			\hline	\hline													
		\end{tabular}															
	\end{center}															
	
	\label{tab:CT1980_WNW}																	
\end{table}

\begin{table}[!htb]																		
	\caption{Joint educational (L,M,H) and racial \textbf{(White, non-White)} distribution of young American  couples in \textbf{1990}}        																	
	
	
	\begin{center}													
		\begin{tabular}{llllrrrrrrr}															
			\hline	\hline													
			& &    & &   \multicolumn{6}{c}{Wife/female partner} &   \\ \cline{5-10}														
			& &    & &   \multicolumn{3}{c}{non-White} & \multicolumn{3}{c}{White} & \\ \cmidrule(lr){5-7} \cmidrule(lr){8-10} 														
			& &    &Edu. & \multicolumn{1}{c}{$L$}   & \multicolumn{1}{c}{$M$} & \multicolumn{1}{c}{$H$} & \multicolumn{1}{c}{$L$}   & \multicolumn{1}{c}{$M$} & \multicolumn{1}{c}{$H$}&         $^{\text{{\normalsize{Total}}}}$ \\		\hline												
			
			\multicolumn{1}{c}{\multirow{6}{*}{\rotatebox[origin=c]{90}{Husband/}}} &  \multicolumn{1}{c}{\multirow{6}{*}{\rotatebox[origin=c]{90}{\underline{male partner}}}}& \multicolumn{1}{c}{\multirow{3}{*}{\rotatebox[origin=c]{90}{\underline{non-W}}}} & 														
			$L$&	32,247	& 	39,317	& 	2,466	&	3,060	& 	5,708	& 	302	&	83,100	\\
			& &\multicolumn{1}{l}{}& 														
			$M$&	34,433	& 	335,016	& 	52,581	&	5,635	& 	46,716	& 	6,732	&	481,113	\\
			& &\multicolumn{1}{l}{}&														
			$H$&	2,399	& 	57,733	& 	100,060	&	331	& 	10,009	& 	11,771	&	182,303	\\
			
			& & \multicolumn{1}{c}{\multirow{3}{*}{\rotatebox[origin=c]{90}{\underline{White}}}} & 														
			$L$&	2,564	& 	4,419	& 	358	&	288,490	& 	326,968	& 	13,244	&	636,043	\\
			& &\multicolumn{1}{l}{}&														
			$M$&	7,809	& 	48,707	& 	8,646	&	273,737	& 	3,012,395	& 	425,235	&	3,776,529	\\
			& &\multicolumn{1}{l}{}& 														
			$H$&	898	& 	11,480	& 	17,995	&	12,978	& 	628,489	& 	877,171	&	1,549,011	\\
			& &\multicolumn{2}{c}{Total}&	80,350	& 	496,672	& 	182,106	&	584,231	& 	4,030,285	& 	1,334,455	&	6,708,099	\\
			\hline	\hline													

		\end{tabular}															
	\end{center}													
	

	\label{tab:CT1990_WNW}																	
\end{table}

\setlength{\tabcolsep}{2pt}																		
\begin{table}[!htb]																		
	\caption{Joint educational (L,M,H) and racial \textbf{(Black, non-Black)} distribution of young American  couples in \textbf{1980}}        																	
	
	
	\begin{center}														
		\begin{tabular}{llllrrrrrrr}															
			\hline	\hline													
			& &    & &   \multicolumn{6}{c}{Wife/female partner} &   \\ \cline{5-10}														
			& &    & &   \multicolumn{3}{c}{Black} & \multicolumn{3}{c}{non-Black} & \\ \cmidrule(lr){5-7} \cmidrule(lr){8-10} 														
			& &    &Edu. & \multicolumn{1}{c}{$L$}   & \multicolumn{1}{c}{$M$} & \multicolumn{1}{c}{$H$} & \multicolumn{1}{c}{$L$}   & \multicolumn{1}{c}{$M$} & \multicolumn{1}{c}{$H$}&         $^{\text{{\normalsize{Total}}}}$ \\		\hline												
			
			\multicolumn{1}{c}{\multirow{6}{*}{\rotatebox[origin=c]{90}{Husband/}}} &  \multicolumn{1}{c}{\multirow{6}{*}{\rotatebox[origin=c]{90}{\underline{male partner}}}}& \multicolumn{1}{c}{\multirow{3}{*}{\rotatebox[origin=c]{90}{\underline{Black}}}} & 														
			$L$&	52,624	& 	57,135	& 	2,882	&	1,521	& 	1,942	& 	201	&	116,305	\\
			& &\multicolumn{1}{l}{}& 														
			$M$&	40,183	& 	215,109	& 	26,318	&	3,081	& 	10,490	& 	1,541	&	296,722	\\
			& &\multicolumn{1}{l}{}&														
			$H$&	2,101	& 	29,703	& 	30,252	&	320	& 	2,885	& 	2,461	&	67,722	\\
			
			& & \multicolumn{1}{c}{\multirow{3}{*}{\rotatebox[origin=c]{90}{\underline{non-B}}}} & 														
			$L$&	560	& 	500	& 	20	&	413,446	& 	369,548	& 	13,314	&	797,388	\\
			& &\multicolumn{1}{l}{}&														
			$M$&	460	& 	2,603	& 	341	&	392,735	& 	2,558,211	& 	268,250	&	3,222,600	\\
			& &\multicolumn{1}{l}{}& 														
			$H$&	140	& 	1,041	& 	802	&	30,674	& 	868,203	& 	888,141	&	1,789,001	\\
			& &\multicolumn{2}{c}{Total}&	96,068	& 	306,091	& 	60,615	&	841,777	& 	3,811,279	& 	1,173,908	&	6,289,738	\\
			\hline	\hline													
		\end{tabular}															
	\end{center}															
	
	\label{tab:CT1980_BNB}																	
\end{table}

\begin{table}[!htb]																		
	\caption{Joint educational (L,M,H) and racial \textbf{(Black, non-Black)} distribution of young American  couples in \textbf{1990}}        																	
	
	
	\begin{center}													
		\begin{tabular}{llllrrrrrrr}															
			\hline	\hline													
			& &    & &   \multicolumn{6}{c}{Wife/female partner} &   \\ \cline{5-10}														
			& &    & &   \multicolumn{3}{c}{Black} & \multicolumn{3}{c}{non-Black} & \\ \cmidrule(lr){5-7} \cmidrule(lr){8-10} 														
			& &    &Edu. & \multicolumn{1}{c}{$L$}   & \multicolumn{1}{c}{$M$} & \multicolumn{1}{c}{$H$} & \multicolumn{1}{c}{$L$}   & \multicolumn{1}{c}{$M$} & \multicolumn{1}{c}{$H$}&         $^{\text{{\normalsize{Total}}}}$ \\		\hline												
			
			\multicolumn{1}{c}{\multirow{6}{*}{\rotatebox[origin=c]{90}{Husband/}}} &  \multicolumn{1}{c}{\multirow{6}{*}{\rotatebox[origin=c]{90}{\underline{male partner}}}}& \multicolumn{1}{c}{\multirow{3}{*}{\rotatebox[origin=c]{90}{\underline{Black}}}} & 														
			$L$&	16,979	& 	31,727	& 	1,831	&	1,137	& 	2,246	& 	24	&	53,944	\\
			& &\multicolumn{1}{l}{}& 														
			$M$&	23,065	& 	266,494	& 	37,943	&	3,258	& 	23,315	& 	3,845	&	357,920	\\
			& &\multicolumn{1}{l}{}&														
			$H$&	747	& 	32,024	& 	37,569	&	208	& 	4,013	& 	4,380	&	78,941	\\
			
			& & \multicolumn{1}{c}{\multirow{3}{*}{\rotatebox[origin=c]{90}{\underline{non-B}}}} & 														
			$L$&	630	& 	877	& 	51	&	307,615	& 	341,562	& 	14,464	&	665,199	\\
			& &\multicolumn{1}{l}{}&														
			$M$&	226	& 	7,657	& 	1,644	&	295,065	& 	3,145,368	& 	449,762	&	3,899,722	\\
			& &\multicolumn{1}{l}{}& 														
			$H$&	197	& 	1,457	& 	2,428	&	15,454	& 	670,217	& 	962,620	&	1,652,373	\\
			& &\multicolumn{2}{c}{Total}&	41,844	& 	340,236	& 	81,466	&	622,737	& 	4,186,721	& 	1,435,095	&	6,708,099	\\
			
			\hline	\hline													
		\end{tabular}															
	\end{center}													
	

	\label{tab:CT1990_BNB}																	
\end{table}

		\FloatBarrier

\subsection*{{Appendix C: Sensitivity analysis with respect to the sequence of sorting along the two traits}}\label{sec:rcheck_seq}


The benchmark analysis assumes that individuals sort into couples along the racial dimension before they sort along the educational dimension.   
The motivation for this assumption is twofold.  
First, the problem is computationally simpler under the assumption of the benchmark analysis.  
It requires to estimate  only eight parameters. By contrast, the alternative sequence of sorting involves the estimation of twelve parameters.  

Second, it is reasonable to assume that the primary trait of sorting was race in 1980, 1990,  2000, and 2015 since the LL-value characterizing the racial segmentation of the marriage market was much higher than any of the cells of the LL-matrix characterizing the economic segmentation of the market.  

In particular, the LL-measure for racial sorting is  $\text{LL}^{\text{sim}}(X K_{1980} Y^T)=0.988$, where \vspace{5mm}\\ 
matrix  $X = \scriptsize{ \begin{bmatrix}
	\bovermat{n}{1    & \cdots &  1} & \bovermat{n}{ 0  & \cdots & 0}  \\
	0    & \cdots  & 0 & 1  & \cdots  & 1  	
	\end{bmatrix} }$   and  matrix 
$Y = \scriptsize{ \begin{bmatrix}
	\bovermat{m}{1    & \cdots & 1} & \bovermat{m}{ 0  & \cdots  & 0}  \\
	0    & \cdots  & 0 & 1  & \cdots  & 1  	
	\end{bmatrix} }$ are defined as before, while $K_{1980}$ is reported by Table \ref{tab:CT1980}. 

Regarding the sorting along education,  $\text{LL}^{\text{gen}} (H K_{1980} H^T)= \scriptsize{ \begin{bmatrix}
	0.425    & 0.911  \\
	0.889 	 & 0.640  	
	\end{bmatrix} }$, where $H=\begin{bmatrix}  1	&	0	&	0	&	1	&	0	&	0	\\
0	&	1	&	0	&	0	&	1	&	0	\\
0	&	0	&	1	&	0	&	0	&	1	\end{bmatrix} $. 
Apparently, $\text{LL}^{\text{sim}}(X K_{1980} Y^T)$ is higher than any of the entries of matrix $\text{LL}^{\text{gen}} (H K_{1980} H^T)$.   

Similarly, in 1990, $\text{LL}^{\text{sim}}(X K_{1990} Y^T)=0.969$ is also higher than any of the elements of  $\text{LL}^{\text{gen}} (H K_{1990} H^T)= \scriptsize{ \begin{bmatrix}
	0.434    & 0.900  \\
	0.910 	 & 0.540  	
	\end{bmatrix} }$ (to check, see $K_{1990}$ reported by Table \ref{tab:CT1990}).	

Also,  
$\text{LL}^{\text{sim}}(X K_{2000} Y^T)=0.957$, and $\text{LL}^{\text{gen}} (H K_{2000} H^T)= \scriptsize{ \begin{bmatrix}
		0.451    & 0.885  \\
		0.922 	 & 0.556  	
		\end{bmatrix} }$. 

In 2015, $\text{LL}^{\text{sim}}(X K_{2015} Y^T)=0.923$, while 	$\text{LL}^{\text{gen}} (H K_{2015} H^T)= \scriptsize{ \begin{bmatrix}
		0.536    & 0.869  \\
		0.865 	 & 0.601  	
		\end{bmatrix} }$.

However,  in 2010, 
the joint distribution of couples along education was more different from the joint distribution under random matching (and therefore, it was more close to the hypothetical outcome of the perfectly positive match) in a certain segment of the market than the joint distribution of couples along race:  $\text{LL}^{\text{sim}}(X K_{2010} Y^T)=0.927$, while 	$\text{LL}^{\text{gen}} (H K_{2010} H^T)= \scriptsize{ \begin{bmatrix}
		0.558    & 0.884  \\
		\textbf{0.947} 	 & 0.618  	
		\end{bmatrix} }$. 
We typeset by bold the cell of the $\text{LL}^{\text{gen}} (H K_{2010} H^T)$ higher than  $\text{LL}^{\text{sim}}(X K_{2010} Y^T)$. 
This fact raises doubt on whether race has  always been the primary trait of sorting.   

\begin{figure}[H]
	
	\caption{Historical trends of the economic divide and the racial inclusiveness based on the MEI-indicator and the MRI-indicator, respectively - \textbf{sorting along race is assumed to follow sorting along education level}}										
	
	\begin{center}
		
	\includegraphics[width=\linewidth]{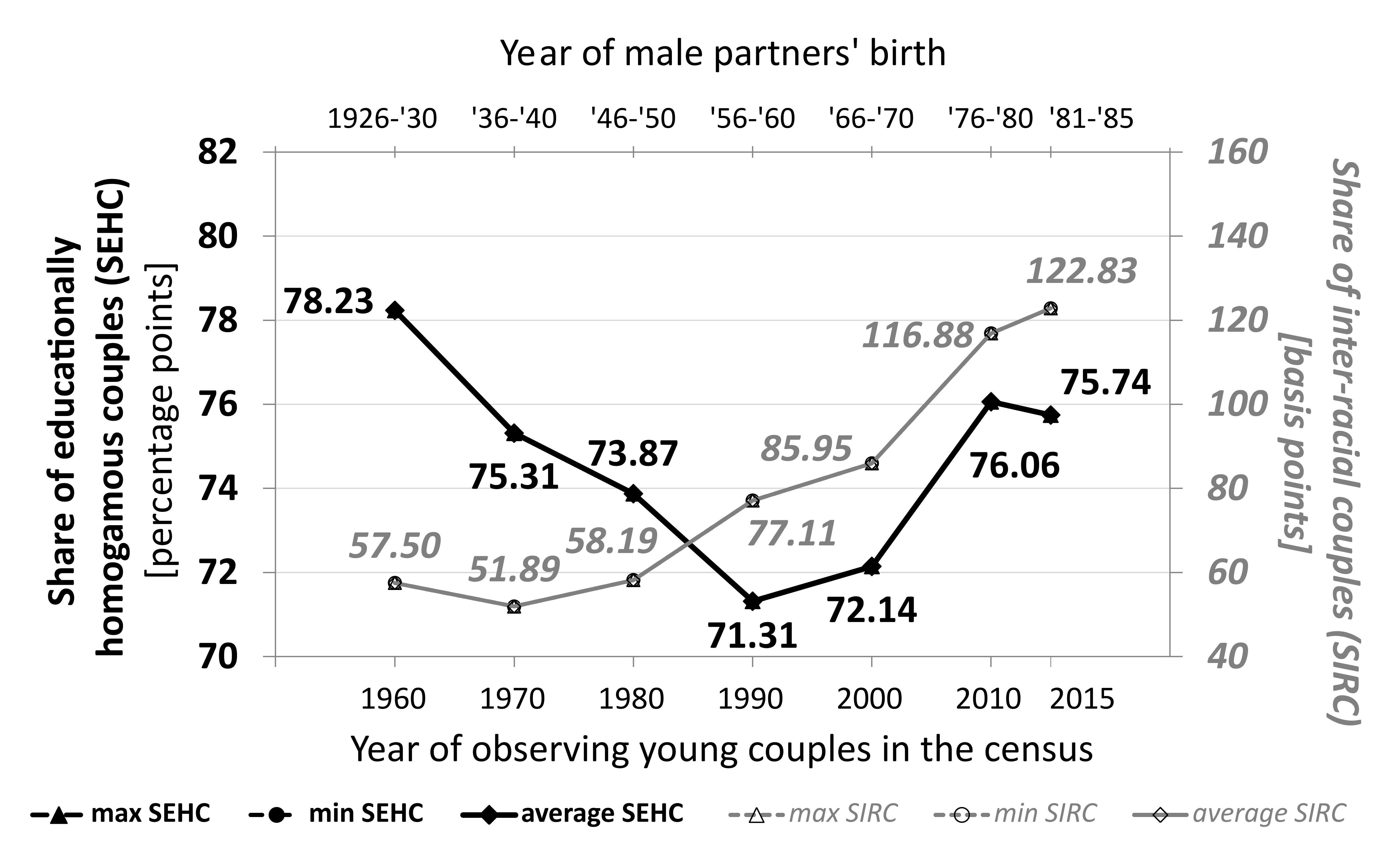}
	
	\end{center}
	\label{fig:aggr_EDURACE}

\textit{Notes}: the benchmark year is 1990. The maximum and the minimum change in SIRC and SEHC attributed to changing racial or educational preferences over each decade (or 5 years in case of the period 2010--2015) is obtained by performing the decompositions with maximizing and  minimizing SEHC and SIRC under the counterfactuals. The point estimates are the averages  of the maximum and minimum values. The three gray lines are coinciding, because the minimum and maximum counterfactual series  of SIRC, as well as their average series,  are hardly different from each other. The detailed results of the decompositions of changes in SIRC and SEHC across the consecutive generations  
are  presented by  Tables  \ref{tab:aggr_SIRC_BW_EDURAC} and  \ref{tab:aggr_SEHC_BW_EDURAC}. 
 \end{figure}

In this appendix, we assume that individuals sort into couples along the educational dimension first.  
The historical counterfactual trend of SIRC and the trend of SEHC are presented by Figure \ref{fig:aggr_EDURACE} 
under the assumption of reversed sequence of sorting (relative to the sequence of sorting assumed in the benchmark analysis).  

By comparing Figures \ref{fig:aggr} and	\ref{fig:aggr_EDURACE}, we find both the historical trend of the economic divide and the historical trend of the racial inclusiveness to be robust to the assumed sequence of sorting along the two traits.

\setlength{\tabcolsep}{2pt}                                                                                                                                                                                                             																		
\begin{table}[!htb]                                                                                                                                                                                                             																		
	\caption{Results of the decompositions - outcome variable: \textit{share of inter-racial couples} (SIRC), period: 1960--2015, racial categories: Black and White, \textbf{sorting is along the educational dimension first}}        																		
	
	\begin{center}                                                                                                                                                                                               																		
		\begin{tabular}{lllrrrrrr}                                                                                                                                                                                    																		
			\hline  \hline                                                                                                                                                                  																		
			&	&	&	  $\;\;$ 1960-$\;$	&	 $\;\;$1970-$\;$	&	 $\;\;$1980-$\;$	&	 $\;\;$1990-$\;$	&	 $\;\;$2000-$\;$	&	 $\;\;$2010-$\;$  \\                                                                                                                                                                             		
			&	&	&	$\;\;$'70$\;\;$	&	$\;\;$'80$\;\;$	&	$\;\;$'90$\;\;$	&	$\;\;$'00$\;\;$	&	$\;\;$'10$\;\;$	&	$\;\;$'15$\;\;$   \\ \cline{4-9}                                                                                                                                                                                		
			& & \multicolumn{1}{l}{Total change in SIRC}  							&	3.65	&	22.99	&	35.78	&	52.45	&	51.30	&	27.99	\\	
			& &  {$\;\;$due to}							&		&		&		&		&		&		\\	
			\multicolumn{1}{c}{\multirow{14}{*}{\rotatebox[origin=c]{90}{Identification:}}} &   \multicolumn{1}{c}{\multirow{7}{*}{\rotatebox[origin=c]{90}{\underline{$\;\;\;$maximization$\;\;\;$}}}} &	$\;$ $\Delta$  distributions ($\Delta$A)  						&	5.58	&	16.98	&	10.13	&	29.04	&	1.62	&	22.19	\\	
			& &	$\;$ $\Delta$  racial preferences ($\Delta$PR) = MRI						&	-5.61	&	6.29	&	18.92	&	8.84	&	30.92	&	5.95	\\	
			& &	$\;$ $\Delta$  educational preferences ($\Delta$PE)						&	-2.31	&	-1.94	&	-0.66	&	-3.90	&	-8.71	&	0.00	\\	
			& &	$\;$ interaction between $\Delta$A and $\Delta$PR						&	3.73	&	-2.37	&	1.40	&	7.09	&	8.10	&	-0.16	\\	
			& &	$\;$ interaction between $\Delta$A and $\Delta$PE						&	2.42	&	3.53	&	0.94	&	6.72	&	8.75	&	0.01	\\	
			& &	$\;$ interaction between $\Delta$PR and $\Delta$PE						&	1.21	&	2.09	&	-1.39	&	3.17	&	10.02	&	0.00	\\	
			& &	$\;$ interaction among $\Delta$A, $\Delta$PE, $\Delta$PR						&	-1.36	&	-1.59	&	6.43	&	1.49	&	0.59	&	-0.01	\\ \cline{3-9}   	
			
			&\multicolumn{1}{c}{\multirow{7}{*}{\rotatebox[origin=c]{90}{\underline{$\;\;\;$minimization$\;\;\;$}}}} &	$\;$ $\Delta$  distributions ($\Delta$A)  						&	5.58	&	16.98	&	10.13	&	29.04	&	1.62	&	22.19	\\	
			& &	$\;$ $\Delta$  racial preferences ($\Delta$PR) = MRI						&	-5.61	&	6.29	&	18.92	&	8.84	&	30.92	&	5.95	\\	
			& &	$\;$ $\Delta$  educational preferences ($\Delta$PE)						&	-2.31	&	-1.94	&	-0.66	&	-3.90	&	-8.71	&	0.00	\\	
			& &	$\;$ interaction between $\Delta$A and $\Delta$PR						&	3.73	&	-2.37	&	1.40	&	7.09	&	8.10	&	-0.16	\\	
			& &	$\;$ interaction between $\Delta$A and $\Delta$PE						&	2.42	&	3.53	&	0.94	&	6.72	&	8.75	&	0.01	\\	
			& &	$\;$ interaction between $\Delta$PR and $\Delta$PE						&	1.21	&	2.09	&	-1.39	&	3.17	&	10.02	&	0.00	\\	
			& &	$\;$ interaction among $\Delta$A, $\Delta$PE, $\Delta$PR						&	-1.36	&	-1.59	&	6.43	&	1.49	&	0.59	&	-0.01	\\	\hline
		\end{tabular}                                                                                                                                                                                   																		
	\end{center}																			
	
	\textit{Source}: Authors' calculation using US census data from IPUMS.\\															
	\textit{Notes}: The decompositions are conducted by using the 																	
	additive decomposition scheme with interaction effects (see Eq.\ref{eq1}), while the counterfactual contingency tables are constructed with the GNM-method (see Subsection \ref{sec:GNM}). 															
	The educational categories are  ``low level of education'' corresponding to not having completed the high school;  															
	``medium level of education'' corresponding to having a high school degree;  															
	and  ``high level of education'' corresponding to holding a tertiary education diploma. 															
	Age of husbands/male partners is between 30 and 34.

	\label{tab:aggr_SIRC_BW_EDURAC}																
\end{table}

\setlength{\tabcolsep}{2pt}                                                                                                                                                                                                             																			
\begin{table}[!htb]                                                                                                                                                                                                             																			
	\caption{Results of the decompositions - outcome variable: \textit{share of educationally homogamous couples} (SEHC), period: 1960--2015, racial categories: Black and White, \textbf{sorting is along the educational dimension first}}        																	
	
	\begin{center}                                                                                                                                                                  																			
		\begin{tabular}{lllrrrrrr}                                                                                                                                                                                    																			
			\hline  \hline                                                                                                                                                                  																			
			&	&	&	  $\;\;$ 1960-$\;$	&	 $\;\;$1970-$\;$	&	 $\;\;$1980-$\;$	&	 $\;\;$1990-$\;$	&	 $\;\;$2000-$\;$	&	 $\;\;$2010-$\;$  \\                                                                                                                                                                             			
			&	&	&	$\;\;$'70$\;\;$	&	$\;\;$'80$\;\;$	&	$\;\;$'90$\;\;$	&	$\;\;$'00$\;\;$	&	$\;\;$'10$\;\;$	&	$\;\;$'15$\;\;$   \\ \cline{4-9}                                                                                                                                                                                			
			& & \multicolumn{1}{l}{Total change in SEHC}  							&	1.82	&	2.78	&	4.72	&	0.63	&	-1.02	&	-0.76	\\		
			& &  {$\;\;$due to}							&		&		&		&		&		&		\\		
			\multicolumn{1}{c}{\multirow{14}{*}{\rotatebox[origin=c]{90}{Identification:}}} &   \multicolumn{1}{c}{\multirow{7}{*}{\rotatebox[origin=c]{90}{\underline{$\;\;\;$maximization$\;\;\;$}}}} &	$\;$ $\Delta$  distributions ($\Delta$A)  						&	4.51	&	4.23	&	7.88	&	-0.26	&	-4.81	&	0.25	\\		
			& &	$\;$ $\Delta$  racial preferences ($\Delta$PR) 						&	0.00	&	0.00	&	0.00	&	0.00	&	0.00	&	0.35	\\		
			& &	$\;$ $\Delta$  edu. preferences ($\Delta$PE) = MEI						&	-2.92	&	-1.44	&	-2.56	&	0.83	&	3.92	&	-0.32	\\		
			& &	$\;$ interaction between $\Delta$A and $\Delta$PR						&	0.00	&	0.00	&	0.00	&	0.00	&	0.00	&	0.02	\\		
			& &	$\;$ interaction between $\Delta$A and $\Delta$PE						&	0.23	&	-0.01	&	-0.60	&	0.06	&	-0.12	&	-0.67	\\		
			& &	$\;$ interaction between $\Delta$PR and $\Delta$PE						&	0.00	&	0.00	&	0.00	&	0.00	&	0.00	&	-0.18	\\		
			& &	$\;$ interaction among $\Delta$A, $\Delta$PE, $\Delta$PR						&	0.00	&	0.00	&	0.00	&	0.00	&	0.00	&	-0.20	\\ \cline{3-9}   		
			
			&\multicolumn{1}{c}{\multirow{7}{*}{\rotatebox[origin=c]{90}{\underline{$\;\;\;$minimization$\;\;\;$}}}} &	$\;$ $\Delta$  distributions ($\Delta$A)  						&	4.51	&	4.23	&	7.88	&	-0.26	&	-4.81	&	0.25	\\		
			& &	$\;$ $\Delta$  racial preferences ($\Delta$PR) 						&	0.00	&	0.00	&	0.00	&	0.00	&	0.00	&	0.35	\\		
			& &	$\;$ $\Delta$  edu. preferences ($\Delta$PE) = MEI						&	-2.92	&	-1.44	&	-2.56	&	0.83	&	3.92	&	-0.32	\\		
			& &	$\;$ interaction between $\Delta$A and $\Delta$PR						&	0.00	&	0.00	&	0.00	&	0.00	&	0.00	&	0.02	\\		
			& &	$\;$ interaction between $\Delta$A and $\Delta$PE						&	0.23	&	-0.01	&	-0.60	&	0.06	&	-0.12	&	-0.67	\\		
			& &	$\;$ interaction between $\Delta$PR and $\Delta$PE						&	0.00	&	0.00	&	0.00	&	0.00	&	0.00	&	-0.18	\\		
			& &	$\;$ interaction among $\Delta$A, $\Delta$PE, $\Delta$PR						&	0.00	&	0.00	&	0.00	&	0.00	&	0.00	&	-0.20	\\	\hline	\hline
		\end{tabular}                                                                                                                                                                                   																			
	\end{center}																				
	
	\textit{Source}: Authors' calculation using US census data from IPUMS.\\																
	\textit{Notes}: Same as under Table \ref{tab:aggr_SIRC_BW_EDURAC}.

	\label{tab:aggr_SEHC_BW_EDURAC}																	
\end{table}

\end{document}